\documentclass[twocolumn,aps,pre,superscriptaddress,nofootinbib,showpacs]{revtex4}

\pdfoutput=1

\usepackage{graphicx}
\usepackage{amsfonts}
\usepackage{hyperref}
\usepackage{amsmath,amssymb}
\usepackage{bm}

\begin{document}

\title{Majorana Fermions in Semiconductor Nanowires}
\author{Tudor D. Stanescu}
\affiliation{Department of Physics, West Virginia University, Morgantown, WV 26506}
\author{Roman M. Lutchyn}
\affiliation{Station Q, Microsoft Research, Santa Barbara, CA 93106-6105}
\author{S. Das Sarma}
\affiliation{Condensed Matter Theory Center, Department of Physics, University of
Maryland, College Park, MD 20742}

\begin{abstract}
We study multiband semiconducting nanowires proximity--coupled with an s--wave superconductor and calculate the topological phase diagram as a function of the chemical potential and magnetic field. The non-trivial topological state corresponds to a superconducting phase supporting an odd number of pairs of Majorana modes localized at the ends of the wire, whereas the non-topological state corresponds to a superconducting phase with no Majoranas or with an even number of pairs of Majorana modes. Our key finding is that multiband occupancy not only lifts the stringent constraint of one-dimensionality, but also allows having higher carrier density in the nanowire. Consequently, multiband nanowires are better-suited for stabilizing the topological superconducting phase and for observing the Majorana physics. We present a detailed study of the parameter space for multiband semiconductor nanowires focusing on understanding the key experimental conditions required for the realization and detection of Majorana fermions in solid-state systems. We include various sources of disorder and characterize their effects on the stability of the topological phase. Finally, we calculate the local density of states as well as the differential tunneling conductance as functions of external parameters and predict the experimental signatures that would establish the existence of emergent Majorana zero--energy modes in solid--state systems.
\end{abstract}

\date{\today}

\maketitle

\section{Introduction}\label{SecI}

The search for Majorana fermions has become an active and exciting pursuit in condensed
matter physics~\cite{Wilczek'09, Stern'10, Franz'10, Nayak'10}. Majorana fermions, particles which are their own antiparticles, were originally envisioned by E. Majorana in 1937~\cite{Majorana'37} in the context of particle physics ({\it i.e.},  the physics of neutrinos). However, the current search for Majorana particles is mostly taking place in condensed matter systems~\cite{levi'11, Science_Majorana'11} where Majorana quasi--particles appear in electronic systems as a result of fractionalization, and as emergent modes occupying non--local zero energy states. The non--locality of these modes  provides the ability to exchange and manipulate fractionalized quasiparticles and leads to non--Abelian braiding statistics~\cite{Nayak_NPB96, read_prb'00, Ivanov_PRL'01, Kitaev'01, Stern_PRB'04, Nayak08, Aliceaetal'10}. Hence, in addition to being of paramount importance for fundamental physics, this property of the Majoranas places them at the heart of topological quantum computing schemes~\cite{Kitaev_AP03, dassarma_prl'05, Stern06a,Bonderson06a, Bravyi'06, Nayak08, tewari_PRL'08, Bonderson'10, Hassler'10, Sau_network,  Bonderson'11, Clarke'10, Flensberg'11, Hassler'11, Preskill'11, Zilberberg_PRA'08}. We mention that solid--state systems, where the Majorana mode emerges as a zero energy state of an effective (but realistic) low--energy Hamiltonian, enable the realization of the Majorana operator itself, not just of the Majorana particle. Consequently, Majorana physics in solid--state systems is in fact much more subtle than originally envisioned by E. Majorana in 1937. For example, in condensed matter systems the non--local non--Abelian topological nature of the Majorana modes that are of interest to us is a purely emergent property.

About ten years ago, Read and Green~\cite{read_prb'00} discovered that Majorana zero-energy modes can appear quite naturally in 2D chiral p--wave superconductors where these quasiparticles, localized at the vortex cores, correspond to an equal superposition of a particle and a hole. A year later, Kitaev~\cite{Kitaev'01} introduced a very simple toy model for a 1D Majorana quantum wire with localized Majorana zero--energy modes at the ends. Both these proposals involve spinless p--wave superconductors where one can explicitly demonstrate the existence of Majorana zero-energy modes by solving the corresponding mean field Hamiltonian. Recently, several groups~\cite{Fu08,Sau'10} suggested a way to engineer spinless p--wave superconductors in the laboratory using a combination of strong spin--orbit coupling and superconducting proximity effect, thus opening the possibility of realizing Majorana fermions in solid--state systems to the experimental field. The basic idea of the semiconductor/superconductor proposal~\cite{Sau'10} is that the interplay of spin--orbit interaction, s--wave superconductivity and Zeeman spin splitting could, in principle, lead to a topological superconducting phase with localized zero--energy Majorana modes in the semiconductor.  Since then, there have been many proposals for realizing solid--state Majoranas in various superconducting heterostructures~\cite{DasSarma_PRB'06, Fu08, FuKane'09, Wimmer_PRL10, Sau'10, Lee'09, Alicea_PRB10, Lutchyn'10, Oreg'10, Sau_long, lutchyn_multi, Mao'11, Potter10, Potter11, Qi_PRB10, Linder_PRL10}. Among them, the most promising ones involve quasi--1D semiconductor nanowires with strong spin--orbit interaction proximity--coupled with an s--wave superconductor~\cite{Lutchyn'10, Oreg'10, lutchyn_multi}. The main advantage of this proposal is its simplicity: it does not require any specialized new materials but rather involves a conventional semiconductor with strong  Rashba coupling such as InAs or InSb, a conventional superconductor such as Al or Nb, and an in-plane magnetic field. High quality semiconductor nanowires can be epitaxially grown (see, for example, Ref.~\cite{Delft_nanowire} for InAs and Ref.~\cite{Nilsson_Nano'09} for InSb) and are known to have a large spin--orbit interaction strength $\alpha$ as well as large Lande $g$-factor ($g_{\rm InAs}\sim 10-25$~\cite{Aleshkin'08} and $g_{\rm InSb}\sim 20-70$~\cite{Nilsson_Nano'09}). Furthermore, these materials are known to form interfaces that are highly transparent  for electrons, allowing one to induce a large superconducting gap $\Delta$~\cite{Doh2005, vanDam2006, Chrestin'97}. Thus, semiconductor nanowires show great promise for realizing and observing Majorana particles~\cite{levi'11, Science_Majorana'11}. It is important to emphasize that in the superconductor--semiconductor heterostructures the Majorana mode is constructed or engineered to exist as a zero--energy state, and as such, it should be experimentally observable in the laboratory under the right conditions.

In a strictly 1D nanowire in contact with a superconductor, the condition for driving the system into a topological superconducting phase~\cite{Lutchyn'10, Oreg'10} is $|V_x| > \sqrt{\Delta^2+\mu^2}$, where $V_x$ is the Zeeman splitting due to the in--plane magnetic field, $\mu$ is the chemical potential and $\Delta$ is the proximity--induced superconducting gap. Thus, the key to the experimental realization of Majorana fermions in this system is the ability to satisfy a certain set of requirements that ensures the stability of the Majorana bound states. The challenging task here is the ability to suppress effects of disorder, control chemical potential fluctuations as well as fluctuations of other parameters.
Given that realizing single channel (or one subband ) nanowire is quite challenging, it is natural to consider semiconductor nanowires in the regime of multi--subband occupancy. It was shown in Ref.~\cite{lutchyn_multi} that this is a promising route and the existence of Majorana fermions does not require strict one dimensionality. In fact, the  stability of the topological superconducting phase is enhanced in multiband nanowires due to the presence of ``sweet spots" (multicritical points in the topological phase diagram, see Ref.~\cite{LutchynFisher'11} for details) in the phase diagram where the system is most robust against chemical potential fluctuations~\cite{lutchyn_multi}. In this paper, we expand on these ideas and explore effects of various perturbations such as disorder in the superconductor and semiconductor, fluctuations in the tunneling matrix elements, etc. on the stability of the topological phase. Our goal is to identify parameter regimes favorable for the exploration of the Majorana physics in the laboratory. We therefore use realistic physical models and realistic values of the parameters throughout this work so that our theoretical results are of direct relevance to experiments looking for Majorana modes in nanowires.

The paper is organized as follows. We begin in Sec.~\ref{SecII} by introducing a tight--binding model for the semiconductor nanowires, and derive the superconducting proximity effect. We show that electron tunneling between semiconductor and superconductor leads to important renormalization of the parameters in the semiconductor. In Sec.~\ref{SecIII}, we calculate the low-energy spectrum in the regime of multiband occupancy and identify the topological phase diagram. In Sec.~\ref{sec:Disorder}, we study disorder effects on the stability of the topological phase. We consider several sources of disorder: short--range impurities in the superconductor, short--range and long--range impurities in the semiconductor as well as fluctuations in the tunneling matrix elements across the interface. In Sec.~\ref{Sec:experimental}, we present results for experimentally observable quantities (e.g., local density of states and tunneling conductance) calculated using realistic assumptions. Finally, we conclude in Sec.~\ref{sec:discussion} with the summary of our main results.

\section{Tight--binding model for semiconductor nanowires}\label{SecII}

\subsection{Spin--orbit interaction and Zeeman terms}\label{SecIIA}

Single--channel semiconductor (SM) nanowires have been recently proposed\cite{Lutchyn'10, Oreg'10} as a possible platform for realizing and observing Majorana physics in solid state systems. Obtaining strictly one-dimensional (1D) nanowires would raise significant practical challenges\cite{Doh2005,vanDam2006}, but this
requirement can be relaxed to a less stringent quasi--1D condition corresponding to multiband occupancy\cite{lutchyn_multi}. The physical system proposed for studying Majorana physics consists of a strongly spin-orbit interacting semiconductor, e.g., InAs and InSb, proximity-coupled to an s-wave superconductor (SC). The quasi--1D SM nanowire is strongly confined in the $\hat{z}$ direction, so that only the lowest corresponding sub-band is occupied, while the weaker confinement in the $\hat{y}$ direction is consistent with a few occupied sub-bands. Consequently, the linear dimensions of the rectangular nanowire satisfy the relation $L_z\ll L_y \ll L_x$ which is the usual physical situation in realistic semiconductor nanowires. The low-energy physics of the SM nanowire is described by the Hamiltonian
\begin{eqnarray}
H_{\rm nw} &=& H_0 +H_{\rm SOI} = \sum_{{\bm i}, {\bm j}, \sigma} t_{{\bm i}{\bm j}}c_{{\bm i}\sigma}^{\dagger}c_{{\bm j}\sigma} -\mu \sum_{{\bm i}, \sigma} c_{{\bm i}\sigma}^{\dagger}c_{{\bm i}\sigma}  \nonumber \\
&+& \frac{i \alpha}{2}\sum_{{\bm i},{\bm \delta}}\left[ c_{{\bm i}+{\bm \delta}_x}^{\dagger}\hat{\sigma}_y c_{{\bm i}} -  c_{{\bm i}+{\bm \delta}_y}^{\dagger}\hat{\sigma}_x c_{{\bm i}} + {\rm h.c.} \right],  \label{Hnw}
\end{eqnarray}
where $H_0$ includes the first two terms and describes hopping on a simple cubic lattice with lattice constant $a$ and the last term represents the Rashba spin-orbit interaction (SOI). We include only nearest-neighbor hopping with $t_{{\bm i}{\bm i}+{\bm \delta}} = -t_0$, where ${\bm \delta}$ are the nearest--neighbor position vectors. In Eq. (\ref{Hnw}), $c_{{\bm i}}^{\dagger}$ represents a spinor $c_{{\bm i}}^{\dagger}=( c_{{\bm i}\uparrow}^{\dagger}, c_{{\bm i}\downarrow}^{\dagger})$ with $c_{{\bm i}\sigma}^{\dagger}$ being electron creation operators with spin $\sigma$, $\mu$ is the chemical potential,   $\alpha$ is the Rashba coupling constant, and $\hat{\bm \sigma}=(\sigma_x,\sigma_y,\sigma_z)$ are Pauli matrices. In the long wavelength limit, ${\bm k} \rightarrow 0$, our model reduces to an effective mass Hamiltonian with $t_0 = \hbar^2a^{-2}/2 m^*$ and Rashba spin-orbit coupling $\alpha_R(k_y\hat{\sigma}_x-k_x\hat{\sigma}_y)$, where $\alpha_R = \alpha a$. In the numerical calculations we use a set of parameters consistent with the properties of InAs, $a=5.3$\AA, $m^*=0.04 m_0$, and $\alpha_R = 0.1$ eV$\cdot$\AA. The position within the cubic lattice is described by ${\bm i} = (i_x, i_y, i_z)$ with $1\leq i_{x(y,z)} \leq N_{x(y,z)}$. In the calculations we have used $N_z=10$, $N_y=250$, and $N_x$ between $10^4$ and $2\cdot 10^4$, which corresponds to a nanowire with dimensions $L_z=5$ nm, $L_y=130$ nm, and $L_x$ between $5$ $\mu$m  and $10$ $\mu$m.

A brute force diagonalization of Hamiltonian (\ref{Hnw}) on a lattice containing more than $10^7$ sites would be numerically very expensive. More importantly, the relevant energy scales in the problem are of the order of a few meV allowing to construct the low-energy model as we show below and significantly reduce the Hilbert space. The largest energy scale is given by the gap between the lowest sub-bands, which, for example, for the first and second subband and the parameters used in our calculation is $\Delta E_{sb} \approx 1.6$meV.  Consequently, we are only  interested in the low-energy  eigenstates of the Hamiltonian (\ref{Hnw}). To obtain these states, we take advantage of the fact that the eigenproblem for $H_0$ can be solved analytically and notice that $H_{\rm SOI}$ can be treated as a small perturbation. Explicitly, the eigenstates of  $H_0$ are
\begin{equation}
\psi_{{\bm n}\sigma}({\bm i}) =\prod_{\lambda=1}^3 \sqrt{\frac{2}{N_\lambda+1}}\sin\frac{\pi n_\lambda i_\lambda}{N_\lambda+1} \chi_\sigma,    \label{psin}
 \end{equation}
 where ${\bm n}=(n_x, n_y, n_z)$ with $1\leq n_\lambda \leq N_\lambda$, and $\chi_\sigma$ is an eigenstate of the $\hat{\sigma}_z$ spin operator. The corresponding eigenvalues are
 \begin{equation}
 \epsilon_{\bm n} \!=\! -2 t_0 \left( \cos\frac{\pi n_x}{N_x\!+\!1} \!+\!  \cos\frac{\pi n_y}{N_y\!+\!1} \!+\!  \cos\frac{\pi n_z}{N_z\!+\!1}-3\right)\!-\!\mu, \label{epsn}
\end{equation}
where the chemical potential in the semiconductor $\mu$ is calculated from the bottom of the band.
We project the quantum problem into the low-energy subspace spanned by the eigenstates $\psi_{{\bm n}\sigma}$ with energies below a certain cutoff value, $\epsilon_{\bm n} < \epsilon_{\rm max}$, where the cutoff energy $\epsilon_{\rm max}$ is typically of the order $15$meV, i.e., one order of magnitude larger than the inter sub--band spacing. The number of states in this low-energy basis is of the order $10^3$ and thus the numerical complexity of the problem is significantly reduced.  The matrix elements of the SOI Hamiltonian are
\begin{eqnarray}
&~&\langle\psi_{{\bm n}\sigma}|H_{\rm SOI}|\psi_{{\bm n^\prime}\sigma^\prime}\rangle = \alpha \delta_{n_z n_z^\prime}\left\{ \frac{1-(-1)^{n_x+n_x^\prime}}{N_x+1}(i\hat{\sigma}_y)_{\sigma \sigma^\prime} \right. \nonumber \\
&~&~~~~~~~~~\times \left. \frac{\sin\frac{\pi n_x}{N_x+1}\sin\frac{\pi n_x^\prime}{N_x+1}}{\cos\frac{\pi n_x}{N_x+1}-\cos\frac{\pi n_x^\prime}{N_x+1}}\delta_{n_y n_y^\prime} - [x \Leftrightarrow y] \right\}, \label{HSOI}
\end{eqnarray}
where the second term in the parentheses is obtained from the first term by exchanging the  $x$ and $y$ indices.

To realize nontrivial topological states that support Majorana modes, it is necessary that an odd number of sub-bands be occupied.~\cite{lutchyn_multi} In the simplest case of a single sub-band nanowire, this can be achieved with the help of a Zeeman field ({\it i.e. } a spin splitting )
\begin{equation}
H_{\rm Zeeman} =\Gamma\sum_{{\bm i}, \sigma,\sigma^{\prime}} c_{{\bm i}\sigma}^\dagger (\hat{\sigma}_x)_{\sigma\sigma^\prime}c_{{\bm i}\sigma^\prime}, \label{HZeeman}
\end{equation}
which opens a gap at small momenta and removes one of the helicities that characterize the spectrum of a Rashba-coupled electron system. In the multi-band nanowires the situation is more complicated (see below) but the Zeeman term is essential to avoid fermion doubling. The Zeeman term can be obtained using an external magnetic field applied along the $\hat{x}$ axis, $\Gamma = g \mu_B B_x /2$. When the chemical potential lies within one of the Zeeman gaps at ${\bm k}=0$, the  condition for odd sub--band occupation is satisfied and thus fermion doubling is avoided, which allows for the existence of Majorana modes at the ends of the nanowire. Note that in SM with a large g-factor, e.g., $g_{\rm InAs}\sim 10$ and $g_{\rm InSb}\sim 50$, relatively small in-plane magnetic fields can open a sizable gap without significantly perturbing superconductivity. This is a crucial ingredient of the present proposal which is particularly important in the context of the effect of disorder on the topological phase as we discuss in Sec.~\ref{sec:Disorder}. For example, in InAs  a magnetic field $B_x\sim 1$T corresponds to $\Gamma \sim 1$meV. Finally, we note that in the basis given by Eq. (\ref{psin}) the Zeeman term has the simple form
\begin{equation}
\langle\psi_{{\bm n}\sigma}|H_{\rm Zeeman}|\psi_{{\bm n^\prime}\sigma^\prime}\rangle =\Gamma \delta_{{\bm n}{\bm n}^\prime} \delta_{\bar{\sigma} \sigma^\prime}, \label{HZeemannn}
\end{equation}
where $\bar{\sigma}=-\sigma$.

\subsection{Proximity--induced superconductivity}\label{SecIIB}

In addition to spin-orbit interaction and Zeeman spin splitting, the only other physical ingredient necessary for creating the Majorana mode is the ordinary s-wave superconductivity which can be induced in the semiconductor by the proximity effect,  through coupling to an s-wave superconductor (SC). A model of the full system that supports the Majorana modes contains, in addition to the nanowire and Zeeman terms, Eqns. (\ref{Hnw}) and (\ref{HZeeman}), respectively, the  Hamiltonian for the superconductor, $H_{\rm SC}$, and a term describing the nanowire--SC tunneling, $H_{\rm nw-SC}$. We note that, to account quantitatively for the superconductivity induced in the nanowire, one should also include possible electron-phonon and electron-electron interactions within the SM itself, $H_{\rm int}$. These interactions may enhance or inhibit the induced effect, depending on the details of the SM material~\cite{deGennes_proximity, proximity_recent}. In this paper we do not take into account effect of interactions in the semiconductor on the proximity effect and use a simple model for the proximity effect using tunneling Hamiltonian approach~\cite{McMillan_proximity} which is appropriate for the sample geometry considered here (thin semiconductor lying on top of the superconductor). The effects of interactions on the topological superconducting phase were recently considered in Refs.~\cite{Loss'11, Sela'11, LutchynFisher'11, Stoudenmire'11}. In addition to affecting the proximity-induced SC gap, the repulsive Coulomb interactions among the SM electrons lead to an effective enhancement of the Zeeman splitting which might be favorable for inducing topological superconductivity~\cite{Stoudenmire'11}.

 Thus, the total Hamiltonian for our model of semiconductor/superconductor heterostructure is given by
\begin{equation}
H_{\rm tot} = H_{\rm nw} + H_{\rm int} + H_{\rm Zeeman} + H_{\rm SC} + H_{\rm nw-SC},  \label{Htot}
\end{equation}
where the tunneling term reads
\begin{equation}
H_{\rm nw-SC} = \sum_{{\bm i}, {\bm j}, \sigma} [\widetilde{t}_{{\bm i}, {\bm j}} c_{{\bm i}\sigma}^\dagger a_{{\bm j}\sigma} + h.c.],    \label{HnwSC}
\end{equation}
with $c_{{\bm i}\sigma}$ and $a_{{\bm j}\sigma}$ being electron destruction operators acting within the SM and SC, respectively. We assume that the matrix elements $\widetilde{t}_{{\bm i}, {\bm j}}$ couple the sites of the SC located at the interface, ${\bm j} = ({\bm r}_\parallel, z_{\rm interface})/a$, where ${\bm r}_\parallel$ is position vector in a plane parallel to the interface and $z_{\rm interface}$ is the coordinate of the interface layer in a slab geometry, to the first layer of the semiconductor wire, ${\bm i} = (i_x, i_y, 1)$.

The proximity effect can be now derived by integrating out superconducting degrees of freedom in Eq. (\ref{Htot}) and considering the resulting effective low-energy theory for the
SM. To identify the form of the effective low-energy Hamiltonian, we consider first the case of a single-band SM coupled to an s--wave SC through an infinite planar interface, then we address the specific issues related to multiband nanowires.

\subsubsection{Infinite planar interface} \label{secIPI}

Within the tunneling Hamiltonian approach, the proximity-effect induced by an s-wave SC can be captured by integrating out the superconducting degrees of freedom and calculating the surface self-energy due to the exchange of electrons between SC and SM~\cite{Sau2010}. We briefly review this approach here and use these results later when discussing the disordered s-wave superconductors. The interface self-energy is given by
\begin{align}
\Sigma_{\sigma \sigma'}(\bm r,\bm r',\omega)={\rm Tr}_{\bm r_1,\bm r_2} \tilde t(\bm r, \bm r_1) G_{\sigma \sigma'}(\bm r_1, \bm r_2, \omega) \tilde t^\dag(\bm r_2, \bm r'),
\end{align}
where $\tilde t(\bm r, \bm r_1)$ is the matrix describing tunneling between semiconductor and superconductor. To illustrate the basic physics, we use here the simplest form for tunneling matrix elements $\tilde t(\bm r, \bm r_1)=\tilde t\delta(z)\delta(z_1)\delta(\bm r^{||}-\bm r_1^{||})$ with $\bm r^{||}$ and $z$ denoting in-plane and out-of-plane coordinates. After some algebra, the surface self-energy $\Sigma(\bm r \!-\! \bm r',\omega)$ is given by
\begin{align}
\Sigma(\bm r - \bm r',\omega)=|\tilde t|^2 \delta(z) \delta(z') \int  \frac{d^3 \bm p}{(2\pi)^3} e^{i \bm p (\bm r\! - \! \bm r')} G(\bm p,\omega),
\end{align}
and finally becomes in the momentum space
\begin{align}
\Sigma(\bm p^{||},\omega)&=|\tilde t|^2 \int \frac{d p_z}{2\pi} G(\bm p,\omega)\\
&=|\tilde t|^2 \int^{\Lambda}_{-\Lambda} d \varepsilon \int \frac{d p_z}{2\pi} \delta(\varepsilon - \xi_{\bm p}) G(\varepsilon,\omega).
\end{align}
where $\Lambda$ is half bandwidth. The density of states $\nu(\varepsilon,\bm p^{||})=\int \frac{d p_z}{2\pi} \delta(\varepsilon - \xi_{\bm p})$ is usually a weakly-dependent function of momenta and energy $\nu(\varepsilon_F,\bm p^{||})\approx \nu(\varepsilon_F) =2\sqrt{1-\zeta^2}/\Lambda$ with $\zeta=(\Lambda-\varepsilon_F)/\Lambda$ and $\varepsilon_F$ being the Fermi level in the superconductor. With these approximations, the surface self-energy becomes
\begin{align}\label{eq:Sigma_clean}
\Sigma(\omega)&=|\tilde t|^2 \nu(\varepsilon_F) \int d \varepsilon G(\varepsilon,\omega)\\
&=-|\tilde t|^2 \nu(\varepsilon_F)\left[ \frac{\omega \tau_0+ \Delta_0 \tau_x}{\sqrt{ \Delta_0^2-\omega^2}}+\frac{\zeta}{1-\zeta^2}\tau_z\right].
\end{align}
In the homogeneous case the last term in Eq.~\eqref{eq:Sigma_clean} represents a shift of the chemical potential and can be neglected as the chemical potential should be determined self-consistently by solving the appropriate equation for the fixed total electron density.

We can now include the surface self-energy $\Sigma(\omega)$ into the SM Hamiltonian and study the effective low-energy model for the semiconductor. This can be done by investigating the poles of the SM Green's function
\begin{eqnarray}
&~&G^{-1}({\bm k},\omega) = \omega\left(1 + \frac{\gamma}{\sqrt{\Delta_0^2-\omega^2}}\right) -  \Gamma\hat{\sigma}_z  \label{G1} \\
&~&~~~~~~~~~~-[\xi_{\bm k} + \alpha_R(k_y\hat{\sigma}_x-k_x\hat{\sigma}_y)]\tau_z - \frac{\gamma\Delta_0}{\sqrt{\Delta_0^2-\omega^2}}\hat{\tau}_x,     \nonumber
\end{eqnarray}
where $\xi_{\bm k}= \hbar^2k^2/2m^*-\mu$, $\alpha_R$ is the Rashba coupling, $\Gamma$ the strength of a Zeeman field oriented perpendicular to the interface, $\Delta_0$ is the value of the superconducting gap inside the SC,  and $\gamma$ is the effective SM-SC coupling. In the calculations, in addition to the values specified in Sec. \ref{SecIIA}, i.e., $m^*=0.04m_0$ and $\alpha_R=0.1$eV$\cdot$\AA, we have $\Delta_0 = 1$meV.  The effecting coupling $\gamma = \widetilde{t}^2 |\psi(i_z=1)|^2 \nu(\varepsilon_F)$ depends on the transparency of the interface, $\widetilde{t}$, the amplitude of the SM wave function at the interface, $\psi(i_z=1)$, and the local density of states of the non-superconducting metal at the interface, which can be expressed in terms of the half-bandwidth $\Lambda$ and the Fermi energy $\varepsilon_F$ of the metal.\cite{Sau2010} Note that the Green's function (\ref{G1}) is written in the Nambu spinor basis $(u_{\uparrow}, u_{\downarrow},  v_{\downarrow}, -v_{\uparrow})^T$ using the Pauli matrices $\hat{\tau}_{\lambda}$ and $\and{\sigma}_\lambda$ that correspond to the Nambu and spin spaces, respectively. The identity matrices $\tau_0$ and $\sigma_0$ are omitted for simplicity.

Explicit comparison between the effective theory described by Eq. (\ref{G1}) and microscopic tight--binding calculations\cite{Sau2010} has shown remarkable agreement.  A similar effective description has proven extremely accurate in describing the proximity effect induced at a topological insulator -- superconductor interface.\cite{Stanescu2010} Can the low--energy physics contained in Eq. (\ref{G1}) be captured by an effective Hamiltonian description? To address this question, we determine the poles of the Green's function at frequencies within the superconducting gap, $\omega<\Delta_0$, i.e., we solve the Bogoliubov--de Gennes (BdG) equation ${\rm det}[G^{-1}]=0$. Explicitly, we have
\begin{eqnarray}
&~&\omega^2\left(1 + \frac{\gamma}{\sqrt{\Delta_0^2-\omega^2}}\right)^2 = \xi_{\bm k}^2 + \lambda_{\bm k}^2 + \Gamma^2 +\frac{\gamma^2\Delta_0^2}{\Delta_0^2-\omega^2} \nonumber \\
&~&~~~~~~~~~~-2\sqrt{\xi_{\bm k}^2(\lambda_{\bm k}^2+\Gamma^2) +\frac{\gamma^2\Delta_0^2\Gamma^2}{\Delta_0^2-\omega^2} }. \label{BdG1}
\end{eqnarray}
Note that we have considered only the lowest energy mode.
In Eq. (\ref{BdG1}) dynamical effects are generated by the frequency dependence of the proximity--induced terms containing the  expression $\gamma/\sqrt{\Delta_0^2-\omega^2}$, with a relative magnitude that depends on the SM-SC coupling strength. In general, we distinguish a weak-coupling regime characterized by $\gamma\ll\Delta_0$ and a strong coupling regime, $\gamma \gg \Delta_0$. In the weak coupling regime we expect negligible dynamical effects at all energies that are not very close to the gap edge, $\omega=\Delta_0$. Neglecting the frequency dependence in the proximity--induced terms, the solution of Eq. (\ref{G1}) becomes
\begin{equation}
E_{\bm k} = Z\sqrt{ \xi_{\bm k}^2 + \lambda_{\bm k}^2 + \Gamma^2+\gamma^2 -2\sqrt{\xi_{\bm k}^2(\lambda_{\bm k}^2+\Gamma^2) + \gamma^2\Gamma^2}},  \label{Ek}
\end{equation}
where $Z=(1+\gamma/\Delta_0)^{-1}<1$ is the quiasiparticle residue at zero energy.

\begin{figure}[tbp]
\begin{center}
\includegraphics[width=0.48\textwidth]{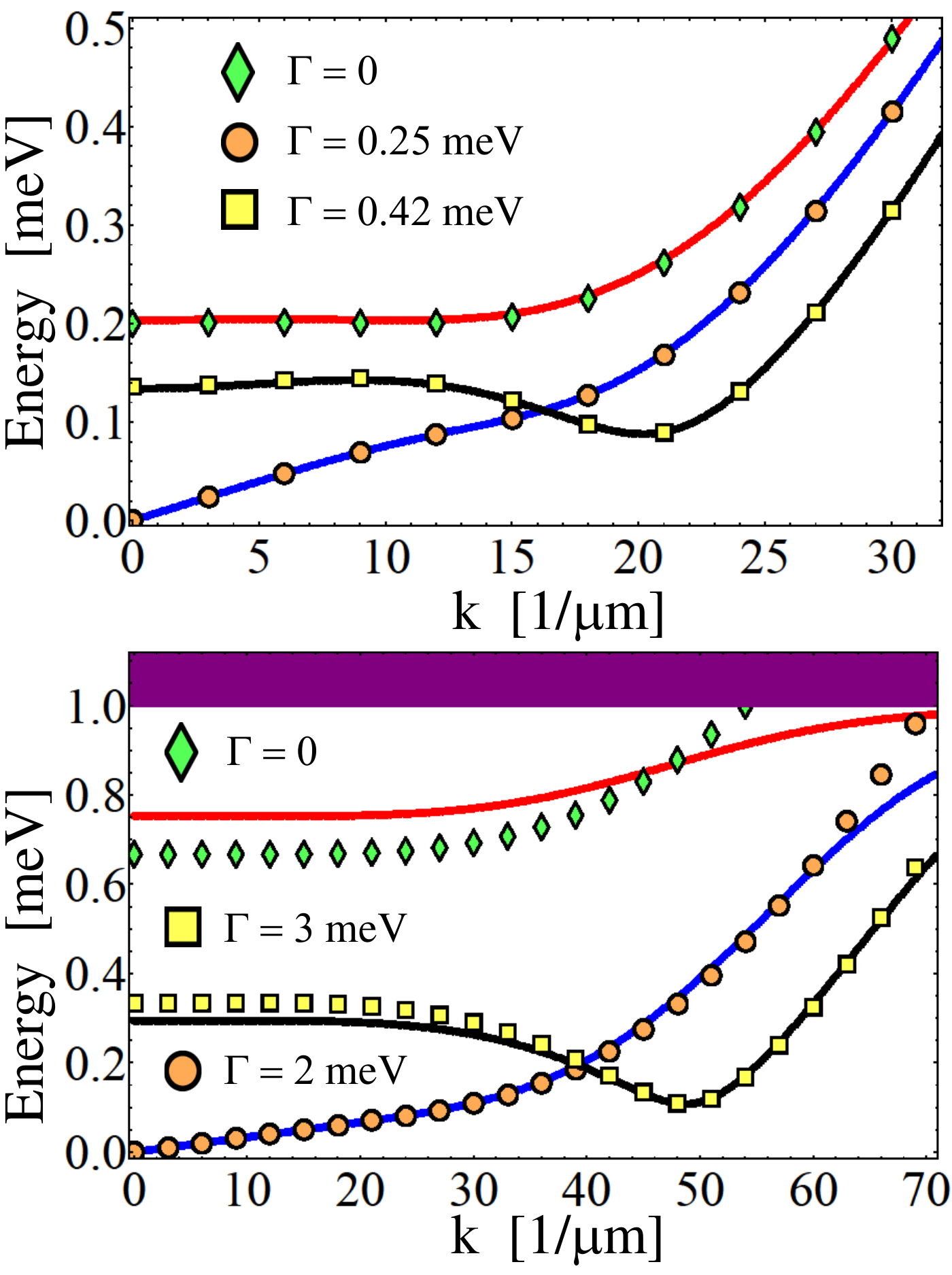}
\vspace{-7mm}
\end{center}
\caption{(Color online) {\it Top}: Low--energy BdG spectrum of a semiconductor with proximity--induced superconductivity. The effective SM--SC coupling is $\gamma=0.25$meV and the chemical potential is $\mu=0$. The induced gap vanishes at $k=0$ in the presence of a Zeeman field $\Gamma=\sqrt{\gamma^2+\mu^2}=0.25$meV. The full lines are obtained by solving Eq. (\ref{BdG1}) , while the symbols represent the spectrum given by Eq. (\ref{Ek}).  {\it Bottom}: Low--energy BdG spectrum for $\gamma=2$meV, $\mu=0$ and three different values of the Zeeman field. The filled area corresponding to energies above $\Delta_0=1$meV  represents the SC continuum. Note that for energies $E_{\bm k}< \Delta_0/2$, Eq. (\ref{Ek}) represents a very good approximation of the BdG spectrum even in the strong--coupling regime.}
\label{Fig1}
\end{figure}

\begin{figure}[tbp]
\begin{center}
\includegraphics[width=0.48\textwidth]{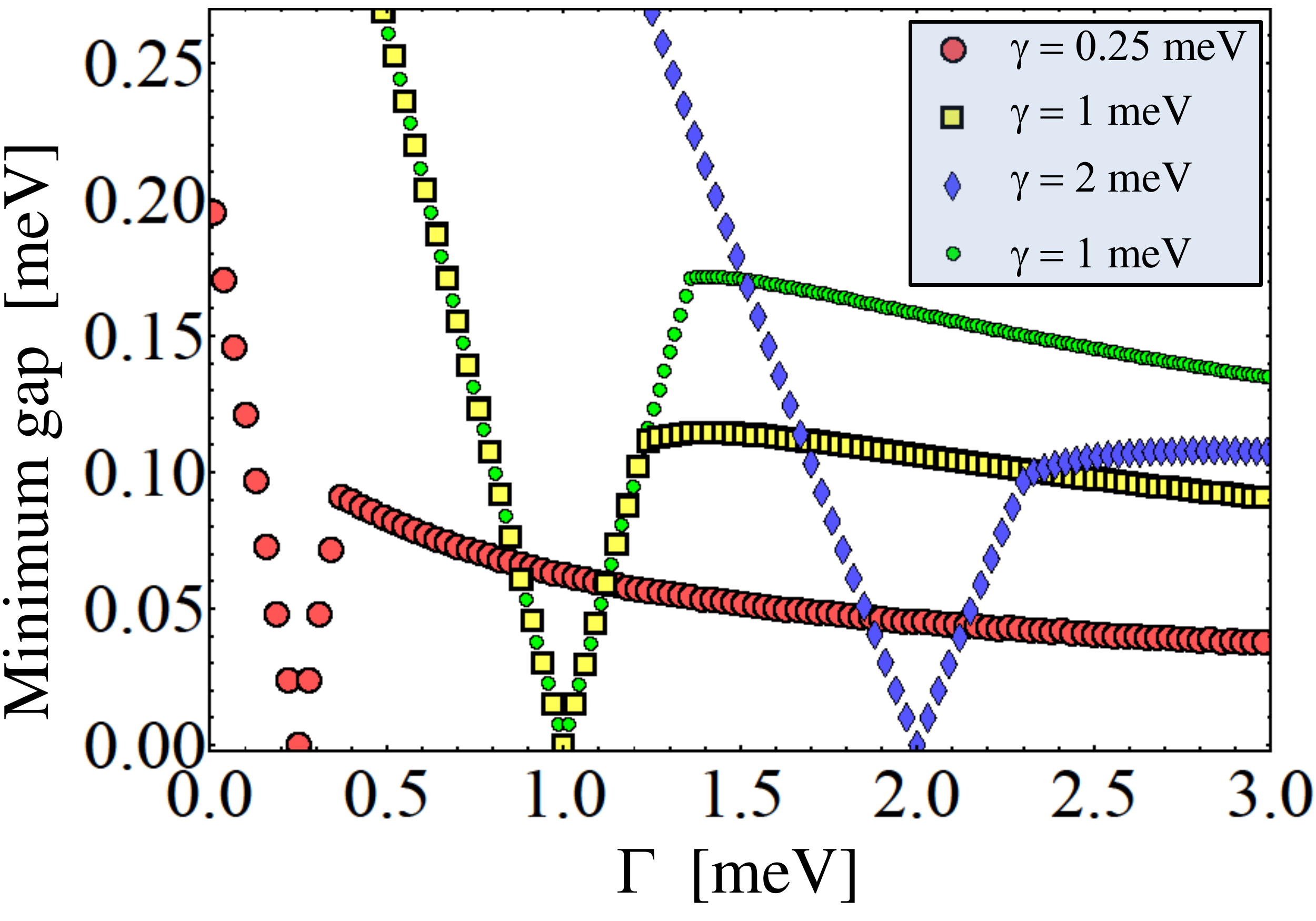}
\vspace{-7mm}
\end{center}
\caption{(Color online) Dependence of the minimum quasiparticle excitation gap in the BdG spectrum given by Eq. (\ref{BdG1}) on the Zeeman field $\Gamma$ for different SM--SC couplings. The chemical potential is $\mu=0$ and the Rashba coefficient is $\alpha_r=0.15$eV$\cdot$\AA for the curve represented by small (green) circles and $\alpha_r=0.1$eV$\cdot$\AA for the other three curves. The system becomes gapless at $\Gamma_c=\sqrt{\gamma^2+\mu^2}$. The superconducting state with $\Gamma<\Gamma_c$ is topologically trivial, while for $\Gamma>\Gamma_c$  one has a topological superconductor that supports Majorana bound states. Note that the optimal quasiparticle gap for the topological SC has a weak dependence on the SM--SC coupling but varies strongly with the strength of the spin--orbit coupling.}
\label{Fig2}
\end{figure}

To evaluate the dynamical effects, we compare the BdG spectrum given by Eq. (\ref{Ek}) with the full solution of Eq. (\ref{BdG1}). The results are shown in Fig. \ref{Fig1} for a weak coupling regime characterized by $\gamma=0.25$meV (top panel) and at large coupling, $\gamma=2$meV (bottom panel). Notice that, even for effective couplings larger than the gap, e.g.,  $\gamma=2\Delta_0$, Eq. (\ref{Ek}) represents a very good approximation of the low energy spectrum. What about couplings that are much larger than $\Delta_0$? To answer this question,  let us consider the dependence of the low--energy spectrum on the coupling constant $\gamma$, and the Zeeman field $\Gamma$. For $\Gamma=0$ the spectrum is gaped and the minimum of the gap is located at $k=0$ (see Fig. \ref{Fig1}). Applying a Zeeman field reduces this minimum gap continuously and at the critical value $\Gamma_c=\sqrt{\gamma^2+\mu^2}$ the spectrum becomes gapless  (see Fig. \ref{Fig1}). For $\Gamma > \Gamma_c$ a gap opens again with a  minimum  at $k=0$ for $\Gamma\geq \Gamma_c$ and at a finite wave vector for large values of the Zeeman field. The dependence of the minimum gap on the Zeeman field is shown in Fig. \ref{Fig2} for three different values of the SM-SC coupling. The vanishing of the gap at the critical point $\Gamma=\Gamma_c$ marks a quantum phase transition from a normal SC at low Zeeman fields to a topological SC  ( when $\Gamma>\Gamma_c$).\cite{Lutchyn'10, Sau2010} The change in the location of the quasiparticle gap  from $k=0$ to a finite wave vector is marked in Fig. \ref{Fig2} by a discontinuity in the slope at $\Gamma^*\geq \Gamma_c$. The optimal value of the excitation gap in the topological phase is obtained for $\Gamma\approx \Gamma^*$. This optimal value depends weakly on the effective coupling $\gamma$, but varies strongly with the spin--orbit coupling.

From the above analysis we conclude that the strong--coupling regime characterized by $\gamma\gg\Delta_0$ is not experimentally desirable, as it would require extremely high magnetic fields to reach the topologically nontrivial phase, i.e., $\Gamma >\sqrt{\gamma^2+\mu^2} \gg \Delta_0$. In addition, it would be difficult to tune the chemical potential and drive the system into a topological superconducting phase for a large SM-SC coupling. Also, as follows from Eq.~\eqref{Ek}, the quasiparticle excitation spectrum decays with increasing $\gamma/\Delta_0$ which leads to the reduced stability of the topological phase against  thermal fluctuations. Hence, an experimentally useful interface should be characterized by an effective coupling  $\gamma$ of order $\Delta_0$ or less, i.e.,  in the intermediate to weak--coupling regime. As shown above, in these regimes the BdG spectrum is accurately approximated by Eq. (\ref{Ek}). Consequently, we can model the low--energy spin--orbit coupled semiconductor with proximity induced superconductivity  using an effective tight--binding model given by the Hamiltonian
\begin{equation}
H_{\rm eff} = H_{\rm SM} + H_{\rm Zeeman} + H_{\Delta},   \label{Heff0}
\end{equation}
where the semiconductor term $H_{\rm SM}$ has the same form as the Hamiltonian $H_{\rm nw}$ for the nanowire given by Eq. (\ref{Hnw}) but with $N_x\rightarrow \infty$ and $N_y\rightarrow \infty$ and the Zeeman term  $H_{\rm Zeeman}$ is given by Eq. (\ref{HZeeman}) with $\hat{\sigma}_x \rightarrow \hat{\sigma}_z$. In addition, all the energy scales are renormalized by a factor $Z=(1+\gamma/\Delta_0)^{-1}<1$, i.e., $t_0 \rightarrow Z t_0$, $\alpha \rightarrow Z \alpha$, etc. The physical meaning of the factor $Z$ written as $Z=\gamma^{-1}/(\gamma^{-1}+\Delta_0^{-1})$ is intuitively clear - it corresponds to a probability to find an electron in the semiconductor. The induced superconductivity is induced by the effective pairing term
\begin{equation}
H_{\Delta} = \sum_{\bm i}\left( \Delta c_{{\bm i}\uparrow}^\dagger c_{{\bm i}\downarrow}^\dagger + h.c.\right),  \label{HDelta}
\end{equation}
with an effective SC order parameter $\Delta = \gamma \Delta_0/(\gamma+\Delta_0)$. With these choices, $H_{\rm eff}$ given by Eq. (\ref{Heff0}) has the same  low energy spectrum as the one  described by Eq. (\ref{Ek}).

\subsubsection{The multiband case} \label{secMC}

To account for the specific aspects that characterize the proximity effect in finite--size systems, we return to the details of deriving the effective low--energy Green's function description, e.g., Eq. (\ref{G1}), starting with the microscopic Hamiltonian (\ref{Htot}). After integrating out the SC degrees of freedom, the effective SM Green's function acquires a self--energy
\begin{eqnarray}
&~&\Sigma_{\rm SC}({\bm n},{\bm n}^\prime) = \sum_{i_x, i_y, j_x, j_y} \psi_{\bm n}(i_x,i_y,1)\widetilde{t}(i_x, i_y)  \label{Sigma} \\
&~&~~~~~~~~~~~~~\times G_{\rm SC}(\omega,{\bm r}_\parallel, z_{\rm interface}) \widetilde{t}(j_x, j_y)\psi_{{\bm n}^\prime}(j_x,j_y,1),     \nonumber \\ ~ \nonumber
\end{eqnarray}
where $\psi_{\bm n}(i_x,i_y,1)$ are the orbital components of the eigenstates described by Eq. (\ref{psin})  and $G_{\rm SC}(\omega,{\bm r}_\parallel, z_{\rm interface})$ is the SC Green's function, both evaluated  at the interface. In Eq. (\ref{Sigma}) ${\bm r}_\parallel = (i_x-j_x, i_y-j_y)$ and $\widetilde{t}(i_x, i_y)$ are matrix elements that couple two sites with in--plane coordinates $(i_x,i_y)$ across the interface. In the discussion of the planar interface we implicitly assumed translational invariance for the $SM-SC$ coupling, i.e., $\widetilde{t}(i_x, i_y) = \widetilde{t}$. Here, we consider position--dependent couplings and argue that engineering interfaces with a transparency that varies {\it across} the wire, i.e., $\widetilde{t}(i_x, i_y) = \widetilde{t}(i_y)$, generates off--diagonal components of the effective SC order parameter that help stabilize the Majorana modes.\cite{lutchyn_multi} We note that variations of the coupling matrix element {\it along} the wire, i.e., in the $x$ direction, act as an effective disorder potential. We will address this issue below in Sec.\ref{sec:tunnelingdisorder}.

\begin{figure}[tbp]
\begin{center}
\includegraphics[width=0.48\textwidth]{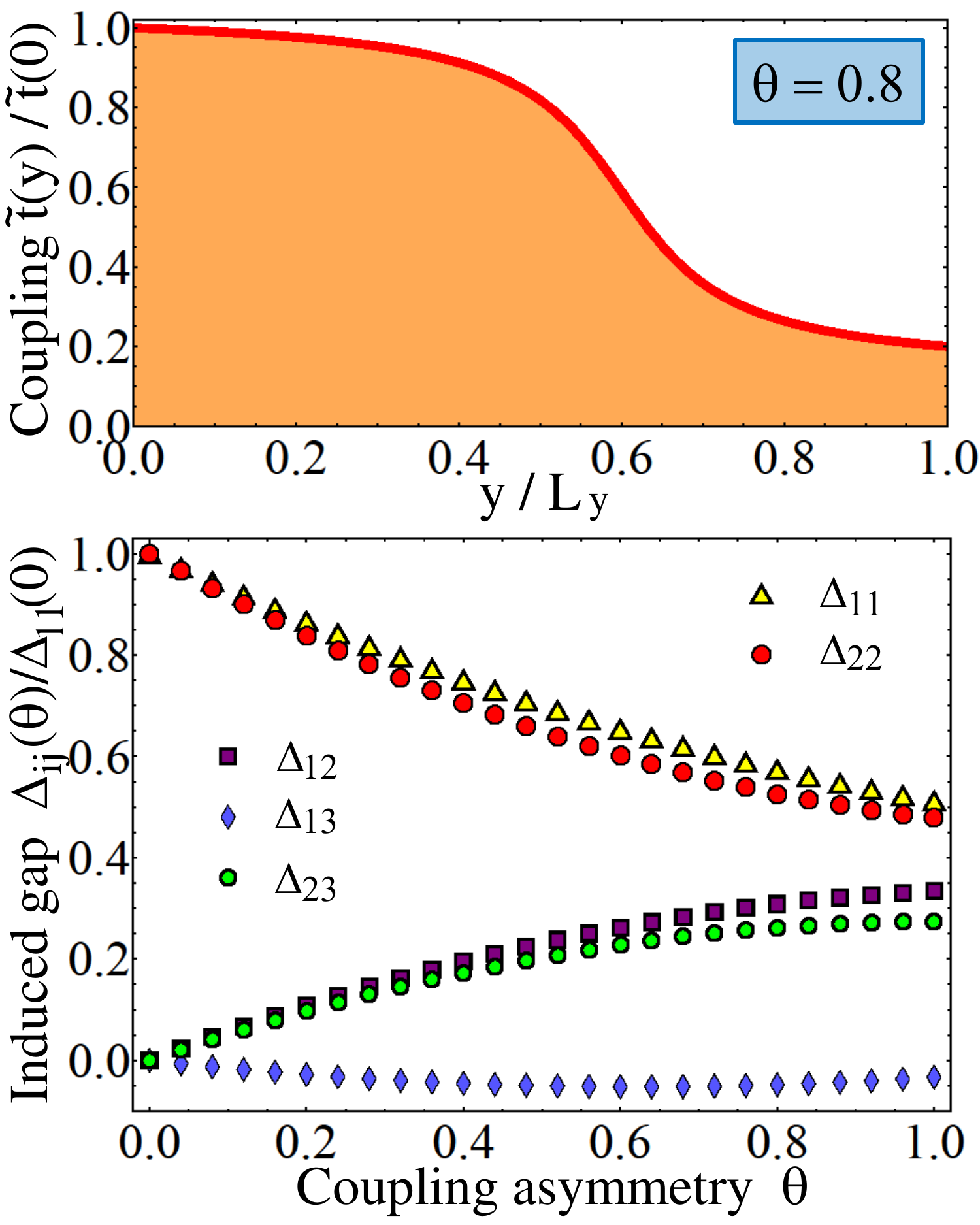}
\vspace{-7mm}
\end{center}
\caption{(Color online) Upper panel: Non--uniform SM--SC coupling $\widetilde{t}(y) = \widetilde{t}_0[1 - \theta p(y)]$ with a smooth profile $p(y)$ and $\theta=[\widetilde{t}(0)-\widetilde{t}(L_y)]/\widetilde{t}(0) = 0.8$. Lower panel: Dependence of the induced gap on the non--uniformity of the coupling across the wire. The inhomogeneous proximity effect induces inter--sub--band pairing  with $\Delta_{n_y n_y\pm1}\approx \Delta_{n_y n_y}/2$ for $\theta>0.7$.}  \label{Fig3}
\end{figure}

The SC Green's function integrated over momenta can be written explicitly as (see Eq.~\eqref{eq:Sigma_clean})
\begin{equation}
G_{\rm SC} = -\nu(\varepsilon_F)\left[ \frac{\omega \tau_0+ \Delta_0 \tau_x}{\sqrt{ \Delta_0^2-\omega^2}}+\frac{\zeta}{1-\zeta^2}\tau_z\right],  \label{GSC}
\end{equation}
where $\zeta=(\Lambda-\varepsilon_F)/\Lambda$ with $\Lambda$ being the half-bandwidth, $\varepsilon_F$ being the Fermi energy and $\Delta_0=1$meV being the s-wave SC gap. In the numerical calculations we have $\varepsilon_F=\Lambda/2$, i.e., $\zeta=0.5$. Using the expression of the SC Green's function given by Eq. (\ref{GSC}), the self-energy (\ref{Sigma}) becomes
\begin{equation}
\Sigma_{\rm SC}({\bm n},{\bm n}^\prime) = -\gamma_{n_y n_y^\prime}\left[\frac{\omega+\Delta_0\hat{\tau}_x}{\sqrt{\Delta_0^2-\omega^2}} + \frac{\zeta\hat{\tau}_z}{\sqrt{1-\zeta^2}} \right]\delta_{n_x n_x^\prime},  \label{SigmaSC}
\end{equation}
with the  implicit assumption that the wire is very thin, $L_z\ll L_y$, and $n_z=n_z^\prime=1$. The coupling matrix in Eq. (\ref{SigmaSC}) is
\begin{eqnarray}
\gamma_{n_y n_y^\prime} &=&\langle n_y|\gamma|n_y^\prime\rangle \label{gammann} \\
&=&  \frac{2}{N_y+1}\sum_{i_y} \gamma(i_y) \sin\left[\frac{n_y i_y \pi}{N_y+1}\right]\sin\left[\frac{n_y^\prime i_y \pi}{N_y+1}\right], \nonumber \\
\gamma(i_y) &=& \frac{4\sqrt{1-\zeta^2}\sin^2\left[\frac{\pi}{N_z+1}\right]}{(N_z+1)\Lambda}~\widetilde{t}^2(i_y). \label{gamma}
\end{eqnarray}
We note that for position--independent SM--SC couplings, $\widetilde{t}(i_y)=\widetilde{t}$, the  matrix $\gamma$ is proportional to the unit matrix and the effective low--energy Hamiltonian can be obtained along the lines of Sec. \ref{secIPI}. However, for non--uniform couplings,  $\gamma_{n_y n_y^\prime}$ acquires off--diagonal elements that generates normal and anomalous inter--sub--band terms in the effective Hamiltonian via the self--energy (\ref{SigmaSC}). The relative magnitude of the off--diagonal terms depends on the non-homogeneity of the SM--SC coupling. To quantify this property, we consider a fixed profile $p(y)$ with the property $p(0)=0$ and $p(L_y)=1$ and the position--dependent tunneling $\widetilde{t}(y) = \widetilde{t}_0[1 - \theta p(y)]$, where $0\leq \theta\leq 1$ is a parameter that measures the degree of non--uniformity of the coupling. Shown in Fig \ref{Fig3} (upper panel) is  $\widetilde{t}(y)$ for $\theta=0.8$.
As we show below, the non--uniform coupling induces an effective pairing $\Delta_{n_y n_y^\prime} = \langle n_y| \gamma \Delta_0/(\gamma+\Delta_0)|n_y^\prime\rangle$, where  $\gamma$ is given by Eq. (\ref{gamma}).  The dependence of $\Delta_{n_y n_y^\prime}$ on the coupling asymmetry parameter $\theta$ is shown in Fig \ref{Fig3} (lower panel).  For uniform tunneling ($\theta = 0$) the off--diagonal elements vanish and the diagonal elements become equal.   By contrast, strongly non-homogeneous tunneling ($\theta\rightarrow 1$) generate off--diagonal terms $\Delta_{n_y n_y^\prime}$ that reach about $50\%$ of the diagonal contributions for neighboring sub--bands, $n_y^\prime=n_y\pm 1$, and are much smaller for $|n_y^\prime-n_y|>1$.

The low--energy effective Hamiltonian for the nanowire can be derived following the scheme described in Sec. \ref{secIPI}. As before, at low energies ($\omega\ll\Delta_0$)  we can neglect the frequency dependence of the dynamically--generated terms, i.e., $1/\sqrt{\Delta_0^2-\omega^2}\approx1/\Delta_0$.  However, due to the inter--sub--band coupling induced by non--homogeneous proximity effect, one cannot simply renormalize the energy by a factor $Z$. This is due to the fact  that the Green's function for the proximity--coupled nanowire, $(G^{-1})_{{\bm n} {\bm n}^\prime} = \omega~\delta_{{\bm n} {\bm n}^\prime} - (H_{\rm nw}+H_{\rm Zeeman})_{{\bm n} {\bm n}^\prime} - \Sigma_{\rm SC}( {\bm n}, {\bm n}^\prime)$, contains a frequency--dependent term $\omega(1 + \gamma_{n_y n_y^\prime}/\Delta_0)$ that is not proportional to the unit matrix. However, we notice that we can define a matrix $Z^{1/2}$ with the property  $Z^{1/2} G^{-1} Z^{1/2} = \omega - H_{\rm eff}$.  Explicitly, we have
\begin{equation}
(Z^{1/2})_{n_y n_y^\prime} = \left\langle n_y\left| \sqrt{\frac{\Delta_0}{\Delta_0 +\gamma}}~\right| n_y^\prime\right\rangle, \label{Z12}
\end{equation}
where $\gamma(i_y)$ is given by Eq.  (\ref{gamma}) and $\langle i_y|n_y\rangle = \sqrt{2/(N_y+1}\sin[i_y n_y \pi/(N_y+1)]$.
Because $\det[Z^{1/2}]>0$, the renormalized Green's function satisfies the same BdG equation as the original Green's function, i.e., $\det[\omega-H_{\rm eff}] = \det[G^{-1}] = 0$. We conclude that the low--energy physics of a nanowire proximity--coupled to an s--wave SC can be described by an effective Hamiltonian $H_{\rm eff}$ that can be  conveniently characterized by its matrix elements in the Nambu  basis $||{\bm n} \sigma\rangle\rangle = (\psi_{{\bm n}\sigma}, -\sigma\psi_{{\bm n}\sigma}^\dagger)^T$  provided by the eigenstates  of $H_0$ given by Eq. (\ref{psin}). Considering only the lowest band, i.e., ${\bm n} = (n_x, n_y, 1)$, we can write explicitly
\begin{eqnarray}\label{eq:26}
&~&~~~~~~~~~~~~~~~\langle\langle {\bm n}\sigma ||H_{\rm eff}|| {\bm n}^\prime \sigma^\prime\rangle\rangle = \label{Heff} \\
&~&(Z^{1/2})_{n_y m_y}\langle {\bm m}\sigma|H_{\rm nw} \hat{\tau}_z + H_{\rm Zeeman} | {\bm m}^\prime\sigma^\prime  \rangle (Z^{1/2})_{m_y^\prime n_y^\prime} \nonumber \\
&~&~~~~- \delta_{n_x n_x^\prime} \delta_{\sigma \sigma^\prime}  \frac{\zeta \Delta_{n_y n_y^\prime}}{\sqrt{1-\zeta^2}} ~\hat{\tau}_z   - \delta_{n_x n_x^\prime} \delta_{\bar{\sigma} \sigma^\prime}  \Delta_{n_y n_y^\prime} ~\hat{\tau}_x,    \nonumber
\end{eqnarray}
where $\bar{\sigma}=-\sigma$, ${\bm m} =(n_x, m_y, 1)$,  ${\bm m}^\prime =(n_x^\prime, m_y\prime, 1)$, and summation over the repeating indices $m_y$, $m_y^\prime$ is implied. In Eq. (\ref{Heff}) the matrix $Z^{1/2}$ is given by Eq. (\ref{Z12}), the Hamiltonian for the nanowire is $H_{\rm nw}=H_0 +H_{\rm SOI}$, and the matrix elements of $H_0$, $H_{\rm SOI}$, and $H_{\rm Zeeman}$ are given by equations (\ref{epsn}), (\ref{HSOI}), and (\ref{HZeemannn}), respectively. We note that for a homogeneous SM--SC interface the normal contribution proportional to $\Delta_{n_y n_y^\prime}$ becomes diagonal and can be absorbed in the chemical potential, but in general it generates inter--sub-band mixing. These induced off--diagonal terms can be significant in the strong--coupling limit. Finally, the effective SC order parameter is
\begin{equation}
\Delta_{n_y n_y^\prime} =  \left\langle n_y\left| \frac{\gamma \Delta_0}{\Delta_0 +\gamma}~\right| n_y^\prime\right\rangle, \label{Deltann}
\end{equation}
where $\gamma$ is given by Eq. (\ref{gamma}) and $|n_y\rangle$ has the same significance as in Eq. (\ref{Z12}).

The effective low--energy BdG Hamiltonian given by Eq. (\ref{Heff}) is the main result of this section. In the  remainder of this work we will study the low--energy physics of a nanowire with  proximity--induced superconductivity by diagonalizing numerically this effective Hamiltonian~\eqref{eq:26}.

\section{Low--energy spectrum, Majorana bound states, and phase diagram}\label{SecIII}

\subsection{General properties of the BdG spectrum}\label{SecIIIA}

The effective  BdG Hamiltonian (\ref{Heff}) can be written as
\begin{equation}
H_{\rm eff} =\widetilde{H}_{\rm nw} \hat{\tau}_z + \widetilde{H}_{\rm Zeeman} + \widetilde{H}_{\Delta} \hat{\tau}_x,
\end{equation}
where $\widetilde{H}_{\rm nw}$ and $\widetilde{H}_{\rm Zeeman}$ are renormalized nanowire and Zeeman Hamiltonians, respectively,  and $\widetilde{H}_{\Delta}$ is the effective pairing with matrix elements $  - \delta_{n_x n_x^\prime} \delta_{\bar{\sigma} \sigma^\prime}  \Delta_{n_y n_y^\prime}$. We note that the normal contribution proportional to $\Delta_{n_y n_y^\prime}$ from Eq. (\ref{Heff}) is included in $\widetilde{H}_{\rm nw}$. As mentioned above, for homogeneous SM--SC coupling the only effect of this term is to generate an overall shift of the energy. For convenience, we eliminate this shift by adding a term  $\delta_{n_x n_x^\prime} \delta_{\sigma \sigma^\prime}\overline{\Delta} {\zeta}/\sqrt{1-\zeta^2}  ~\hat{\tau}_z$ to the effective Hamiltonian  (\ref{Heff}), where
\begin{equation}
\overline{\Delta} = \frac{\Delta_0}{N_y}\sum_{i_y}\frac{\gamma(i_y)}{\Delta_0+\gamma(i_y)}
\end{equation}
is  the ``average'' effective pairing. Note that $\Delta_{n_y n_y}\rightarrow \overline{\Delta}$ for $n_y \gg 1$ and the diagonal contribution to the energy--shifting term is partially canceled even in the non--homogeneous case. Finally, to be able to compare results corresponding to various degrees of inhomogeneity in the coupling, i.e., different $\theta$ parameters, we define the average coupling strength as
\begin{equation}
\overline{\gamma} =\frac{1}{N_y} \sum_{i_y} \gamma(i_y).   \label{gammabar}
\end{equation}

\begin{figure}[tbp]
\begin{center}
\includegraphics[width=0.48\textwidth]{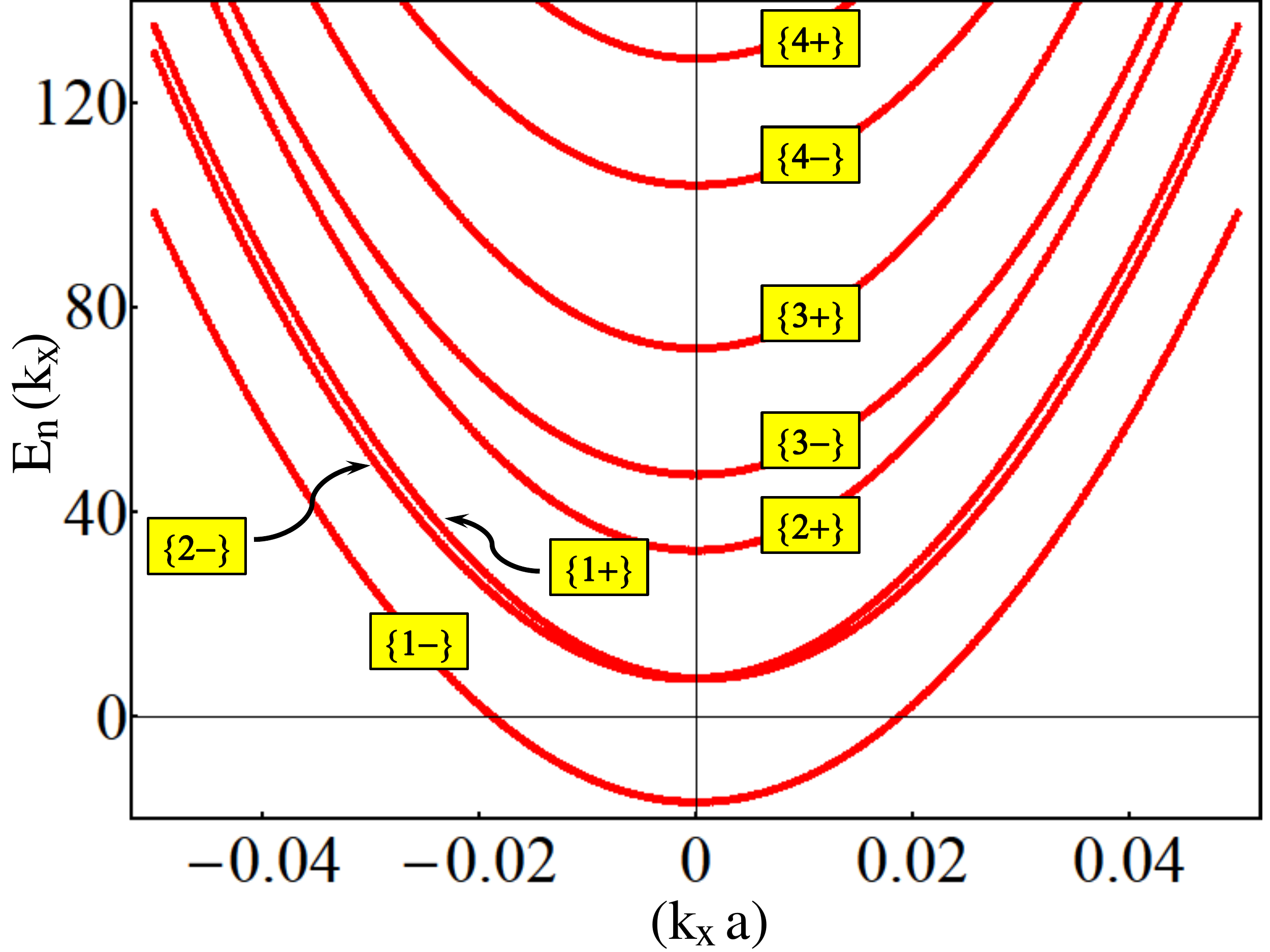}
\vspace{-7mm}
\end{center}
\caption{(Color online) Spectrum of an infinite non--superconducting  wire in the presence of a Zeeman field. The energies  are obtained by diagonalizing the Hamiltonian  $\widetilde{H}_{\rm nw} + \widetilde{H}_{\rm Zeeman}$ for $\mu=5E_{\alpha}$ and $\Gamma \approx 15 E_\alpha$ and are measured relative to the chemical potential. Due to the presence of the Zeeman field, each $n_y$ band is split into two sub--bands marked $\{n_y -\}$ and  $\{n_y +\}$. Note that, for this value of $\Gamma$,  the sub--bands  $\{1 +\}$ and  $\{2 -\}$ are degenerate at $k_x = 0$.  }  \label{Fig4}
\end{figure}

To obtain better insight into the properties of the BdG Hamiltonian, we start with a non-superconducting system described by the Hamiltonian $\widetilde{H}_{\rm nw} + \widetilde{H}_{\rm Zeeman}$ and $L_x\rightarrow \infty$, i.e.,  an infinitely long renormalized nanowire placed into an effective magnetic field. The spectrum of the renormalized wire is shown in Fig. \ref{Fig4} for a chemical potential $\mu/E_\alpha = 5$ and a Zeeman field $\Gamma/E_\alpha \approx 15$, where $E_\alpha = m^* \alpha_{\rm R}\approx 0.6K$ is the characteristic  spin-orbit coupling energy. Here and below we systematically use $E_\alpha$ as  energy unit. The bare nanowire spectrum is renormalized due to a weak inhomogeneous coupling with a profile given in Fig. \ref{Fig3}, $\theta = 0.8$,  and $\overline{\gamma} = 0.25 \Delta_0$. For $\Gamma=0$ the sub--bands with a given  value of $n_y$ are double degenerate at $k_x=0$, but this degeneracy is removed by the Zeeman field. However, for special values of $\Gamma$, sub--bands corresponding to different values of $n_y$ may become degenerate at $k_x=0$ (see Fig. \ref{Fig4}). If the chemical potential has a value such that the degeneracy point occurs at zero energy, $(\Gamma, \mu)$ represent a so--called ``sweet spot''.\cite{lutchyn_multi} Adding superconductivity, will now open a gap in the spectrum near $E=0$. In the weak--coupling limit, one can determine the topological nature of the induced superconductivity by simply counting the number of sub--bands crossed by the chemical potential: an odd number corresponds to a topologically non-trivial SC, while an even number results in a standard superconductor.\cite{lutchyn_multi} Within this simplified picture, the ``sweet spots'' represent critical points.  As will be shown below, the properties of the system in the interesting parameter regimes near the ``sweet spots'' are determined by the effective inter--band pairing, i.e., by the non--homogeneity of the SM--SC proximity effect.

\begin{figure}[tbp]
\begin{center}
\includegraphics[width=0.48\textwidth]{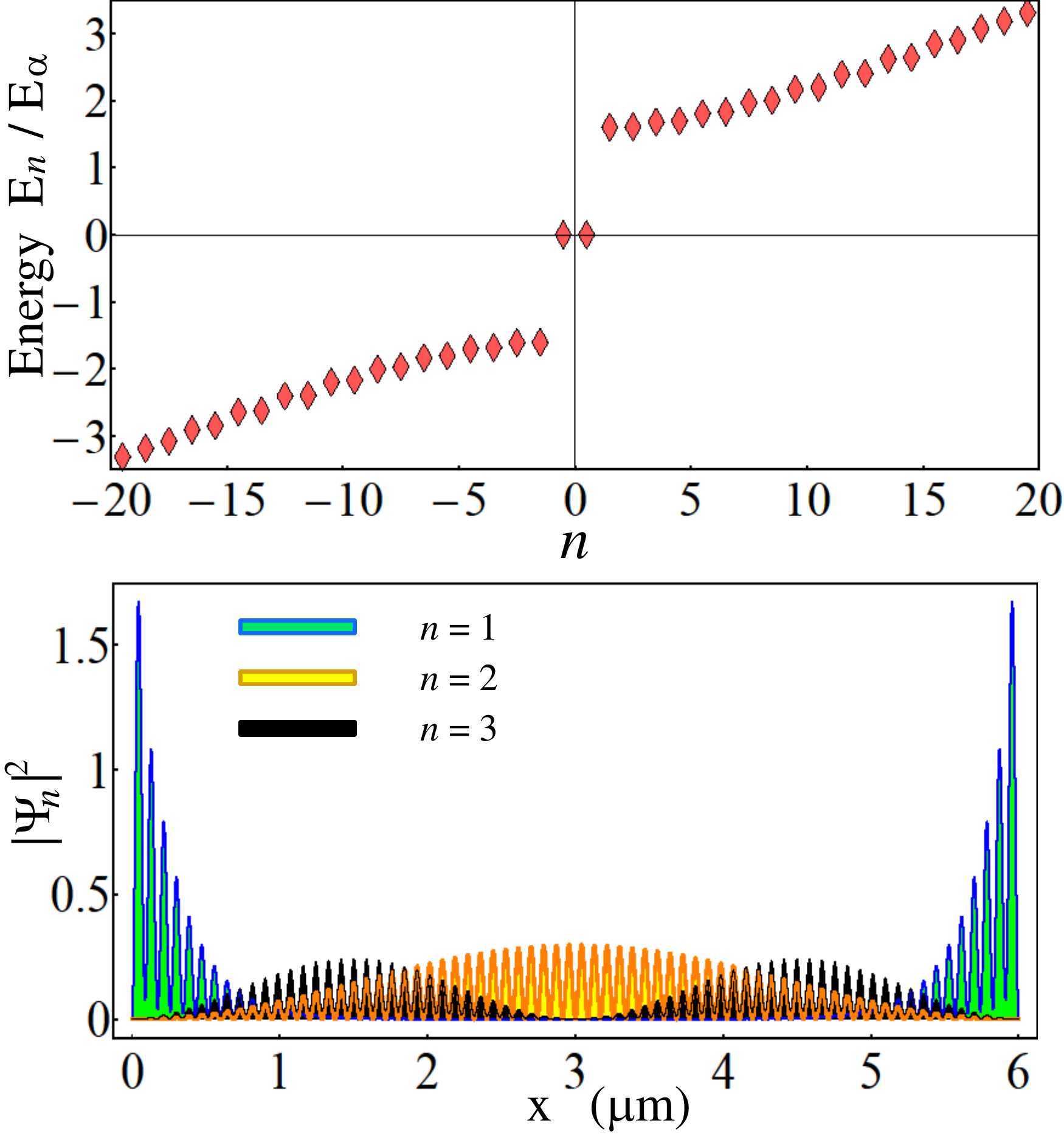}
\vspace{-7mm}
\end{center}
\caption{(Color online) Upper panel: BdG energy spectrum for a finite superconducting wire obtained by numerical diagonalization of $H_{\rm eff}$. The parameters $\mu$ and $\Gamma$ are the same as in Fig. \ref{Fig4} and ${\it n}$ labels the eigenvalues of  $H_{\rm eff}$ staring with the lowest energy and has the same sign as $E_{\it n}$. The in--gap states are Majorana  zero--energy  modes. Lower panel: The particle--component of the wave--function amplitude for the lowest energy states. The Majorana modes (${\it n}=1$) are localized at the ends of the wire, while the finite energy states extend over the entire wire.}  \label{Fig5}
\end{figure}

Next, we diagonalize numerically  the full effective Hamiltonian (\ref{Heff}) for a finite wire using the same set of control parameters, $\mu/E_\alpha = 5$ and $\Gamma/E_\alpha \approx 15$. The corresponding low--energy spectrum is shown in the upper panel of Fig. \ref{Fig5}. The eigenstates are labeled by an integer number {\it n} that has the same sign as the corresponding eigenvalue $E_{\it n}$. The spectrum is characterized by a gap $\Delta^*\approx 1.8E_\alpha$ and a pair of zero--energy Majorana bound states. To prove the localized nature of these states, we calculate the wave function amplitude  and show that the  Majorana modes are localized near the ends of the wire (Fig. \ref{Fig5} lower panel). By contrast, finite energy states extend over the entire system. The oscillations of the  wave function amplitudes is associated with the Fermi momentum  $k_F\approx 0.02/a$, as shown in Fig. \ref{Fig4}. We note that amplitudes shown in Fig. \ref{Fig5} represent the particle component of the BdG wave-functions, i.e., $|u_{{\it n}\uparrow}|^2+|u_{{\it n}\downarrow}|^2$. For a finite-energy state, e.g., ${\it n} = 2$ the corresponding total spectral weight is about $1/2$, with the other half coming from  from ${\it n} = -2$. The Majorana modes have a total weight of one, which corresponds to one physical particle, but this weight is spatially separated into two contributions localized near the ends of the wire. Removing the Majorana pair would require overlapping the two components, which cannot be done by  local perturbations. This is, of course, the topological immunity of the Majorana modes, which is crucial for topological quantum computation. The characteristic length scale for the localized modes is controlled by the minimum value of the quasi--particle gap in a wire with no ends, e.g., with periodic boundary conditions, $\Delta_\infty^*$. As will be shown below (see Sec \ref{SecIIIC}), in a finite wire it is possible that bound states localized near the ends of the system  have energies within the gap. When in--gap states are present, the lowest--energy localized state sets the value of the mini--gap, $\Delta^* < \Delta_\infty^*$.  We emphasize that for a set of control parameters $(\Gamma,  \mu)$ corresponding to a non--vanishing minimum quasi--particle gap $\Delta_\infty^*$, the mini--gap  $\Delta^*$ is always nonzero. The characteristic length scale for the localized modes diverges in the limit $\Delta^*\rightarrow 0$. Hence, the topological phase is protected as long as the quasi--particle gap remains finite.  Consequently,  to determine the stability of the Majorana bound states, our key task is to determine the dependence of  $\Delta_\infty^*$ on various physical parameters and experimentally--relevant perturbations, e.g., chemical potential, Zeeeman field, SM--SC coupling, and charged impurity and coupling--induced disorder.

At this point in our analysis it is important to clarify the role played by the parameters that incorporate the SM--SC proximity effect into the low--energy effective theory. In particular, we address the following question: how does the low--energy spectrum depend on the strength of the SM--SC coupling (i.e., on $\overline{\gamma}$), on the non-homogeneity of the coupling ($\theta$), and on the dynamical effects included in the effective description ($Z^{1/2}$)? The non-homogeneity of the coupling is responsible for generating inter sub--band pairing in Eq. (\ref{Deltann}). These off--diagonal contributions play a minor role away from the ``sweet spots'', where they generate a small quantitative change of the quasi--particle gap. However, in the vicinity of the ``sweet spots'' $\Delta_\infty^*$ vanishes in the absence of inter sub--band pairing and the non-homogeneity of the coupling (i.e., $\theta >0$) becomes crucial.
\begin{figure}[tbp]
\begin{center}
\includegraphics[width=0.48\textwidth]{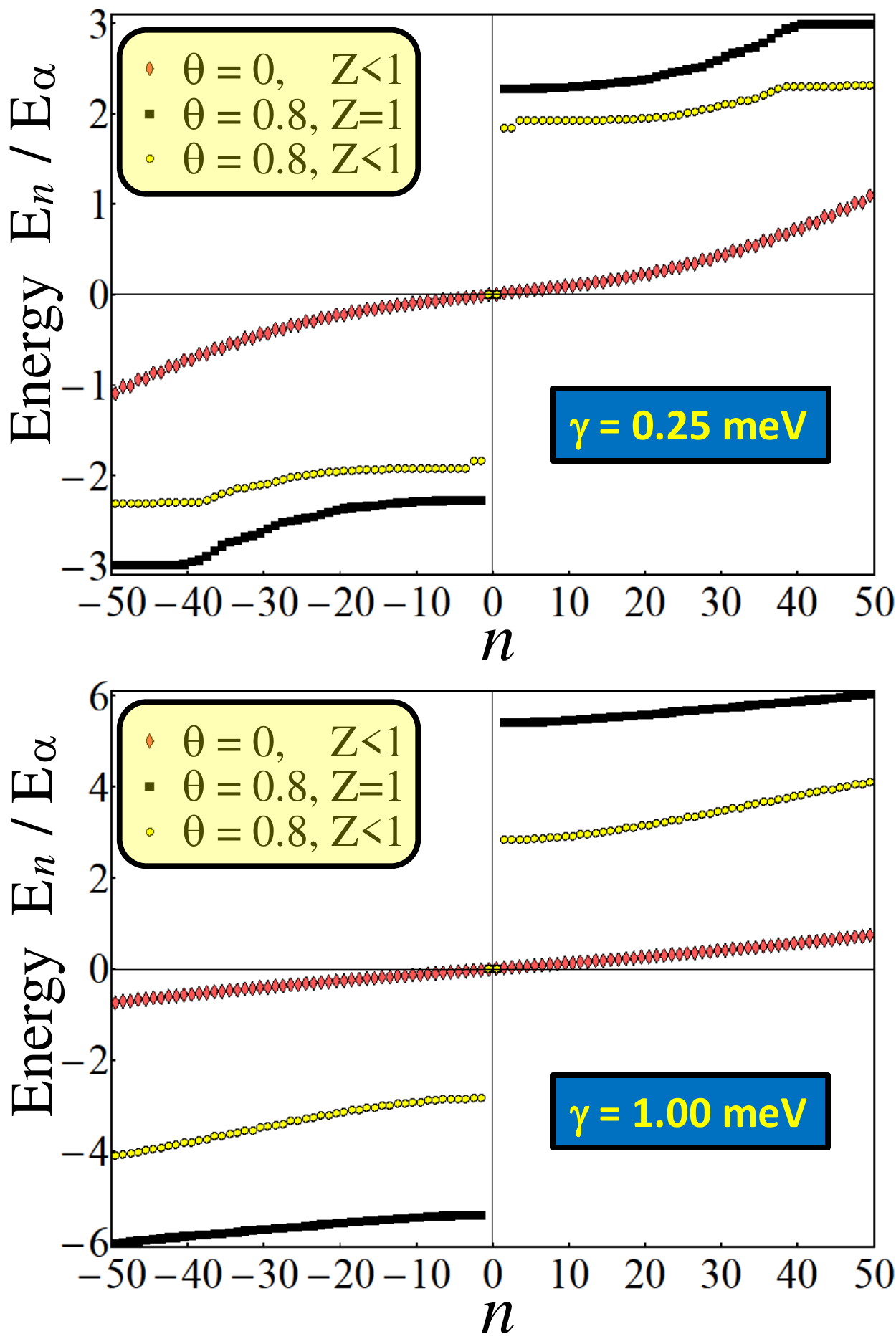}
\vspace{-7mm}
\end{center}
\caption{(Color online) Upper panel: Low--energy spectra  for weak SM--SC coupling ($\overline{\gamma}=0.25\Delta_0$) in the vicinity of the ``sweet spot'' ($\mu=14.5E_\alpha$, $\Gamma=15.3E_\alpha$). The yellow circles correspond to inhomogeneous coupling with $\theta=0.8$ and takes into account dynamical effects, the  black squares are obtained by neglecting dynamical effects, i.e., $(Z^{1/2})_{n_y n_y^\prime}=\delta_{n_y n_y^\prime}$, and the red diamonds are for a homogeneous coupling.  Lower panel: Same as in the upper panel for $\overline{\gamma}=\Delta_0$ near the ``sweet spot'' ($\mu=14.5E_\alpha$, $\Gamma=24.2E_\alpha$). Note that inclusion of dynamical effects through $Z^{1/2}$ renormalizes the energy scales, while inhomogeneous coupling play a critical role in establishing a finite gap near the ``sweet spot''. The location of a given ``sweet spot'' depends on the coupling strength.}  \label{Fig6}
\end{figure}
As expected, inclusion of dynamical effects through $Z^{1/2}$ renormalizes the energy scales.  Without these effects, the mini--gap would increase monotonically with the coupling strength, but dynamical effects limit its maximum value. The optimal quasi--particle gap obtains in the intermediate coupling regime $\overline{\gamma}\sim \Delta_0$. Further increase of the coupling leads to a decrease of the gap.  To illustrate the features described above, we show in Fig. \ref{Fig6} low--energy spectra in the vicinity of the ``sweet spot'' ($\mu=14.5E_\alpha$, $\Gamma=15.3E_\alpha$) at  weak coupling ($\overline{\gamma}=0.25\Delta_0$, top panel) and intermediate coupling ($\overline{\gamma}=\Delta_0$, bottom panel). Note that in the absence of inter--band pairing, i.e., for homogeneous SM--SC coupling, the gap near the ``sweet spot'' collapses. Also, inclusion of dynamical effects at intermediate and strong coupling is the key for obtaining the correct energy scales.  Finally, the location of the ``sweet spots'' in the $\Gamma-\mu$ plane depends on the coupling strength and, more generally,  the location of phase boundaries depends on the strength of the SC proximity effect.

\begin{figure}[tbp]
\begin{center}
\includegraphics[width=0.48\textwidth]{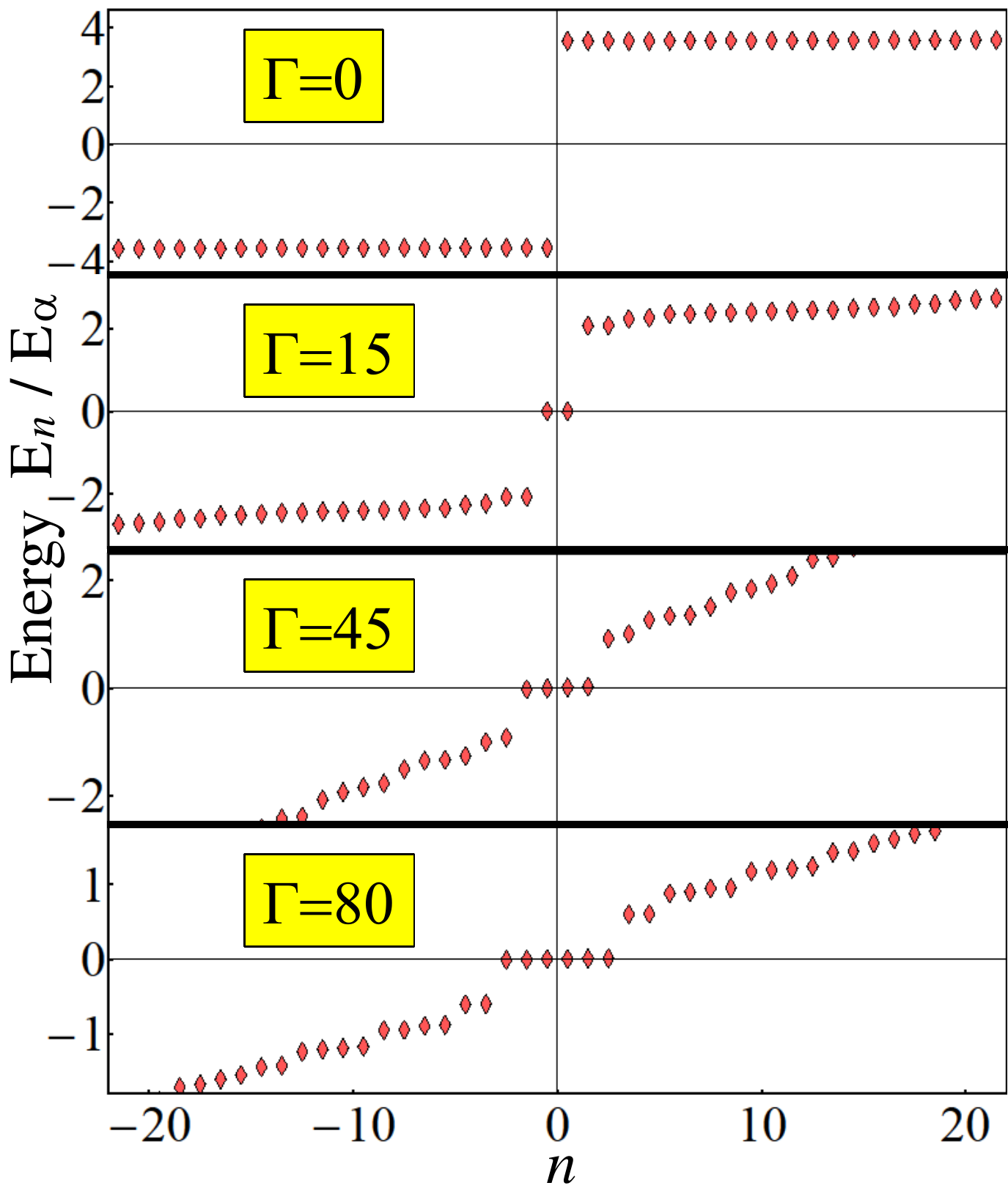}
\vspace{-7mm}
\end{center}
\caption{(Color online) Sequence of low-energy spectra obtained for four different values of the Zeeman field separated by points with a vanishing mini-gap. The spectra with an odd number of pairs of zero-energy modes ($N$) characterize topological SC phases, while those with an even $N$ correspond to trivial SC phases. Note the overall decrease of the mini-gap with the Zeeman field. The system is characterized by the following parameters: $\mu=30E_\alpha$, $\overline{\gamma}=0.25\Delta_0$, and $\theta=0.8$.}  \label{Fig7}
\end{figure}

\subsection{Phase diagram for multi--band superconducting nanowires}\label{SecIIIB}

Topological superconductivity and, implicitly, the Majorana bound states are protected by the quasi--particle gap $\Delta_\infty^*$, as discussed above. The vanishing of  $\Delta_\infty^*$ signals a transition between topologically nontrivial and topologically trivial superconductivity. In a multi--band system, such transitions can be caused, for example, by varying the Zeeman field while maintaining a fixed value of the chemical potential. The vanishing of $\Delta_\infty^*$ at certain specific values of $\Gamma$ reveals a sequence of alternating SC phases with trivial and nontrivial topological properties.  A natural question is whether different topologically nontrivial (or trivial) phases have exactly the same low--energy properties.  While topologically identical, these phases may have have some distinct features, at least in  certain parameter regimes.

\begin{figure}[tbp]
\begin{center}
\includegraphics[width=0.48\textwidth]{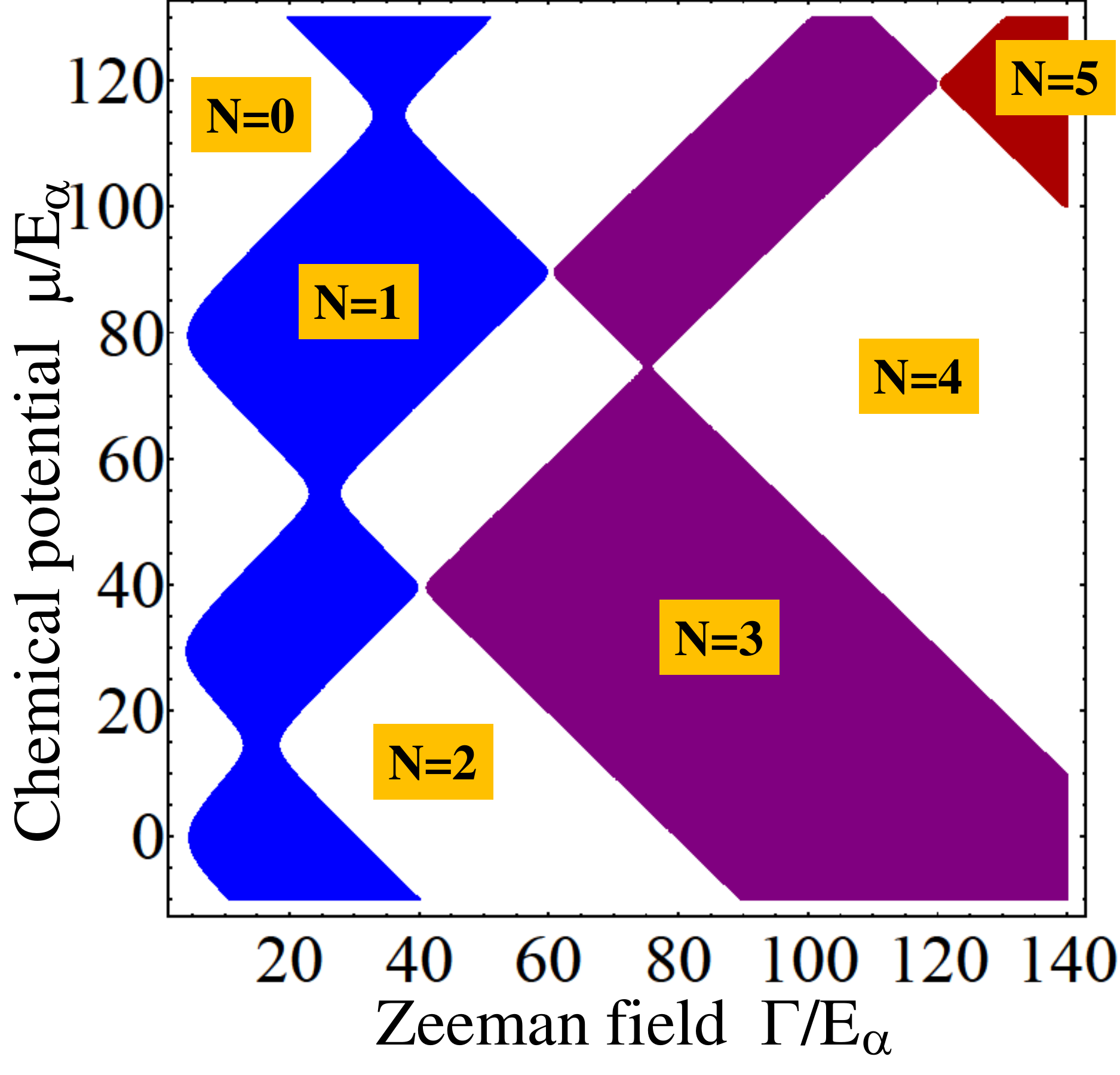}
\vspace{-7mm}
\end{center}
\caption{(Color online) Phase diagram of the multiband nanowire as function of the Zeeman field $\Gamma$ and the chemical potential $\mu$.  The quasi--particle gap $\Delta_\infty^*$ of the effective low--energy Hamiltonian (\ref{Heff}) vanishes at the phase boundaries. Superconducting phases characterized by an odd (even) number of pairs of zero--energy Majorana modes are topologically nontrivial (trivial). The coupling of the SM nanowire to the s--wave SC is characterized by $\overline{\gamma}=0.25\Delta_0$ and $\theta=0.8$. The inhomogeneous coupling induces off--diagonal pairing $\Delta_{n_y n_y^\prime}$ which removes the x-points creating regions with stable nontrivial (near the ``sweet spots'') or trivial SC. }  \label{Fig8}
\end{figure}

To address this question, we calculate the low-energy spectrum  for a system with a fixed chemical potential, $\mu = 30E_\alpha$, and different  values of $\Gamma$, one  from each of four intervals separated by points characterized by a vanishing gap. The results are shown in Fig. \ref{Fig7}. The distinctive feature of the spectra in Fig. \ref{Fig7} is their number of zero--energy modes. In the presence of a very weak Zeeman field, the SC phase has trivial topology and no zero--energy modes. Increasing $\Gamma$ generates a transition to a topological SC phase characterized by one pair of Majorana bound states. The further increase of the applied Zeeman field produces alternating topologically trivial and nontrivial SC phases with increasing number of zero--energy bound states. Topological SC phases are characterized by an odd number of pairs of Majorana bound states, while trivial SC phases have an even number of pairs.

\begin{figure}[tbp]
\begin{center}
\includegraphics[width=0.48\textwidth]{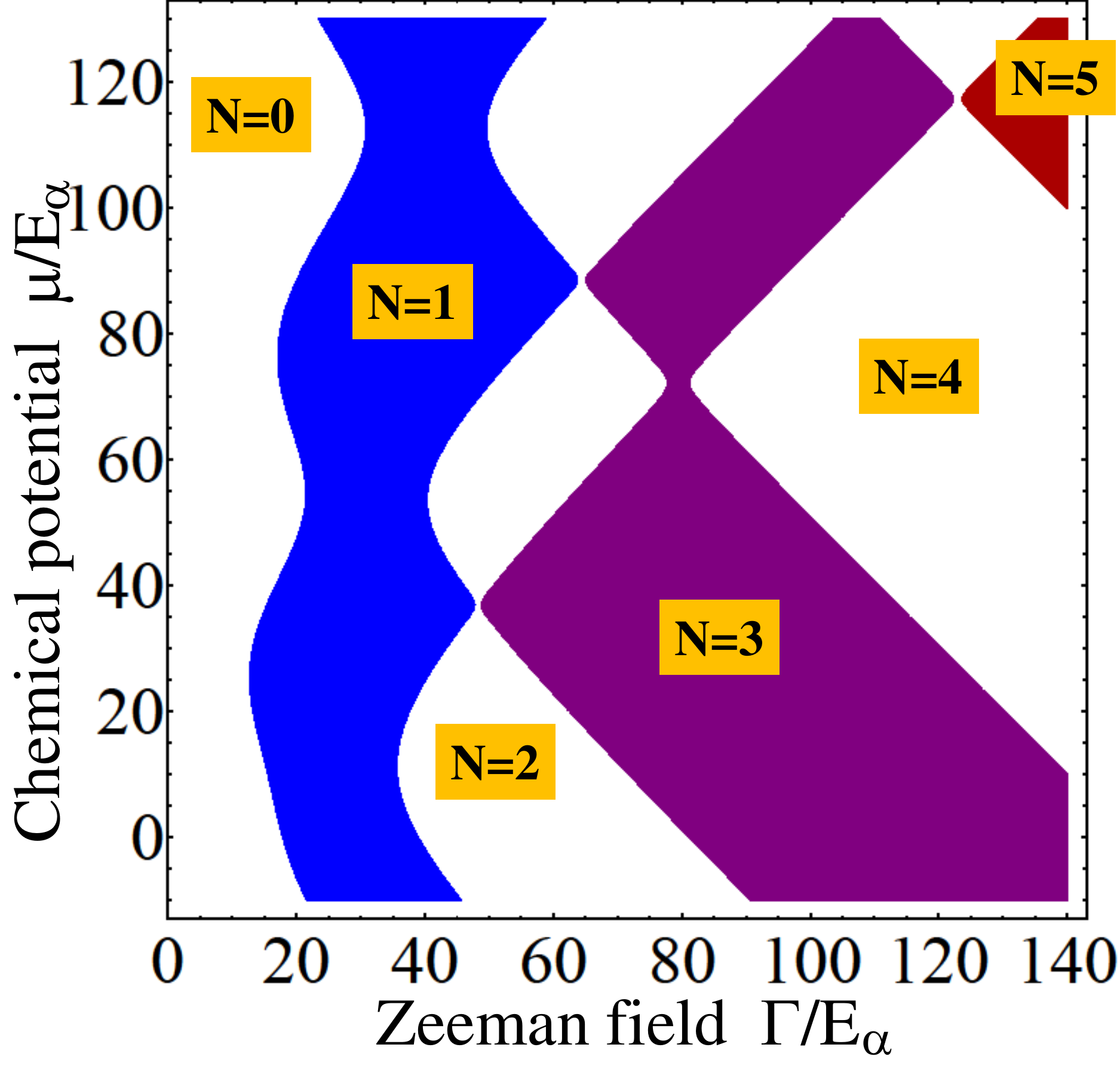}
\vspace{-7mm}
\end{center}
\caption{(Color online) Phase diagram of the multiband nanowire at intermediate coupling. The SM--SC coupling is $\overline{\gamma}=\Delta_0$, while the other parameters are the same as in Fig. \ref{Fig8}.  Note that the ``sweet spots'' of the topological phase characterized by $N=1$ are significantly expanded as compared with the weak--coupling case shown in Fig. \ref{Fig8}. As a result, tuning the Zeeman field in the vicinity of $\Gamma=30E_\alpha$ allows for huge variations of the chemical potential without crossing a phase boundary, i.e., without closing the gap.}  \label{Fig9}
\end{figure}

This type of sequence of alternating SC phases is independent of the chemical potential or the strength of the SM--SC coupling. As discussed above, in the weak--coupling limit  $\overline{\gamma}\rightarrow 0$ this behavior can be directly related to the number of sub--bands of an infinite wire that cross the chemical potential: an odd (even)  number corresponds to a nontrivial (trivial) topological superconductor. At the phase boundary between two SC phases with different topologies, two sub--bands become degenerate at $k_x=0$. In addition, at certain special values of the control parameters $\Gamma$ and $\mu$ two different phase boundaries intersect leading to multicritical points. We will call these special crossing points x-points. The ``sweet spots'' mentioned above are examples of such x-points. The key question is how the phase boundaries evolve when we turn on the SM--SC coupling and, in particular, what the physics is in the vicinity of the x-points, see also Ref.~\cite{LutchynFisher'11}.
To obtain the global phase diagram in the $\Gamma-\mu$ plane, we determine the parameters that satisfy the condition $\Delta_\infty^*(\Gamma, \mu)=0$, i.e., we identify the phase boundaries for transitions between topologically trivial and non--trivial phases by imposing the condition of a vanishing quasi--particle gap. We show below that this approach is consistent with the calculation of the topological index $\cal M$ (Majorana number) which distinguishes topologically trivial and non-trivial SC phases~\cite{Kitaev'01, lutchyn_multi}.
The results are shown in Fig.\ref{Fig8} for weak coupling ($\overline{\gamma}=0.25\Delta_0$) and in Fig.\ref{Fig9} for intermediate coupling ($\overline{\gamma}=\Delta_0$).

For $\overline{\gamma}= 0.25\Delta_0$ (see Fig. \ref{Fig8}), the only significant difference as compared to the weak--coupling picture presented above is the disappearance of the phase boundary crossings at the x-points. Instead, the region in the vicinity of these points is occupied by the phase that is robust against variations of the chemical potential. Near the ``sweet spots'', this phase is the nontrivial topological SC. We note that the disappearance  of phase boundary crossings is a direct result of the off--diagonal pairing induced by a non-uniform SM--SC coupling. Uniform tunneling ($\theta=0$) does not eliminate the x-points, independent of the coupling strength,  but pushes them to higher values of the Zeeman field as $\overline{\gamma}$ increases.
Also, we note that the characteristic width of the
stable phase in a given avoided crossing region is controlled  by a specific off--diagonal component $\Delta_{n_y, n_y^\prime}$. For example, the ``sweet spots'' inside the phase characterized by $N=1$ (see Fig. \ref{Fig8}) are controlled by the dominant matrix elements $\Delta_{n_y n_y+1}$ (see Fig. \ref{Fig3}), while the  avoided  crossings  within the $N=2$ topologically trivial phase are controlled by matrix elements $\Delta_{n_y n_y+2}$, that are typically smaller. Hence, we expect the strongest effect within the topological phase characterized by one pair of Majorana bound states. As this phase also requires relatively low Zeeman fields, it is the experimentally relevant phase for realizing and observing Majorana fermions. At intermediate couplings, the $N=1$ phase is pushed to slightly higher fields (see Fig. \ref{Fig9}). However, as a result of the effective phase space of the ``sweet spots`` expanding significantly, this regime presents the remarkable possibility of being able to vary the chemical potential over energy scales of the order $10$meV  without crossing a phase boundary. As we will show below, this feature has major experimental implications in the sense that the elusive Majorana mode is most likely to be experimentally realized in the laboratory in this particular physically realistic parameter regime. We note that this interesting parameter regime exists for the multiband situation.

\begin{figure}[tbp]
\begin{center}
\includegraphics[width=0.48\textwidth]{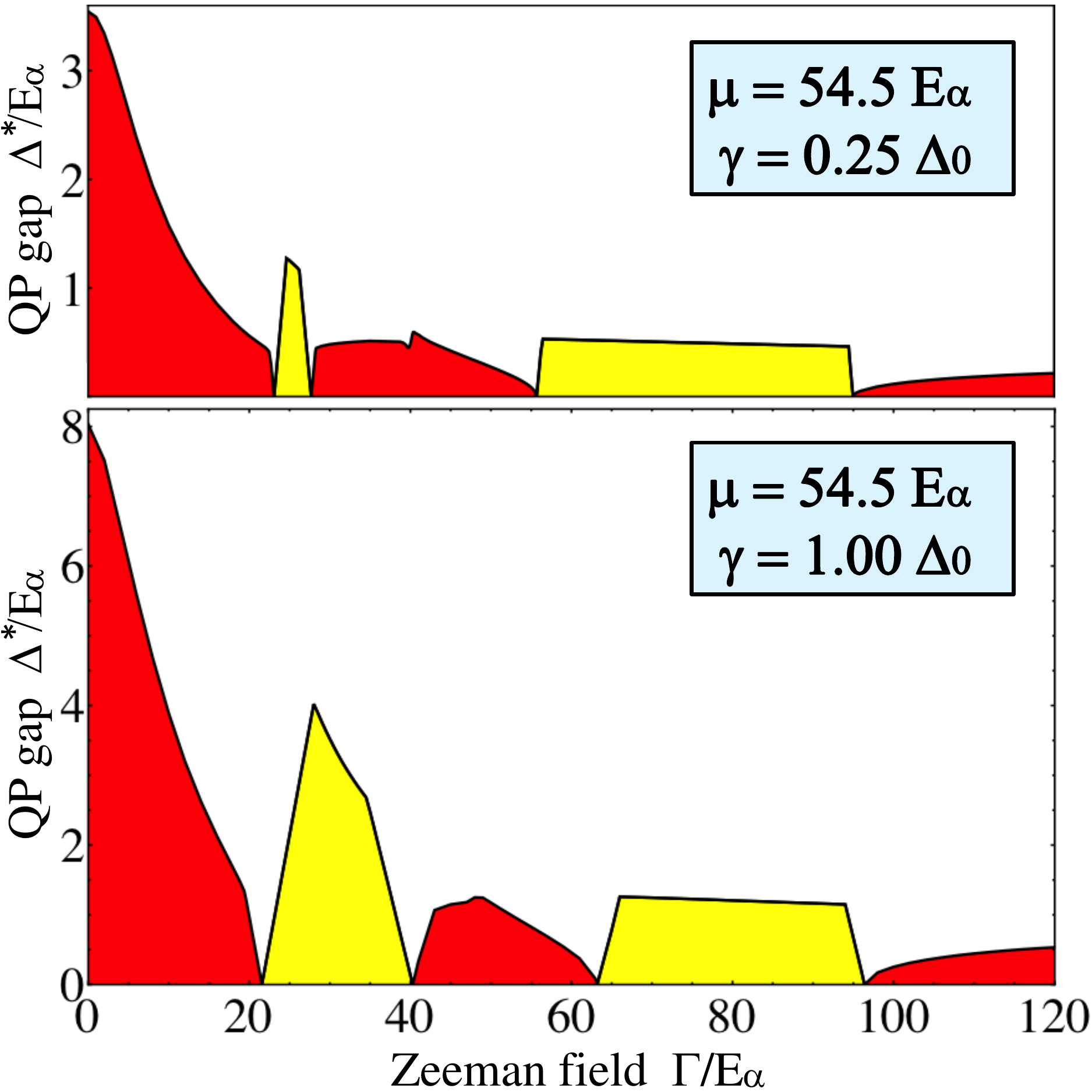}
\vspace{-7mm}
\end{center}
\caption{(Color online) Dependence of the gap on the Zeeman field for a fixed chemical potential. The upper panel corresponds to weak SM--SC coupling with $\overline{\gamma}=0.25\Delta_0$ and $\theta=0.8$, while the lower panel is obtained for an intermediate coupling with $\overline{\gamma}=\Delta_0$ and $\theta=0.8$. Within the red (darker gray) regions superconductivity is topologically trivial, while in the yellow (light gray) regions  the system is topologically nontrivial. These curves correspond to horizontal cuts in the phase diagrams shown in figures \ref{Fig8} and \ref{Fig9}, respectively,  through a sweet spot of the phase $N=1$. Note that in the limit $\theta\rightarrow 0$, i.e., for uniform SM--SC coupling, the width of the lower field topologically nontrivial region shrinks to zero for both coupling strengths.}  \label{Fig10}
\end{figure}

\subsection{Dependence of the gap on the Zeeman field and the chemical potential}\label{SecIIIC}

\begin{figure}[tbp]
\begin{center}
\includegraphics[width=0.48\textwidth]{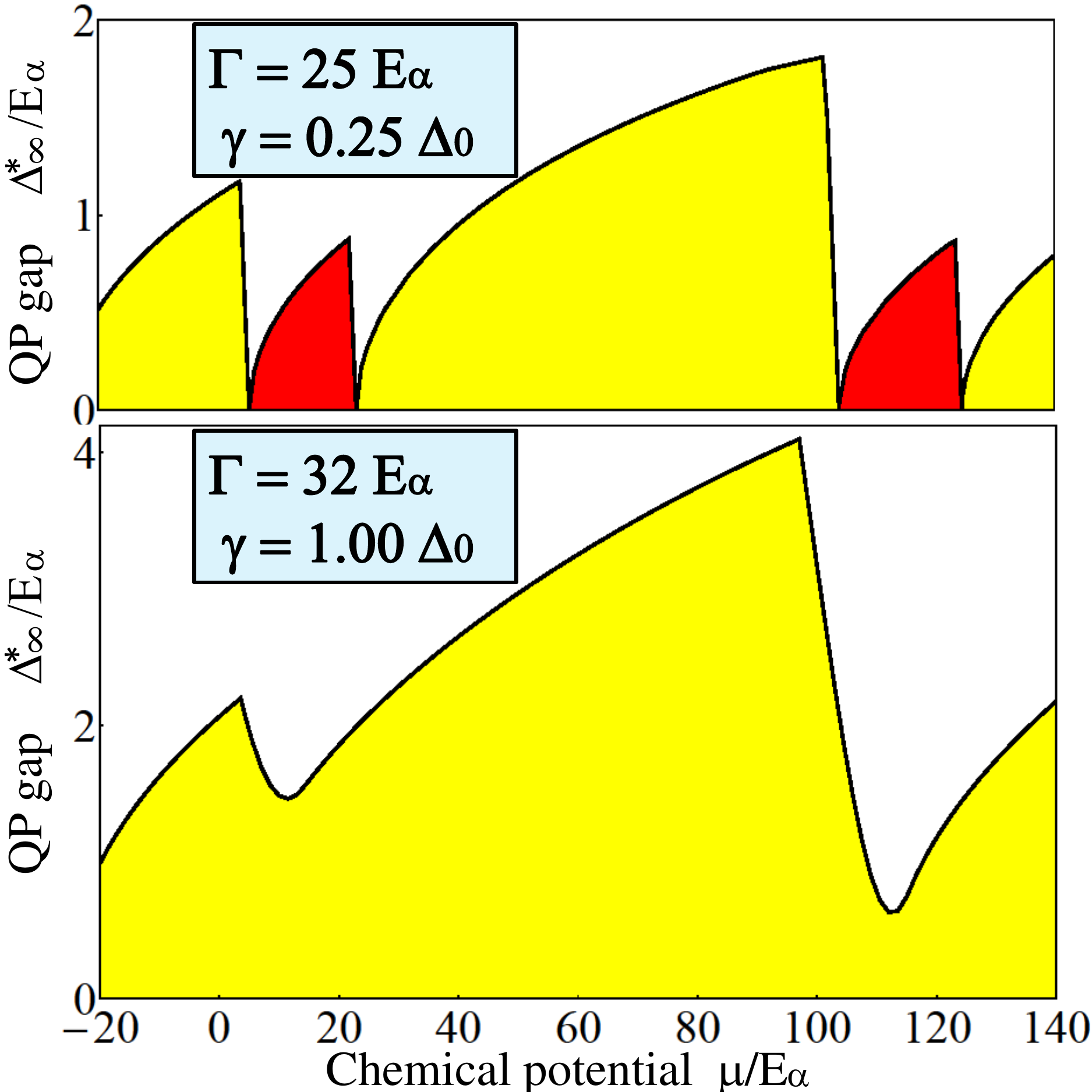}
\vspace{-7mm}
\end{center}
\caption{(Color online) Dependence of the  quasi--particle gap on the chemical potential for a fixed Zeeman field. The upper panel corresponds to weak coupling, $\overline{\gamma}=0.25\Delta_0$, and $\Gamma=25E_\alpha$, while the lower panel is obtained at intermediate coupling, $\overline{\gamma}=\Delta_0$, and $\Gamma=32E_\alpha$. The curves represent vertical cuts through the $N=1$ topological phase in the phase diagrams shown in figures \ref{Fig8} and \ref{Fig9}. At intermediate coupling (lower panel) the gap is finite over the entire chemical potential range, making the topological SC phase in a system with average chemical potential $\bar{\mu}\approx 60E_\alpha$ is robust against variations of the chemical potential of the order $\delta\mu = 5$meV.} \label{Fig11}
\end{figure}

The phase diagram provides information about the topological nature of the phase characterized by given sets of control parameters $(\Gamma, \mu)$. However, we are ultimately interested in the stability of the Majorana bound states that occur at the ends of a nanowire within the topologically nontrivial phase. As the Majorana zero--energy modes are protected by the quasi--particle gap, knowing the size of the gap and the dependence of $\Delta_\infty^*$ on the control parameters is critical.  To address this issue, we determine the dependence of the gap on both the Zeeman field at fixed chemical potential and on $\mu$ at fixed $\Gamma$. The results are shown in figures \ref{Fig10} and \ref{Fig11}. In general, the gap is non-zero everywhere except at points corresponding to phase boundaries. In the vicinity of a point $(\Gamma_c, \mu_c)$ with $\Delta^*(\Gamma_c, \mu_c)=0$ the minimum gap in an infinitely long wire occurs at $k_x=0$. The dependence of this minimum on the Zeeman field is approximately linear in the deviation from the critical field, $|\Gamma-\Gamma_c|$ (see Fig. \ref{Fig10}). This generalizes the single--band results shown in Fig. \ref{Fig2}. Note that outside this linear regime, the minimum gap occurs at finite wave vectors. Finally, we note that, at intermediate couplings, the Zeeman field can be tuned so that the gap remains finite over a large range of chemical potentials (Fig. \ref{Fig11}, bottom panel). Such regimes are extremely stable against fluctuations of the chemical potential produced by disorder or other perturbations, as we will show explicitly in the next section. We emphasize that the critical ingredients for realizing  this regime are: i) the off-diagonal pairing obtained by a non--uniform SM--SC coupling, and ii) an effective average coupling $\overline{\gamma}$ of the order of the bare SC gap $\Delta_0$.  Note that the coupling strength $\overline{\gamma}\propto \tilde{t}^2/\Lambda L_z^3$ can be controlled by either modifying the tunneling $\tilde{t}$, or by changing the width $L_z$ of the nanowire in the direction perpendicular to the interface with the superconductor.

\begin{figure}[tbp]
\begin{center}
\includegraphics[width=0.48\textwidth]{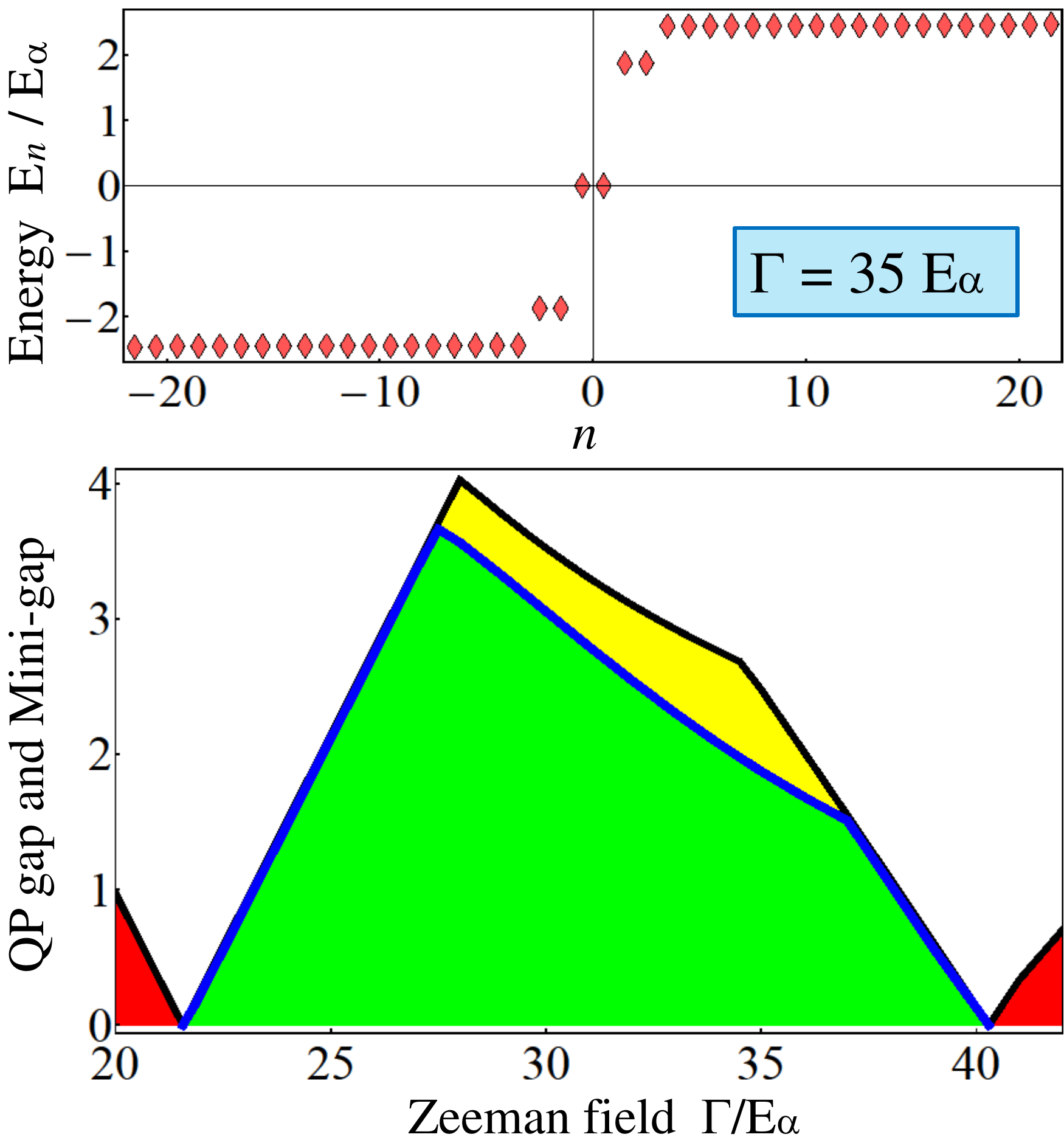}
\vspace{-7mm}
\end{center}
\caption{(Color online) Top panel: Spectrum  of a system characterized by $\mu=54.5E_\alpha$, $\Gamma=35E_\alpha$, $\overline{\gamma}=\Delta_0$, and $\theta=0.8$. The states with ${\it n}=\pm 1$ are Majorana zero--modes,  ${\it n}=2, 3$ correspond to states localized near the ends of the wire, and states with ${\it n}\geq 4$ are extended states. The mini--gap is $\Delta^*=E_2$ and the minimum quasi--particle (QP) gap is $\Delta_\infty^*=E_4$ The finite--energy localized states can be viewed as precursors of the extra--pair of Majorana bound states that obtains for $\Gamma>40E_\alpha$ . Bottom panel: Dependence of the quasi--particle (QP) gap  $\Delta_\infty^*$ and mini--gap  $\Delta^*$ on the Zeeman field for a system with the same parameters as in Fig. \ref{Fig10}. Note that  $\Delta^*=\Delta_\infty^*$  in the vicinity of phase transition points. The quasi--particle gap   $\Delta_\infty^*$  has the same values as in Fig.  \ref{Fig10}.} \label{Fig12}
\end{figure}

As we mentioned above, the value of the gap in a finite system is, in general, smaller that the minimum gap in a system with the same parameters but no ends, e.g., with periodic boundary conditions. We emphasize that this is not a finite size effect, but is due to the appearance  of in--gap states that are localized near the ends of the wire. In the topologically non--trivial phase, the characteristic length scale for these states is the same as that of the Majorana zero modes and, for wires with lengths larger than this scale, their energy is independent of $L_x$.  To illustrate this behavior, we show in Fig. \ref{Fig12} (top panel) the spectrum of a system characterized by $\mu=54.5E_\alpha$, $\Gamma=35E_\alpha$, $\overline{\gamma}=\Delta_0$, and $\theta=0.8$.   We notice a pair of zero--energy Majorana modes and a number of states with almost the same energy that are extended over the entire length of the wire, as we checked explicitly. In addition, we notice pair of states with energy within the bulk gap. An analysis of the position dependence of $|\psi(x)|^2$ reveals that these states are localized near the ends of the wire.  The minimum energy of the extended states is equal to the quasi--particle gap $\Delta_\infty^*$, while the lowest energy of the bound states is equal to the mini--gap $\Delta^*$. We can understand these bound states as precursors of the extra--pair of Majorana zero--modes that characterize the topologically trivial phase that obtains for $\Gamma > 40 E_\alpha$. Increasing the magnetic field pushes down the energy of the bound states until it vanishes at the transition, when   $\Delta^*=\Delta_\infty^*=0$. On the other side of the transition, an extra--pair of localized states will be characterized by zero energy (see Fig. \ref{Fig7}). The evolution of the mini--gap with the Zeeman field is shown in the lower panel of Fig. \ref{Fig12}. Note that in the vicinity of the transition points $\Delta^*=\Delta_\infty^*$, while deep inside the topological phase  $\Delta^*<\Delta_\infty^*$. Similar behavior can be observed throughout the phase diagram, including the topologically trivial phases.

\subsection{Phase diagram for an effective three--band model using the topological invariant.}

In this section we consider an effective three-band toy model which allows one to qualitatively understand several features observed in the detailed numerical simulations discussed in the previous sections.

The topological phase diagram for the multiband nanowire can be obtained analytically using topological arguments due to Kitaev~\cite{Kitaev'01}. Namely, the superconducting phase hosting Majorana fermions has an odd fermion parity whereas the non-topological phase has even fermion parity. Thus, the phase diagram can be obtained by calculating the $Z_2$ topological index $\cal M$ (Majorana number) defined as
\begin{align}\label{eq:Majorana}
{\cal M}={\rm sgn}[{\rm Pf} B(0)]{\rm sgn}[{\rm Pf} B(\pi/a)]=\pm 1.
\end{align}
The change of $\cal M$ signals the transition between trivial (${\cal M}=1$) and non-trivial phases (${\cal M}=-1$). The antisymmetric matrix $B$ in Eq.\eqref{eq:Majorana} represents the Hamiltonian of the system in the Majorana basis~\cite{Kitaev'01} and can be constructed by the by the virtue of particle-hole symmetry~\cite{Lutchyn'10, Ghosh'10}. Specifically, the particle-hole symmetry of the BdG Hamiltonian is defined as
\begin{align}
\Theta H_{\rm BdG}(p)\Theta^{-1}=-H_{\rm BdG}(-p),
\end{align}
where $\Theta=UK$ is an anti-unitary operator with $U$ and $K$ representing unitary transformation and complex conjugation, respectively. One can check that the matrix $B(P)\!=\!H_{\rm BdG}(P)U$ calculated at the particle-hole invariant points where $H_{\rm
BdG}(P)\!=\!H_{\rm BdG}(-P)$ is indeed antisymmetric $B^T(P)\!=\!-B(P)$. In 1D there are two such points: $P\!=\!0,\!\frac{\pi}{a}$ with $\frac{\pi}{a}$ being the momentum at the end of the Brillouin zone and $a$ being the lattice spacing. The function $\rm Pf$ in Eq.~\eqref{eq:Majorana} denotes Pfaffian of the antisymmetric matrix $B$. In the continuum approximation, where the lattice spacing $a \rightarrow 0$ and $P\!=\!\pi/a\!\rightarrow\! \infty$, the value of ${\rm sgn}[{\rm Pf} B(\pi/a)]\!=\!+1$. Thus, the topological phase boundary given by the change in the topological index is determined by ${\rm Pf} B(0)$, and, thus, is accompanied by the gap closing at zero momentum. Note that the topological reconstruction of the spectrum is always accompanied by closing of the bulk gap~\cite{read_prb'00} since ${\rm Det} \, H_{\rm BdG}\!=\!{\rm Pf} B^2$. Our approach of calculating the TP invariant relies on translational symmetry. Recently, the expression for the TP invariant was generalized to spatially inhomogeneous case~\cite{Akhmerov'11, Beenakker'11}.

We now calculate ${\rm Pf} B(0)$ for a simplified three-subband model and compare the phase diagram with the numerical one presented in the previous section. To make progress we assume that $\Delta_{i,i}=\Delta$, $\Delta_{i,i+1}=\Delta'$ and consider only diagonal in the subband index spin-orbit coupling terms in Eq.~\eqref{HSOI}. With these approximations, ${\rm Pf} B(0)$ becomes
\begin{widetext}
\begin{align}
{\rm Pf} B(0)&=\left(\delta E_{12}^2 \left(-V_x^2+\Delta^2+(\delta E_{13}-\mu)^2\right) \left(V_x^2-\Delta ^2-\mu ^2\right)-\delta E_{13}^2 \left(V_x^2-(\Delta -\Delta')^2-\mu^2\right) \left(V_x^2-(\Delta +\Delta')^2-\mu ^2\right)\right. \nonumber\\
&\left. +\left(V_x^2-\Delta ^2-\mu ^2\right) \left(V_x^4+\left(\Delta ^2-2\Delta'^2\right)^2+2 \left(\Delta ^2+2\Delta'^2\right) \mu ^2+\mu ^4-2 V_x^2 \left(\Delta^2+2\Delta'^2+\mu ^2\right)\right) \right. \nonumber\\
&\left. +2\delta E_{13} \mu  \left(V_x^4+\Delta^4-\Delta ^2 \Delta'^2+2\Delta'^4+\left(2 \Delta ^2+3\Delta'^2\right) \mu ^2+\mu ^4-V_x^2 \left(2 \Delta ^2+3\Delta'^2+2 \mu ^2\right)\right) \right. \nonumber\\
&\left. +2 \delta E_{12} \left(\delta E_{13}^2 \mu  \left(-V_x^2+\Delta ^2+\Delta'^2+\mu ^2\right)+\mu  \left(-V_x^2+\Delta ^2+\mu ^2\right) \left(-V_x^2+\Delta ^2+2\Delta'^2+\mu ^2\right) \right. \right. \nonumber \\
& \left. \left. +\delta E_{13} \left((V_x^2-\Delta^2)\Delta'^2+\left(2 V_x^2-2 \Delta ^2-3\Delta'^2\right) \mu ^2-2 \mu ^4\right)\right)\right). \label{PfB0}
\end{align}
\end{widetext}
Here $\delta E_{12(3)}$ represents the energy difference between first and second (third) subbands due to the transverse confinement. The superconducting gaps are related to the nominal bulk gap $\Delta_0$ via relations $\Delta=\gamma \Delta_0 /(\gamma+\Delta_0)$ and $\Delta'=0.25\gamma \Delta_0 /(\gamma+\Delta_0)$ which take into account the dependence of the induced parameters on tunneling strength. Here we have chosen a reasonable ratio of $\Delta'/\Delta=0.25$. Similarly, all other energies are renormalized in the following way $E \rightarrow E \frac{\Delta_0}{\gamma+\Delta_0}$ as explained in the previous sections.

\begin{figure}[tbp]
\begin{center}
\includegraphics[width=0.49\textwidth]{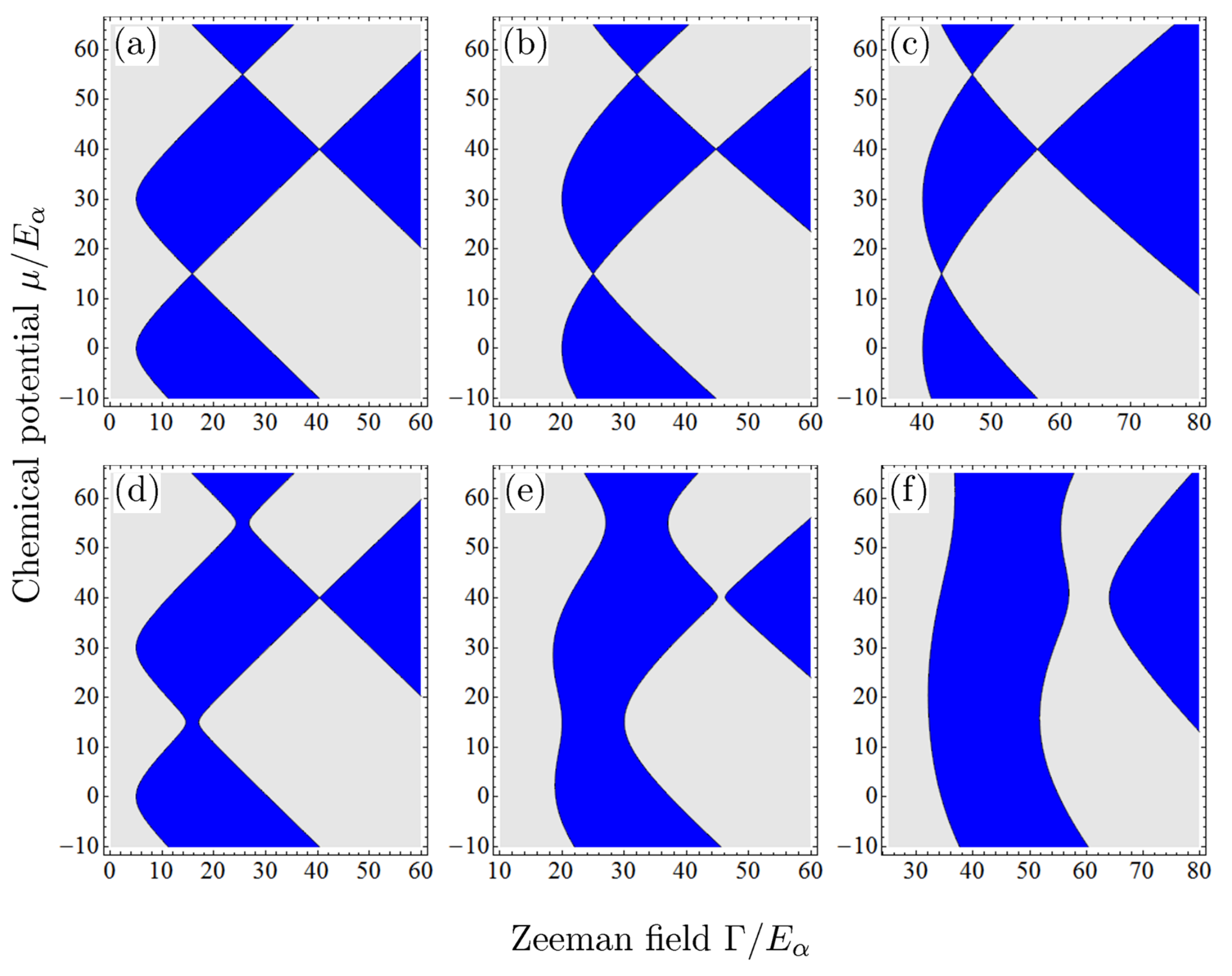}
\vspace{-7mm}
\end{center}
\caption{(Color online) Phase diagram of the multiband nanowire at different SM-SC tunneling strengths $\gamma$ and interband pairing $\Delta_{12}=\Delta_{23}=\Delta'$ calculated analytically for the effective three-band model. (a)-(c) correspond to $\gamma=5,20,40$, respectively, and superconducting gaps $\Delta/E_{\alpha}=4$ and $\Delta'/E_{\alpha}=0$. (d)-(f) correspond to $\gamma=5,20,40$, respectively, and superconducting gaps $\Delta/E_{\alpha}=4$ and $\Delta'/E_{\alpha}=1$. We used here $\delta E_{12}/E_{\alpha}=30$ and $\delta E_{13}/E_{\alpha}=80$.} \label{Fig13}
\end{figure}

The phase diagram showing a sequence of topological phase transitions for the three sub-band toy model is shown in Fig.~\ref{Fig13}. The panels (a)-(c) represent the phase diagram with no interband mixing terms (i.e. $\Delta'=0$). One can clearly see crossings in the phase diagram which represent the ``sweet spots". One can also notice the effect of the renormalization due to SM-SC tunneling as we increase $\gamma$ - the superconducting and normal terms are rescaled in a different way as explained above. In the weak-coupling regime $\gamma \ll \Delta_0$, the normal terms are not significantly renormalized since $\frac{\Delta_0}{\gamma+\Delta_0}\approx 1$ whereas induced pairing is small $\Delta\approx \gamma$ and is entirely determined by the normal state level broadening $\gamma$. On the other hand, in the strong-coupling regime $\gamma \gg \Delta_0$,  the normal terms are decreasing with $\gamma$ because $\frac{\Delta_0}{\gamma+\Delta_0}\approx \Delta_0/\gamma \ll 1$ whereas $\Delta$ saturates at $\Delta_0$. Thus, strong SM-SC tunneling leads to important quantitative effect which should be taken into account in a realistic model for the proximity effect.

The panels (d)-(f) represent the phase diagram with finite interband mixing terms $\Delta' \neq 0$. Here we find qualitative agreement with the numerical results presented in the previous sections, compare Figs.~\ref{Fig8} and \ref{Fig9} with \ref{Fig13} (d) and (e). Different renormalization of the normal and SC terms has a two-fold effect on the phase diagram: a) with the increase of tunneling the topological phase is effectively ``pushed" towards higher magnetic fields; b) even small interband pairing term opens a large gap at the sweet spot leading to extended vertical topological regions. This insensitivity of the topological phase against chemical potential fluctuations can be exploited for the protection against disorder in the multisubband occupied nanowires with no such situation arising in the single channel case.


\subsection{Phase diagram in the presence of a transverse external potential.}

\begin{figure}[tbp]
\begin{center}
\includegraphics[width=0.49\textwidth]{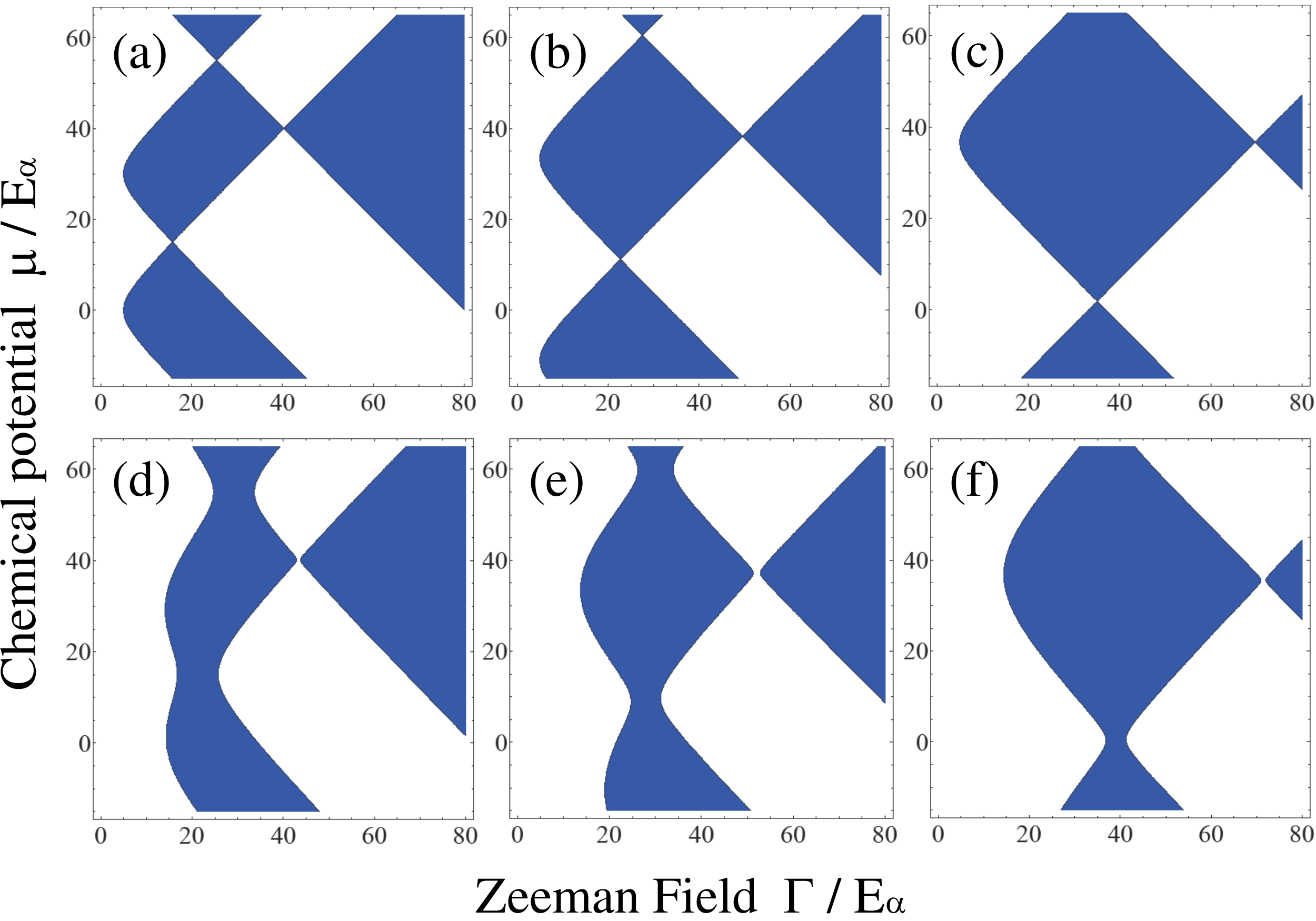}
\vspace{-7mm}
\end{center}
\caption{(Color online) Phase diagram of a multi--band wire  in a transverse external field. The strength of the SM--SC coupling is $\overline{\gamma}=\Delta_0$, with uniform coupling ($\theta=0$, i.e., $\Delta^\prime=0$) for the top panels and non--uniform tunneling ($\theta\approx 0.4$,  $\Delta^\prime=\Delta/4$) corresponding to the bottom panels. The amplitude of the transverse potential is: $V_0=0$ for panels (a) and (d), $V_0=50E_\alpha$ for (b) and (e), and $V_0=100E_\alpha$ for (c) and (f).  Note that the topologically nontrivial phase has a vanishing width at the sweet spots in the absence of inter--sub--band pairing (a-c) for all values of the external potential. For non--uniform coupling (d-e), the transverse potential reduces the width of the topological phase near the sweet spots. .} \label{Fig13x}
\end{figure}

A key feature of the phase diagrams shown above is represented by the  hot spots. In the presence of inter sub--band pairing, the topologically nontrivial phases expand in the vicinity of these hot spots, and, for certain values of the Zeeman field, become stable over a wide range of values for the chemical potential. As we show in the subsequent sections, this property is critical for stabilizing the topological superconducting phase and, ultimately, for realizing and observing the Majorana zero modes. The necessary ingredient for obtaining this expansion of the hot spots is a non--vanishing  inter--sub--band pairing, which is obtained by a non--uniform SM--SC coupling characterized by a strong position dependence along the direction transverse to the wire (see Fig. \ref{Fig3}). The natural question is whether this effect can be obtained by breaking the symmetry in the transverse direction using an external field, e.g., generated by a gate potential, instead of a non--uniform coupling. This would constitute an alternative to engineering non--uniform SM--SC interfaces that would be much simpler to implement and would allow better control. To investigate this possibility, we consider an external potential that varies  linearly  across the nanowire,
\begin{equation}
V_{\rm ext}(y) = V_0(2y/Ly-1),  \label{Vy}
\end{equation}
where $V_0$ is the amplitude of the transverse external potential. The matrix elements for the external potential are strictly off--diagonal and couple strongly the neighboring sub--bands, while other contributions are at least one order of magnitude smaller. Consequently, to a first approximation we have , $\langle n_y|V_{\rm ext}|n_y^\prime\rangle\approx  0.4V_0 \delta_{n_y^\prime, n_y\pm 1}$. Even in the presence of this inter--band coupling, the Pfaffian  ${\rm Pf} B(0)$ for the effective three--band model can be determined analytically and it is given by an expression that generalizes Eq. (\ref{PfB0}).
The corresponding phase diagrams for both uniform and non--uniform SM--SC couplings and different values of the external potential are shown in Fig.  \ref{Fig13x}.

The key conclusion of this calculation is that inter--sub--band mixing due to an external transverse potential does not lead to an expansion of the topological phase in the vicinity of the sweet spots, but rather determine a shift in their position. This is illustrated by  diagrams (a-c) in Fig. \ref{Fig13x}, which correspond to homogeneous SM-SC coupling, i.e., $\theta=0$, and different values of the external field. Notice that, to a first approximation, the effect of the transverse potential is equivalent to increasing the inter--sub--band spacing $\delta E_{n_y n_y^\prime}$.  If the transverse potential is applied across a wire that is non--uniformly coupled to the SC, in addition to shifting the position of the sweet spots, it reduces the stability of the topological phase in their vicinity, as illustrated in panels (d-f) for $\theta\approx0.4$. In addition, as a consequence of effectively increasing the inter--sub--band spacing, the regions between successive  sweets spots occupied by the topological SC phase expands with increasing the transverse potential, as shown in panels (c) and (f). Note that, in the presence of non--uniform SM--SC coupling, the  phase diagrams for $V_0>0$ and $V_0<0$ (not shown in Fig. \ref{Fig13x} are slightly different. The strong dependence of the phase boundaries on the transverse potential, especially in the vicinity of the sweet spots, could be used experimentally for driving  the system across the topological phase transition by tuning a gate potential, instead of changing the magnetic field.

\section{Effect of disorder on the topological superconducting phase}\label{sec:Disorder}

In this section we consider the effect of disorder on the stability of the topological superconducting phase harboring Majorana fermions. In a realistic system, disorder comes in various ways that affect the topological phase very differently. In this paper we consider three types of disorder: impurities in the s--wave superconductor, short-- and long--range disorder in the semiconductor wire, and random nonuniform coupling between the semiconductor wire and the superconductor, which mimics a rough interface and the imprecision in engineering inhomogeneous $y$--dependent couplings. We begin by considering short--range impurities in the bulk superconductor, followed by the consideration of the other two types of disorder which are both more complex to treat theoretically and more detrimental to the existence of the Majorana modes.

\subsection{Short-range impurities in the bulk superconductor.}

In order to understand the effect of non--magnetic impurities on the induced superconductivity in the semiconductor, we first review the results on the proximity effect for the infinite planar interface presented in the previous section. The basic idea is that the presence of short--range non--magnetic impurities in the metal modifies the bulk Green's function,  which  then is used to derive the appropriate superconducting proximity effect.

\begin{figure}[tbp]
\begin{center}
\includegraphics[width=0.48\textwidth]{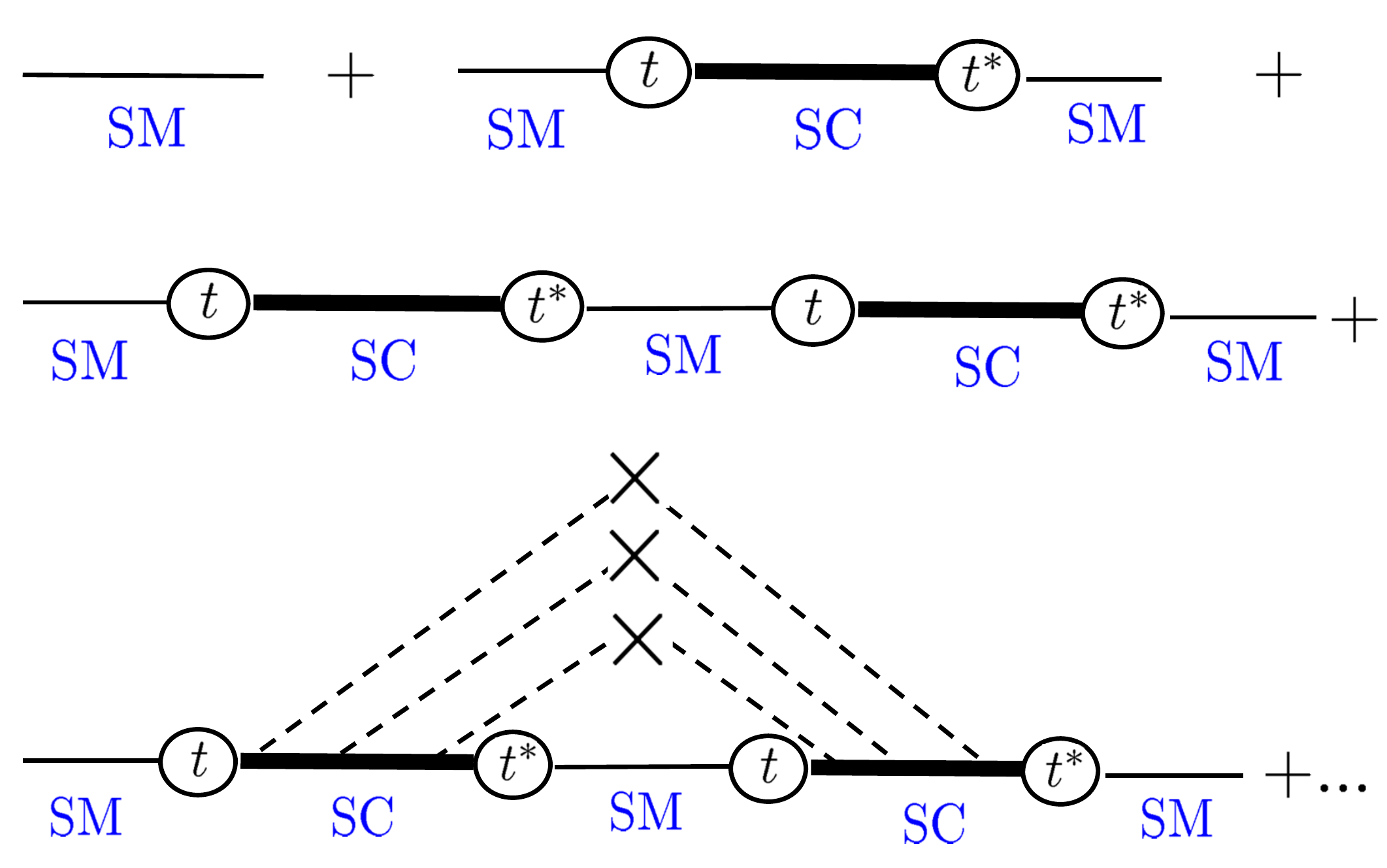}
\vspace{-7mm}
\end{center}
\caption{(Color online) Diagrammatic perturbation theory in the tunneling between semiconductor and superconductor. Disorder-averaging is performed at each order in tunneling $t$. The thick solid line represents disorder-averaged Green's function in the superconductor $\bar{G}(\bm p,\omega)$. The last diagram corresponds to irreducible contributions, as far as disorder averaging is concerned.  } \label{fig:disorder}
\end{figure}

We begin our discussion by considering the perturbation theory in the tunneling $t$, which is justified in the limit of low interface transparency. The lowest order contributions of the diagrammatic expansion in $t$ are shown in Fig.~\ref{fig:disorder}. One can notice that the self-energy at the second order in $t$ is determined by the disorder-averaged Green's function in the superconductor. Since typical s-wave superconductors are disordered (i.e. $\tau \Delta_0 \ll 1$ with $\tau$ being momentum relaxation time), the effect of impurity scattering is non-perturbative. In general, this yields a non-trivial problem because of the self-consistency condition which now has to be solved in the presence of disorder. The problem, however, can be substantially simplified if we neglect the effect of the in-plane magnetic field on the s-wave superconductivity. This condition can be justified due to the vast difference of the $g$-factors in the superconductor and semiconductor. We also assume here that the superconductor is thin enough so that we can neglect orbital effects. At this level of approximation, the problem reduces to understanding the effect of non-magnetic impurities on the bulk s-wave superconductivity, which has been investigated long time ago by Abrikosov and Gor'kov~\cite{AG'61}. The main result of Abrikosov-Gor'kov's theory is that disorder does not affect s-wave superconductivity, {\it i.e.}, the superconducting gap is not suppressed by non-magnetic impurities, in agreement with Anderson's arguments invoking time-reversal symmetry. Specifically, the disorder at the single-particle Green's function level merely leads to the renormalization of the parameters $\tilde\Delta_0 = \eta_{\omega} \Delta_0$ and $\tilde \omega = \eta_{\omega} \Delta_0$:
\begin{align}\label{eq:Greens_dis}
\bar{G}(\bm p,\omega)&=\frac{\tilde \omega  \tau_0 +\xi_{\bm p}\tau_3 +\tilde \Delta_0 \tau_x}{\tilde \Delta_0^2+\xi_{\bm p}^2-\tilde \omega^2},\\
\eta_{\omega}&=1+\frac{1}{2\tau \sqrt{\Delta_0^2-\omega^2}}.
\end{align}
Here bar represents disorder--averaged Green's function and the disorder strength is parameterized by the impurity scattering time $\tau$ defined as $1/\tau=\nu(\varepsilon_F)n_i u_0^2$ with $\nu(\varepsilon_F)$, $n_i$ and $u_0$ being the density of states at the Fermi level, the impurity concentration, and the scattering potential, respectively. Thus, at this level of the perturbation theory, the proximity effect can be included using the formalism developed for the clean case, see Eq.~\eqref{eq:Sigma_clean}. By integrating out the degrees of freedom corresponding to the superconductor, one obtains the interface self-energy:
\begin{align}
\Sigma(\omega)&=|\tilde t|^2 \nu(\varepsilon_F) \int d \varepsilon \bar{G}(\varepsilon,\omega)\\
&=- |\tilde t|^2 \nu(\varepsilon_F) \left[\frac{\tilde \omega \tau_0+\tilde \Delta_0 \tau_x}{\sqrt{\tilde \Delta_0^2-\tilde \omega^2}}+\frac{\zeta}{1-\zeta^2}\tau_z\right].
\end{align}
Finally, upon substituting the expressions for $\tilde \omega$ and $\tilde \Delta_0$, we find that proximity--induced superconductivity is not affected by disorder in the s--wave superconductor
\begin{align}
\Sigma(\omega)&=- |\tilde t|^2 \nu(\varepsilon_F) \left[ \frac{\omega \tau_0+\Delta_0 \tau_x}{\sqrt{ \Delta_0^2-\omega^2}}+\frac{\zeta}{1-\zeta^2}\tau_z \right].
\end{align}
Moreover, one can see that, unlike impurities in the semiconductor, the disorder in the superconductor does not lead to momentum relaxation in the semiconductor at this order of the perturbation theory.

As the interface transparency is increased, one needs to consider higher--order terms in tunneling. These terms involve reducible and irreducible contributions, see Fig.~\ref{fig:disorder}. The former depend only on the disorder--averaged Green's function, and, in a sense, are easy to take into account, whereas the latter involve non--trivial higher-order correlation functions (i.e. diffusons and Cooperons) and lead to momentum relaxation in the semiconductor
. In this work, we consider only reducible contributions and neglect higher order correlation functions with respect to disorder-averaging which, one can show, are much smaller than the reducible ones~\cite{Potter_disorder}. Therefore, our minimal treatment of the disorder in the superconductor is justified and we believe that the superconducting disorder is irrelevant for the topological superconductivity. However, the disorder in the semiconductor (and at the interface) is relevant, as we discuss next.

This conclusion holds as long as we neglect the effect of external magnetic field on the s--wave superconductivity, which is allowed as long as the field is not too large to destroy the superconducting state. This is realistic due to the large difference between the $g$-factors in superconductors ($g_{\rm SC}\sim 1$) and semiconductors ($g_{\rm SM}\sim 10-70$), allowing one to always find a parameter regime where the magnetic field opens a large spin gap in the spectrum without destroying superconductivity. Also, note that our conclusion regarding the short--range impurities in the superconductor is valid when the strength of the disorder is much larger than the superconducting gap: $\tau \Delta_0 \gg 1$ but is small enough not to affect the density of states. We have restricted our analysis to the experimentally relevant regime $\gamma \lesssim \Delta_0$. In the opposite limit $\gamma \gg \Delta_0$, electrons spend most of the time in the s-wave superconductor where their dynamics is not governed by the helical Hamiltonian required to have a topological phase and, thus, this limit is not experimentally desirable, see discussion in Sec.~\ref{secIPI}.

We emphasize that our finding that short--range impurities in the bulk superconductor do not affect the topological superconducting phase emerging in the semiconductor is quite general since the electrons in the nanowire, being spatially separated from the bulk superconductor, simply do not interact directly with the short--range disorder in the bulk superconductor. Our use of the short--range impurity model to characterize the disorder in the bulk superconductor is justified, since strong metallic screening inside the bulk superconductor would render all bare long--range disorder into screened short--range one. Thus, as long as the applied magnetic field does not adversely affect the s--wave superconductivity, our conclusion regarding the validity of the Anderson theorem (i.e., no adverse effect from non--magnetic impurities) to the whole superconductor--semiconductor heterostructure system applies. In this context, we mention the recent theoretical analysis of Ref.~\cite{Tewari'11}, where it was explicitly established that,  in the structure we are considering,  the proximity--induced superconducting pair potential remains s--wave both in the topological and in the non--topological phase, in spite of the non--vanishing Zeeman field. The existence of the s--wave pairing potential, even in the presence of a parallel magnetic field (provided it is not too large), is the key reason for the short--range disorder in the superconductor not having any effect on the topological phase.

\subsection{Disorder in the semiconductor nanowire}

In sharp contrast to disorder in the superconductor, disorder in the nanowire can have significant adverse effects
on the stability of the topological SC phase. There are several different potential sources of disorder in the semiconductor. We focus on two sources that are the most relevant experimentally: random variations of the width of the SM nanowire and random potentials created by charged impurities. The first type of disorder is generally long--ranged, while the second type can be either long-- or short--ranged.  How to account for the effects of disorder depends crucially on the type of physical quantity that one is interested in. Here we focus on the low--energy spectrum, which carries information about the stability of the Majorana bound states, and on thermodynamic quantities such as the local density of states, which are experimentally relevant. The effect of disorder on these quantities is sample--dependent. We emphasize that averaging over different disorder realizations is equivalent in this case with sample averaging. As the goal is to observe stable Majorana fermions in a given nanowire, we investigate here the spectrum of the system for a given disorder realization and focus on establishing the general parameter regimes (e.g., amplitudes and length scales of the disorder potential) consistent with realistic experimental conditions that ensure the stability of the topological SC phase. More specifically, we study several different disorder realizations characterized by a given set of parameters and extract the generic features associated with that type of disorder.

For a single channel Majorana nanowire, one can obtain some analytical results by performing disorder averaging~\cite{Motrunich'01, Gruzberg'05, Brouwer'11a}. Specifically, for a model of the spinless p-wave superconductor it has been shown~\cite{Motrunich'01, Gruzberg'05} that disorder drives the transition into non-topological phase when impurity scattering rate becomes comparable with the induced superconducting gap. In more realistic spinful models involving semiconductor nanowires, the physics is richer and depends on the strength of the magnetic field. We refer the reader to Ref.~\cite{Brouwer'11a} for more details. The generalization of these results to the case of disordered multi-band semiconductor nanowires is an interesting open problem.

\subsubsection{Semiconductor nanowires with random edges}

The dimensions of the nanowire in the transverse direction satisfy the relation $L_y\gg L_z$. The small thickness $L_z$ is critical for the effectiveness of the superconducting proximity effect, as  the SM--SC effective coupling $\overline{\gamma}$ scales, approximately, as $1/L_z^3$. On the other hand, the much larger width $L_y$ is required by the multichannel condition. Atomic--scale variations of $L_z$ generate huge local potential variations (of the order $500 E_\alpha$) that  would effectively cut the wire in several disconnected pieces. Topological SC phases may exist inside each of these pieces, but the Majorana states will be localized at  the boundaries separating different segments and, in general, tunneling between them will be nonzero. To realize a single pair of Majorana zero--energy states localized at the ends of the wire, $L_z$ should be uniform along the system. (We mention in passing that modern MBE growth is consistent with very small variations in $L_z$ as necessary for the realization for the Majorana. )

\begin{figure}[tbp]
\begin{center}
\includegraphics[width=0.48\textwidth]{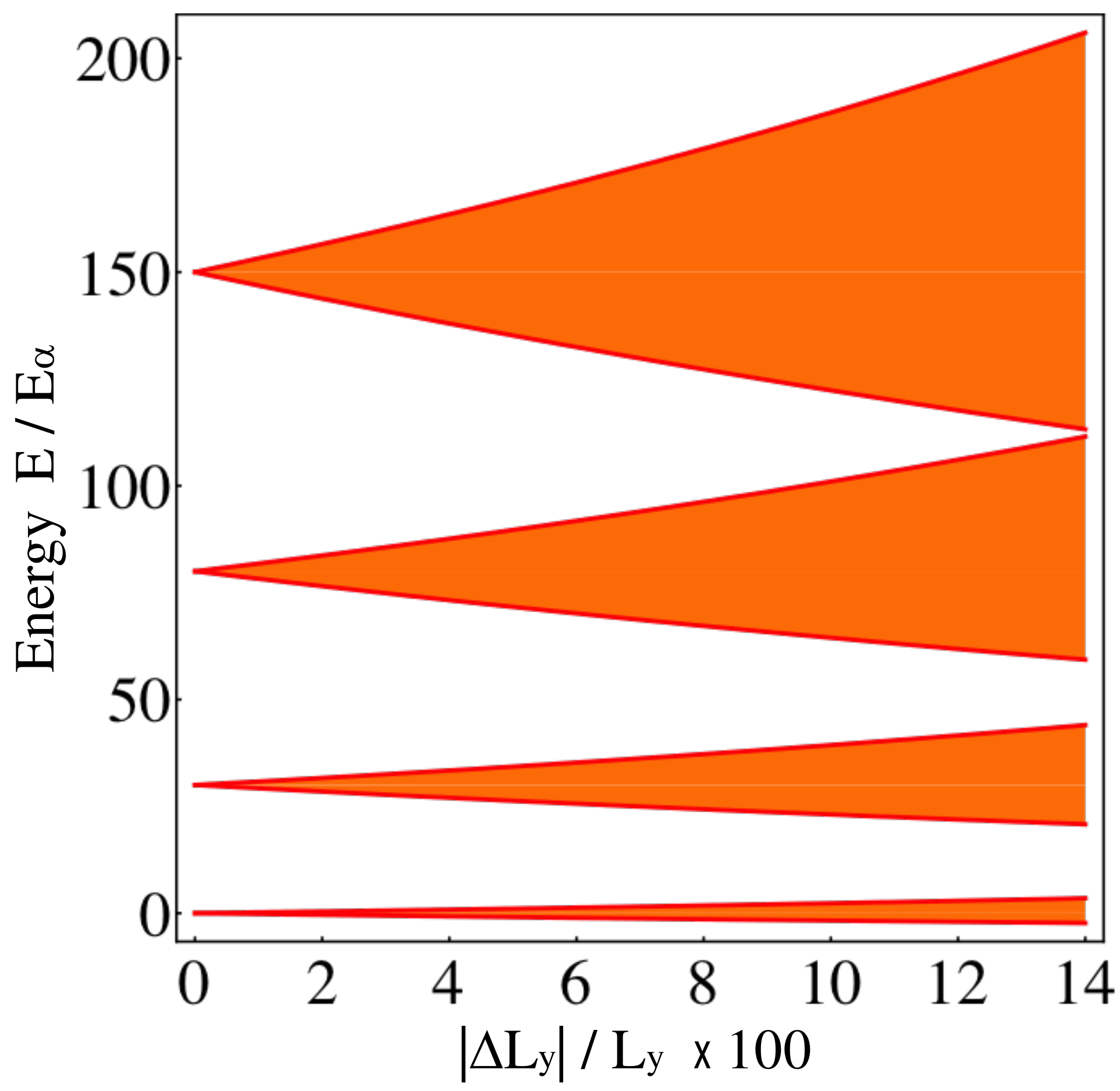}
\vspace{-7mm}
\end{center}
\caption{(Color online) Broadening of the sub--bands in a nanowire with random edges.  For $n_y>2$ the broadening becomes comparable with the inter--band gap in the presence of fluctuations $\Delta L_y$ representing a few percent of the width $L_y$, i.e., the sub--bands loose their identity.} \label{Fig14}
\end{figure}

Engineering a long wire with constant width $L_y$ may be, on the other hand, extremely challenging and less relevant for the stability of the topological SC phase. We assume that $L_y$ varies along the wire randomly in atomic--size steps that extend along hundreds of lattice sites in the x--direction, resulting in a function $L_y(x)$ that varies over length scales larger than the width of the wire. These fluctuations  generate local variations of the bare sub--band energies given by Eq. (\ref{epsn}) of the order of
\begin{equation}
\Delta\epsilon_{n_y} = -2 t_0 \left[\cos\left(\frac{n_y \pi a}{L_y +\Delta L_y}\right) -  \cos\left(\frac{n_y \pi a}{L_y }\right)\right],
\end{equation}
where $L_y$ is the average width of the nanowire, $\Delta L_y$ the value of the local variation, and $a$ the lattice spacing. Note that different  $n_y$ sub--bands are shifted differently, i.e., the random edge is not equivalent to long-range chemical potential fluctuations. To provide a quantitative measure of this effect, we show in Fig. \ref{Fig14} the evolution of the   nanowire sub--bands with the size of the fluctuations $\Delta L_y$. Note that the sub--bands loose their identity in the presence of fluctuations representing a few percent of the wire width, as the broadening becomes comparable with inter--sub-band gaps. The natural question is how this broadening affects the low--energy physics of the nanowire and, in particular, the topological SC phase. Intuition based on the weak--coupling picture would suggest that the topological phase might become unstable, as the parity of the number of sub--bands crossing the chemical potential becomes an ill defined quantity.

\begin{figure}[tbp]
\begin{center}
\includegraphics[width=0.48\textwidth]{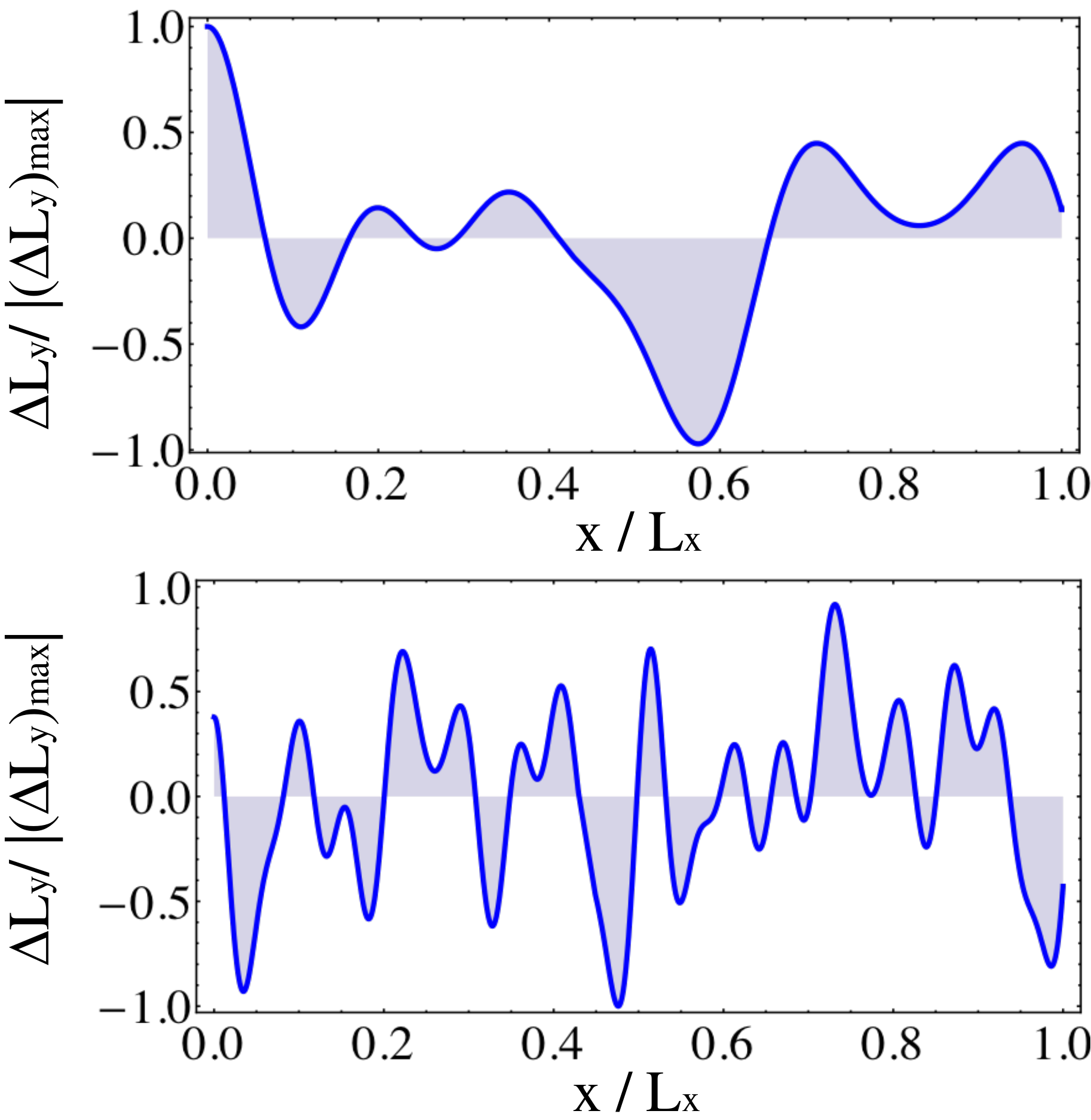}
\vspace{-7mm}
\end{center}
\caption{(Color online) Profile of the variation of the nanowire width, $\Delta L_y$, as a function of position along the wire. The characteristic length scale for profile I (upper panel) is about $10\%$ of the nanowire length $L_x$, wile for profile II (lower panel) it is approximately $5\%$. Each profile generates a series of particular disorder realizations characterized by  different values of the maximum amplitude $|\Delta L_y|_{\rm max}$. Random edge profiles corresponding to a given characteristic length and having the same maximum amplitude have similar effect on the low--energy physics of the nanowire.} \label{Fig15}
\end{figure}

To quantify the effect of random edges on the low--energy physics, we first parametrize this type of disorder. We consider a nanowire of width $L_y(x) = L_y +\Delta L_y(x)$, where $ \Delta L_y(x)$ is a random function characterized by
by a certain maximum amplitude and a characteristic wavelength.  Two examples of random edge profiles are shown in Fig. \ref{Fig15}. We note that the actual width of the wire varies in atomic steps, i.e,  $|\Delta L_y|_{\rm min}=a$ and we assume that the characteristic length scale of these variations is much larger than the atomic scale. For example, for a nanowire with $L_x=5\mu m$, a random edge profile like in the upper panel of Fig. \ref{Fig15} and a maximum amplitude of $10\%$, the atomic edge steps extend over hundreds of unit cells. In the calculations we explore the effect of random edges with maximum amplitudes up to $10\%$ of $L_y$ and various characteristic wavelengths.
\begin{figure}[tbp]
\begin{center}
\includegraphics[width=0.48\textwidth]{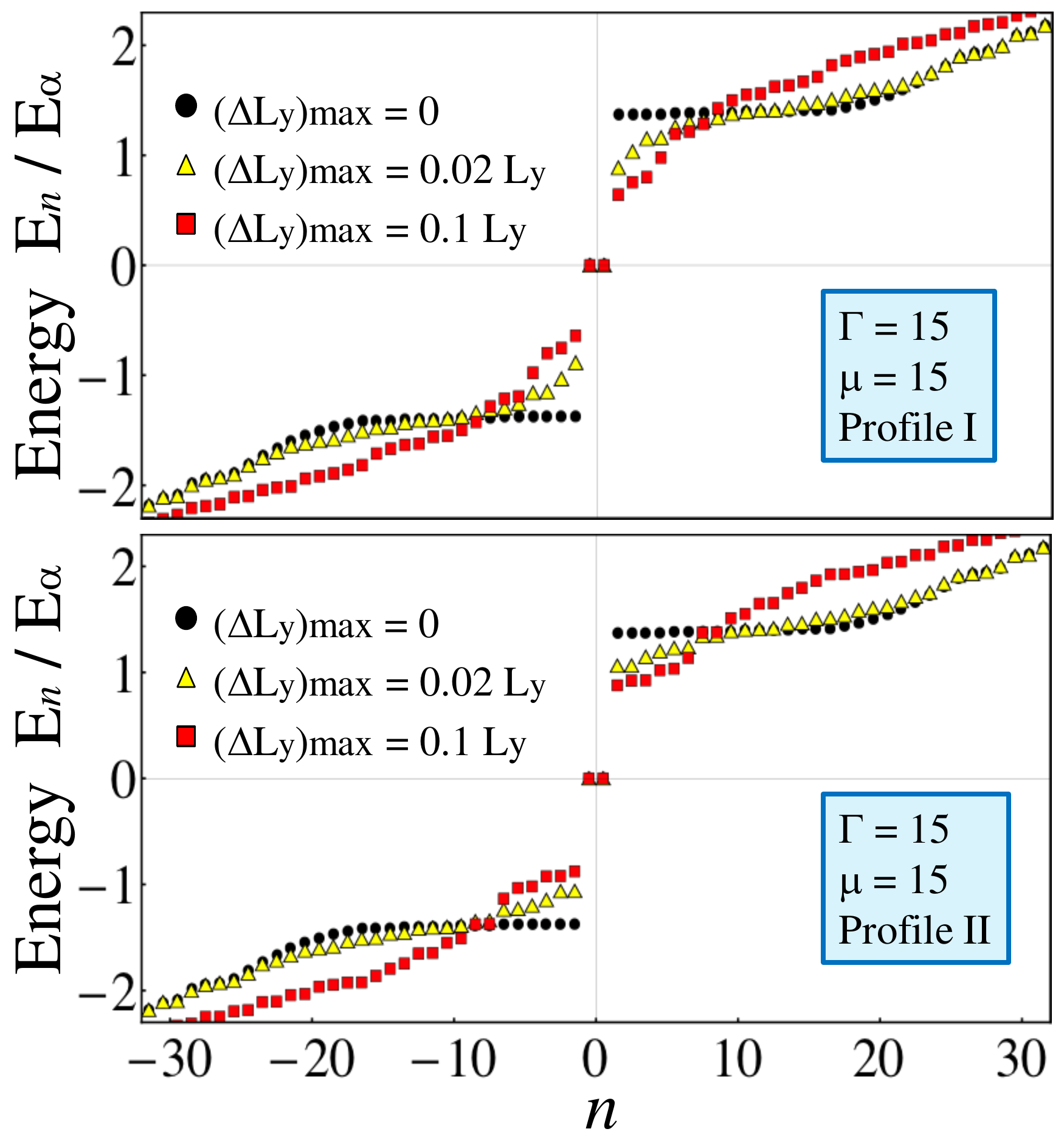}
\vspace{-7mm}
\end{center}
\caption{(Color online) Spectrum of a weakly coupled  nanowire ($\overline{\gamma}=0.25\Delta_0$) with random edges in the topological phase with $N=1$ (see Fig. \ref{Fig8}). In the presence of disorder the mini--gap $\Delta^*$ decreases, but remains finite. The calculation was done using the random edge profiles shown in Fig. \ref{Fig15} for maximum amplitudes of $2\%$ and $10\%$. Longer range disorder (upper panel) has stronger effects than short range disorder (lower panel). Small amplitude variations of the nanowire width of the order $2\%$ generate a reduction of $\Delta^*$ up to  $30\%$, but further increasing the amplitude has a weak effect at low energies.} \label{Fig16}
\end{figure}

\begin{figure}[tbp]
\begin{center}
\includegraphics[width=0.48\textwidth]{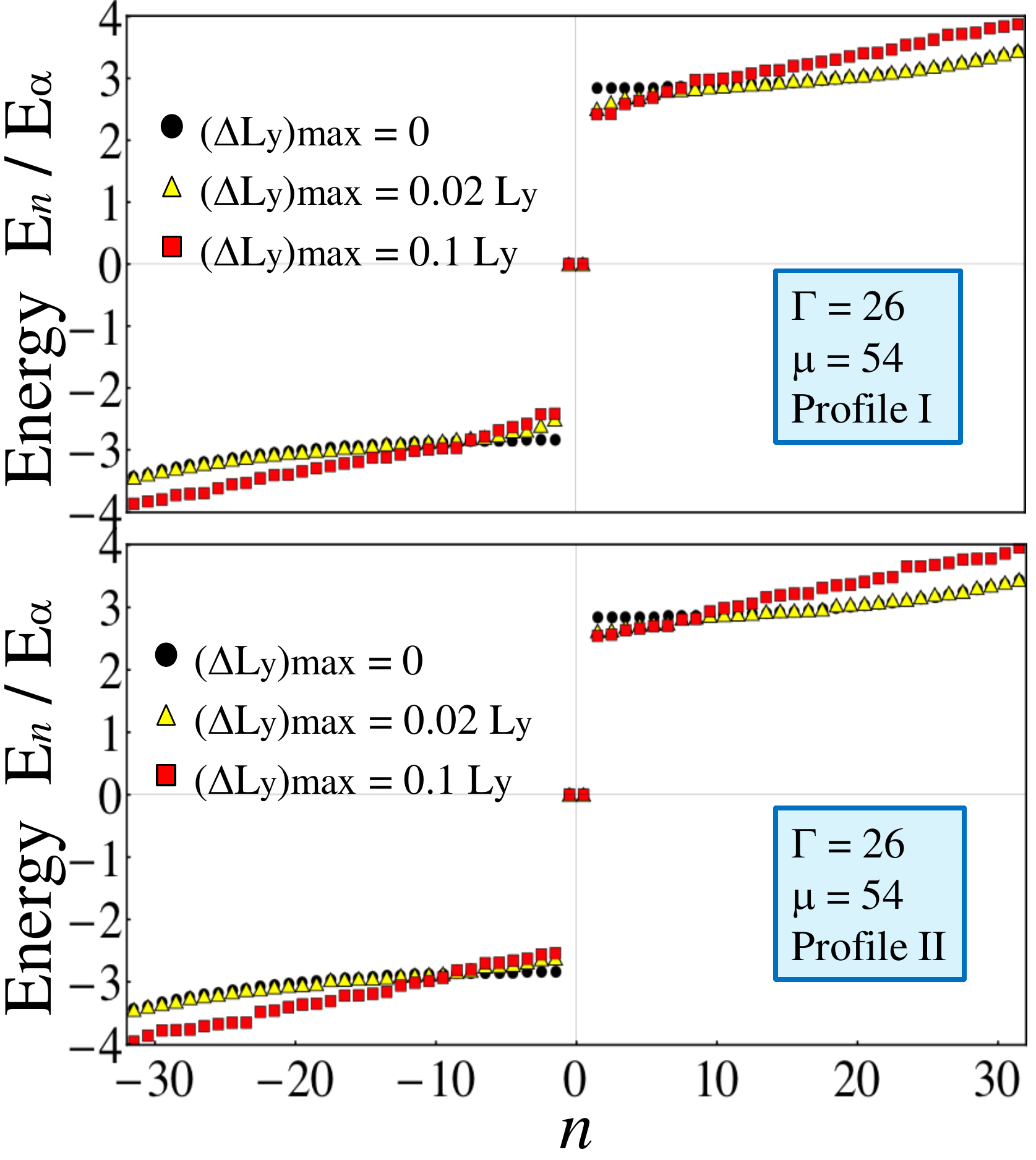}
\vspace{-7mm}
\end{center}
\caption{(Color online)  Spectrum of  a nanowire with random edges at intermediate coupling, $\overline{\gamma}=\Delta_0$. The control parameters correspond to a point in the phase diagram inside the topological phase with $N=1$ (see Fig. \ref{Fig9}).  The random edges are given by the profiles in Fig. \ref{Fig15}.} \label{Fig17}
\end{figure}

The mini--gap $\Delta^*$ is reduced by the presence of random edges. However, for control parameters corresponding to points in the phase diagram away from phase boundaries, the amplitude of  the variations of $L_y$ required for a complete collapse of the gap is well above $10\%$. Examples of the spectra for systems with random edges are shown in figures \ref{Fig16} and \ref{Fig17}. Based on a number of similar calculations for different disorder realizations and control parameters, we have  established the following general conclusions:

i) The details of the low--energy spectrum of a disordered nanowire depend on the particular disorder realization. Nonetheless, different disorder profiles characterized by a given  amplitude and having  the same characteristic length scale are likely to generate similar values of the mini--gap $\Delta^*$, with the exception of a few ``rare events'', which are  characterized by significantly lower gap values. A calculation that involves averaging over disorder will capture these ``rare events'' and will predict a value of the gap much lower than the typical value. Such a calculation would be relevant for an extremely long wire, i.e., in the limit $L_x\rightarrow \infty$ , or for systems with a very large density of states at the relevant energies (e.g., a metal). However, in a typical semiconductor wire the number of states that control the low-energy physics is of the order of $100$. How the energies of these states are modified in the presence of disorder, depends on the specific details of the disorder profile.  Hence, any experimentally relevant conclusion  regarding the low--energy spectrum or the local density of states of a disordered nanowire should be based on calculations involving specific disorder realizations. A direct consequence of these considerations is that nominally identical samples with the same average disorder (e.g. same mobility) may have very different Majorana minigaps since they are likely to have different disorder configurations. The distribution of the minigaps was recently studied in Ref.~\cite{Brouwer'11b}.

ii) Long range disorder has a stronger effect than short range disorder. This is general characteristic of disordered nanowires, regardless of the source of disorder, and will be studied in more detail in the next section. We note that the effects of variations of wire width $\Delta L_y(x)$ at atomic length scales are negligible.

iii) Intermediate coupling $\overline{\gamma} \sim \Delta_0$ represents the optimal SM--SC coupling regime.  The large gap that characterizes the topological $N=1$ phase in this regime is robust against fluctuations of the wire width of the order $\pm 10\%$ for any set of parameters that are not in the immediate vicinity of a phase boundary. Most importantly, this condition can be satisfied for specific values of the Zeeman field (e.g., of the order $30E_\alpha$ for the parameters corresponding to the phase diagram in Fig. \ref{Fig9}) over a large range of chemical potentials.

\subsubsection{Nanowires with charged impurities}

A major source of disorder in the nanowire consists  of charged impurities.  Because the carrier density in the SM nanowire is small, charged impurities are not effectively screened and their presence can potentially have significant effects. The nature of these charged impurities, the values of their effective charge and their exact locations depend on the details of the nanowire engineering process and will have to be determined by future experimental studies. Here, we are interested in the fundamental question regarding the stability of the topological SC phase. In particular we address the following question: Is it possible to realize stable zero--energy Majorana modes in a nanowire with charged impurities within a realistic scenario?  To answer this general question, we focus on four key aspects of the problem: {\it a}) The screening of charged impurities by the electron gas in the SM nanowire, {\it b}) The dependence of the low--energy physics on the concentration of impurities,  {\it c}) The dependence  of the low--energy spectrum on the Zeeman field, and {\it d}) The effect of long range random potentials.

\paragraph{Screening of charged impurities.}

\begin{figure}[tbp]
\begin{center}
\includegraphics[width=0.48\textwidth]{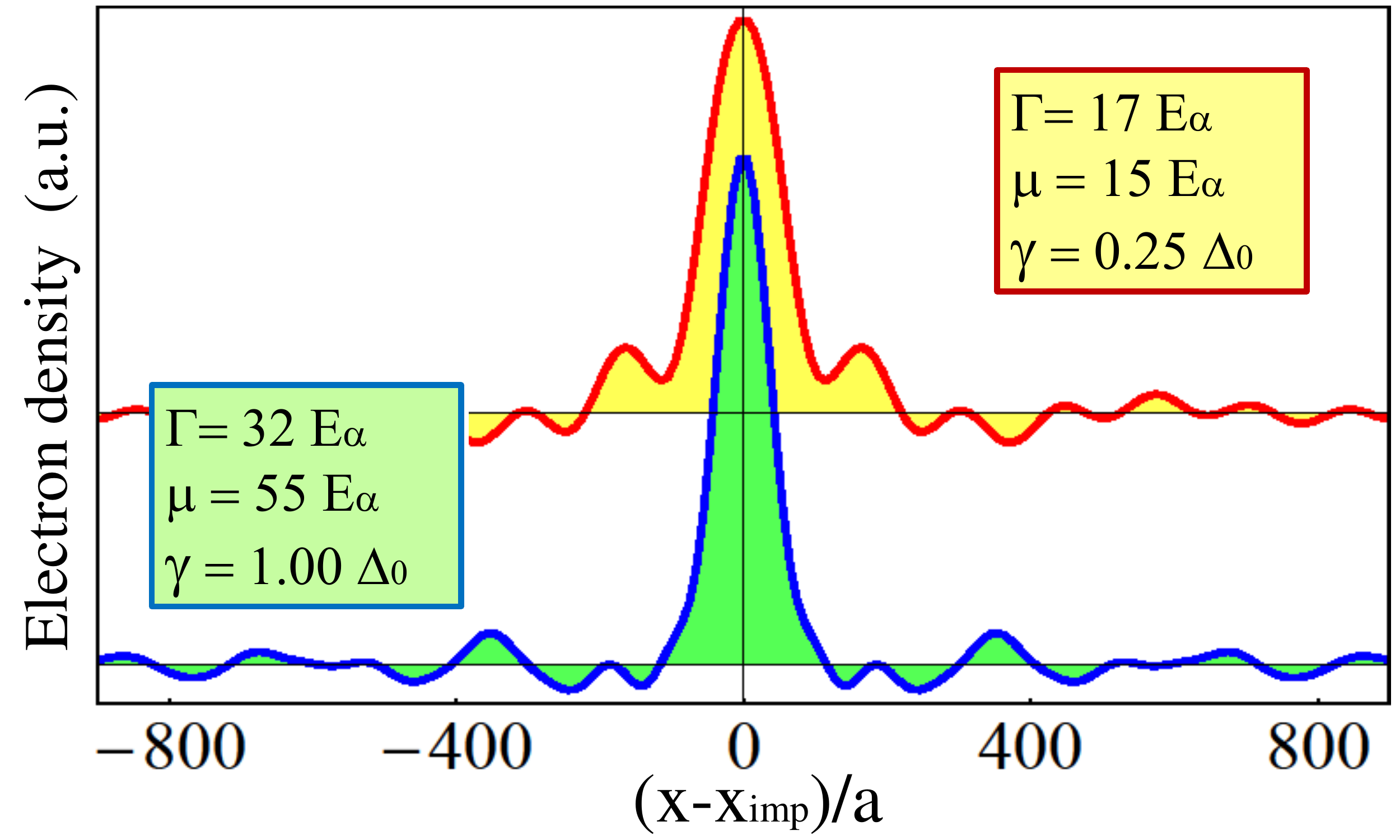}
\vspace{-7mm}
\end{center}
\caption{(Color online)  Induced carrier density as a function of distance from the impurity. The  two curves, corresponding to different external sets of parameters for the nanowire, have been shifted for clarity. The upper curve corresponds to a system with a single occupied sub--band and is characterized by $\lambda \approx 45 a$ (i.e., $63\%$ of the induced charge is in a disk of radius $45 a$), while the lower curve is for a system with three occupied sub--bands and has  $\lambda \approx 30 a$. Note that  the effective impurity potentials that generate these density profiles are given by Eq. (\ref{Vs}) with the corresponding values of $\lambda$.} \label{Fig18}
\end{figure}

We start by considering a single charge $q$ inside or in the vicinity of the nanowire. For concreteness we assume that the charge is positioned near the middle of the wire and one lattice spacing away from its surface, i.e., for a wire that occupies the volume defined by $0\leq x_j \leq L_j$, with $j\in \{x, y, z\}$, the position of the impurity is given by $(x_{imp}, y_{imp}, z_{imp}) = (L_x/2, L_y/2, -a)$. This corresponds, for example,  to a charged impurity localized at the interface between the SM and the SC. We consider the extreme case $q=\pm e$, where $e$ is the elementary charge, although in practice it is likely that the effective charge is only a fraction of this value due to screening by electrons in the SC. We neglect the screening due to the presence of the superconductor, which may significantly reduce the effective potential created by the charge. A simple estimate of the screening effects due to the electrons in the semiconductor within the Thomas-Fermi approximation is highly inaccurate due to the low carrier density. We checked this property explicitly by calculating numerically the carrier density induced by a given effective potential $V({\bm r})$, e.g., a screened Coulomb potential. We find that the relationship between the induced local carrier density $\delta n({\bm r})$ and the local effective potential is highly non-linear.   In addition, the density is characterized by strong Friedel--type oscillations (see Fig. \ref{Fig18}). Hence, solving quantitatively the screening problem for the nanowire would require a self-consistent calculation that includes electron--electron interactions at the Hartree-Fock level.  This calculation is beyond the scope of the present study and will be addressed elsewhere. Here we address a more limited question: What is the characteristic length scale over which the external charge $q$ is screened?  We define this length scale $\lambda$ as the characteristic length of the volume that contains $63\%$ (i.e., a fraction equal to $1-1/e$) of the induced charge.
Specifically, the potential created by the charge $q$ at a point ${\bm r}$ inside the semiconductor is
\begin{equation}
V({\bm r}-{\bm r}_{imp}) = \frac{-e q}{4 \pi \epsilon_0\epsilon_r |{\bm r}-{\bm r}_{\rm imp}|},
\end{equation}
where ${\bm r}_{\rm imp} = (x_{imp}, y_{imp}, z_{imp})$ is the position of the impurity and $\epsilon_r$ is the relative dielectric constant of the SM. For InAs  $\epsilon_r=14.6$. Next, we take into account the fact that the nanowire is extremely thin in the z direction ($L_z\approx 10a$), hence the wave function profile along this direction is very little affected by the presence of the impurity.  On the other hand, the induced charge has a strong dependence on x (the direction along the wire) and y (the transverse direction). As mentioned above, the impurity potential is screened by the induced charge outside a region with  a characteristic length scale $\lambda$ that contains a fraction equal to $1-1/e$ of the induced charge. We assume that the screened potential is qualitatively described by an expression of the form
\begin{equation}
V_{s} =V({\bm r}- {\bm r}_{imp})\exp\left[-\frac{\sqrt{(x-x_{\rm imp})^2+(y-y_{\rm imp})^2}}{\lambda} \right]. \label{Vs}
\end{equation}
The parameter $\lambda$ from Eq. (\ref{Vs}) is determined self--consistently by imposing the condition that $63\%$ of the induced charge be in a disk of radius $\lambda$ and thickness $L_z$ centered at $(x_{imp}, y_{imp}, L_z/2)$. The results for two different sets of parameters are shown in Fig. \ref{Fig18}. As expected, nanowires with multiple occupied sub--bands provide a more effective screening, which is reflected in a lower value of the screening length $\lambda$. We emphasize that the present approach is not fully self--consistent and is therefore only of qualitative validity. An effective screened potential that includes exactly  the contribution of the induced charge with details (e.g., oscillatory components)  that are not captured by Eq. (\ref{Vs}) should give results very similar to what we obtain here. Nonetheless, we expect these details to have a weak effect on the final results. We also note that the self--consistent calculation of the screening length is done for a non--superconducting nanowire, then the effective  impurity potential $V_s$ is added to the total
nanowire Hamiltonian before including the proximity effects due to the SM--SC coupling.

\begin{figure}[tbp]
\begin{center}
\includegraphics[width=0.48\textwidth]{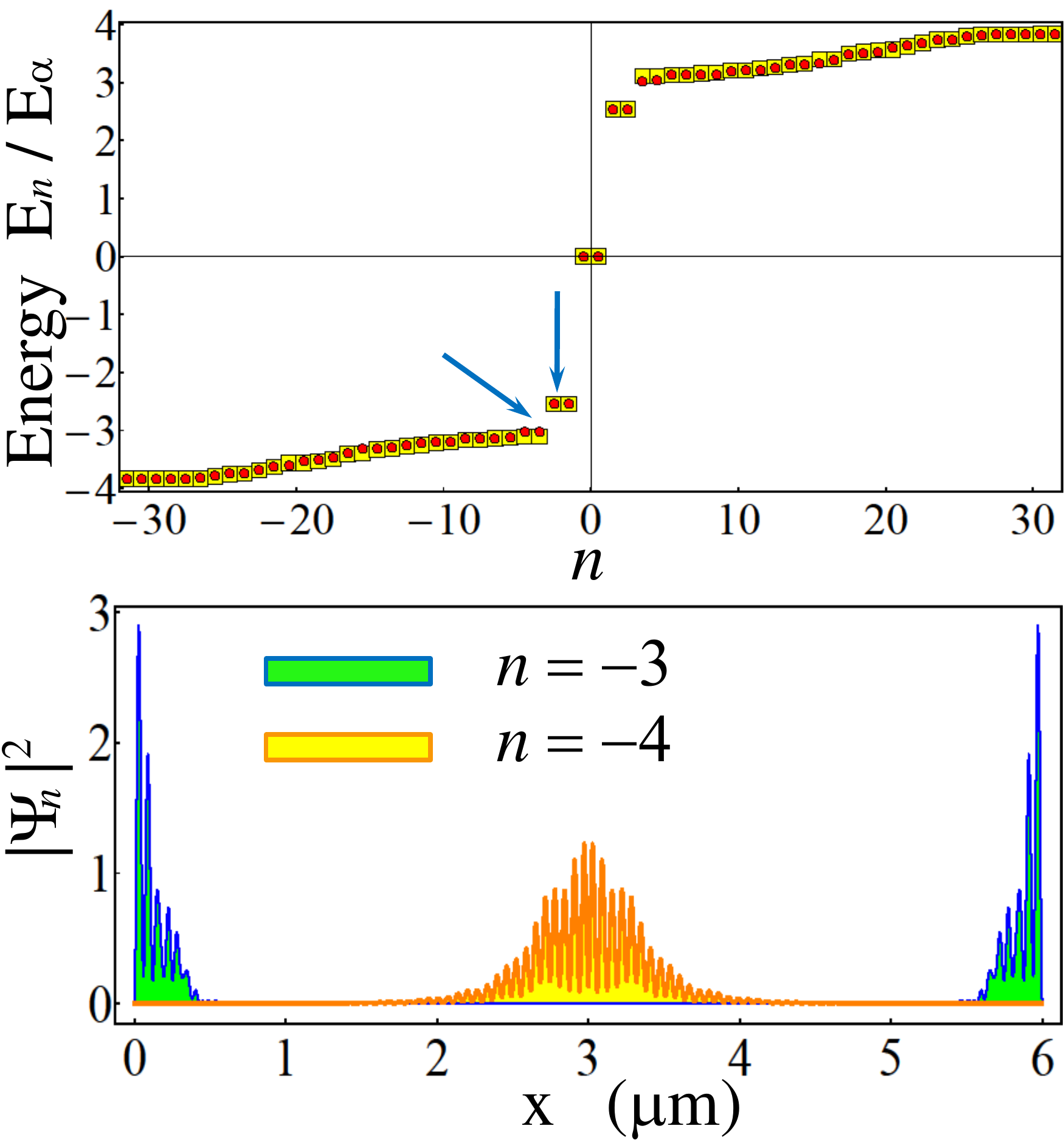}
\vspace{-7mm}
\end{center}
\caption{(Color online)  Top panel: Comparison between the low--energy spectrum of a clean nanowire (yellow squares) and the spectrum of a nanowire with a charged impurity with $q=e$ (red circles). Note that the two spectra are almost on top of each other. Bottom panel: Electron wave function amplitudes for two low--energy states (marked by arrows in the top panel) in the presence of the charged impurity. The Majorana zero modes (${\it n}=\pm 1$, not shown) and the in--gap modes (${\it n}=\pm 2, \pm 3$, see Fig. \ref{Fig12} and the corresponding discussion)  are localized near the ends of the wire and are not affected by the presence of the impurity. The low--energy bulk states $|{\it n}|>4$ that are extended in a clean nanowire become localized near the impurity.} \label{Fig19}
\end{figure}

What is the effect of the charged impurity on the low-energy spectrum of the superconducting nanowire? The effective potential in the vicinity of the impurity is extremely large, e.g., $V_s(L_x/2, L_y/2, L_z/2)\approx 640$meV  (i.e., $\approx 12000 E_\alpha$), much larger than any other relevant energy scale in the problem, hence one would naively expect significant effects.   However, the low-energy physics of the SM nanowire is controlled by single particle states with low wave numbers and the matrix elements of the impurity potential between these state are relatively small. In other words,  the impurity will strongly affect the spectrum at intermediate and  high energies, but will have a relatively small  effect at low energies. To illustrate this property, we show  in Fig. \ref{Fig19} a comparison between the low--energy spectra of a clean nanowire and of a nanowire with a charged impurity. We note that the presence of the external charge induces the formation of localized states in the vicinity of the impurity (see Fig. \ref{Fig19}).

Charge impurities in one-dimensional quantum wires produce weakly long-range disorder since the Coulomb potential decays as $\ln |q a|$ in the momentum space for $q\rightarrow 0$ with $a$ being the short-distance cut-off associated with the transverse dimensions of the nanowire~\cite{DasSarma'85}. This is to be contrasted with the much stronger $q^{-1}$ ($q^{-2}$)
long wavelength divergence of the bare Coulomb disorder in two (three) dimensions. The weakly long-range nature of 1D Coulomb potential suggests that any regularization of the long-range disorder would be a reasonable approximation in spite of the fact that Thomas-Fermi screening itself is weak in 1D. In particular, the presence of interband scattering in the multisubband situation would essentially lead to effective 2D screening in the system, which should suffice to regularize the singular Coulomb disorder. At very low densities, where the nanowire is strictly in 1D limit with only the lowest subband occupied, weak screening will lead to the formation of the inhomogeneous electron puddles in the system around the charge impurities due to the failure of screening. This situation is detrimental to the Majorana formation and must be avoided. It is clear that higher density and multisubband occupancy would be favorable for the experimental realization of the Majorana modes in the semiconductor nanowires.


\begin{figure}[tbp]
\begin{center}
\includegraphics[width=0.48\textwidth]{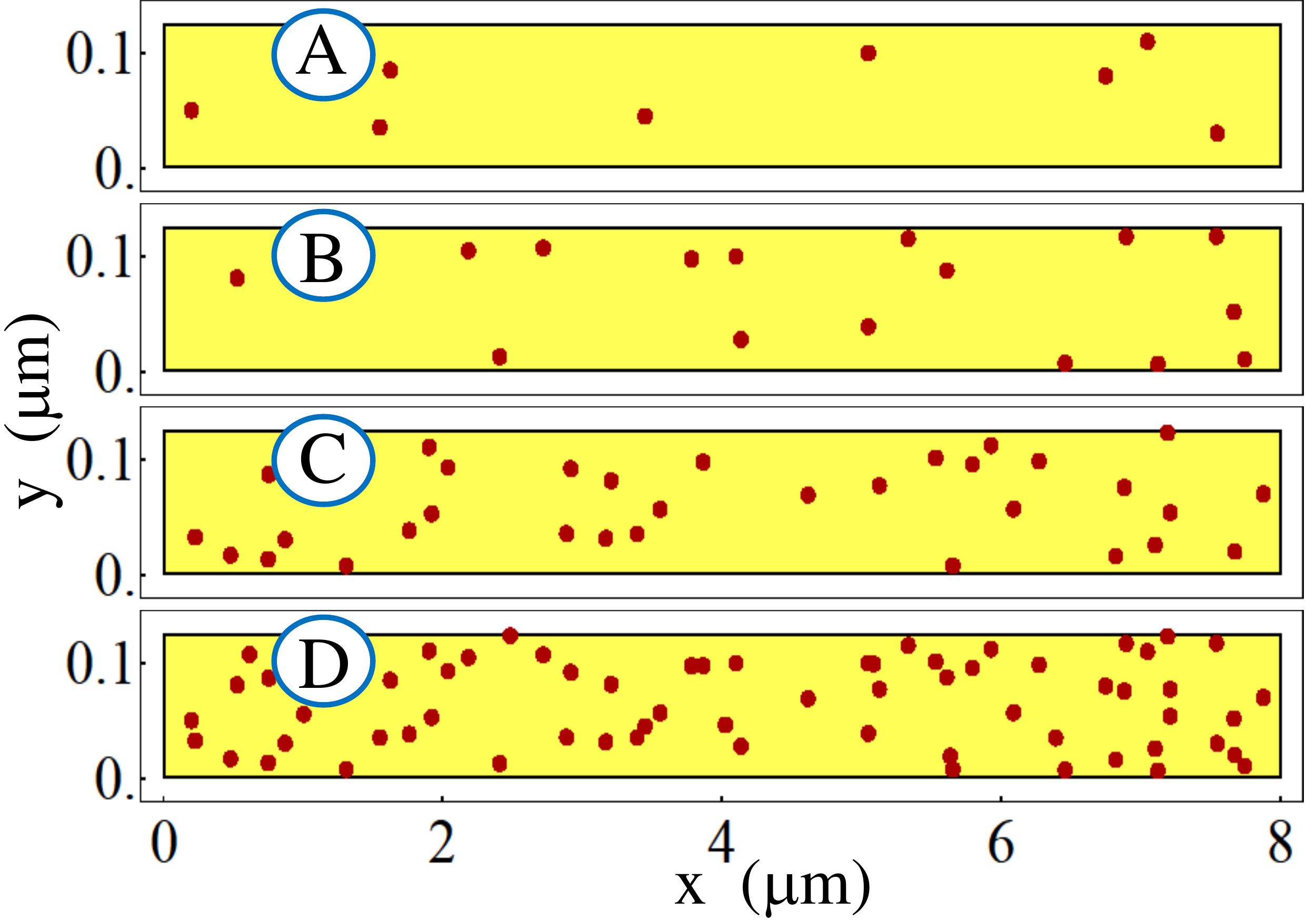}
\vspace{-7mm}
\end{center}
\caption{(Color online)  Specific disorder realizations in nanowires with charged impurities. The dots represent the locations of the impurities. Each impurity has a charge $q=\pm e$ and is positioned one lattice spacing away from the surface of the nanowire along the z direction. The linear impurity densities $n_{\rm imp}$ are: $1 \mu m^{-1}$ (A),   $2 \mu m^{-1}$ (B), $4 \mu m^{-1}$ (C),  and $8 \mu m^{-1}$ (D). } \label{Fig20}
\end{figure}

\paragraph{Charged impurity disorder.}

The next question that we address concerns the dependence of the low energy spectrum of a nanowire with charged disorder on the concentration of impurities. We emphasize that the details of the low--energy physics depend on the specific disorder realization. In particular, the low--energy states are localized in the vicinity of the impurities (see Fig. \ref{Fig19}) and their energies depend on the specifics of the real--space disorder configuration. Hence, as discussed above, calculations of single-particle quantities, e.g., the local density of states, should involve specific disorder realizations, rather than disorder averaging.
 In Fig. \ref{Fig20} we show four different specific disorder realizations,  corresponding to linear impurity densities ranging from $1 \mu m^{-1}$ to  $8 \mu m^{-1}$, which are reasonably realistic impurity densities ($\sim 10^{15}$ cm$^{-3}$) in high quality semiconductor structures. The impurities carry charge $q=e$ and are positioned at a distance of one lattice constant away from the SM surface. The effect of these impurities is incorporated through an impurity potential of the form
\begin{equation}
V_{\rm imp}({\bm r}) = \sum_{j} V_s({\bm r}, {\bm r}_{j}),  \label{Vimp}
\end{equation}
where $V_s$ is given by Eq. (\ref{Vs}) and ${\bm r}_{j}$ are the impurity position vectors. We note that in Eq. (\ref{Vimp}) the screened potential is characterized by a screening length $\lambda$ determined as described above for a single impurity. This approximation does not take into account the impact of the dependence of the effective potential associated with a given charged impurity on the location of the charge and on the presence of other impurities. It also neglects the effect of screening by the SC itself which should strongly suppress the effective disorder.

\begin{figure}[tbp]
\begin{center}
\includegraphics[width=0.48\textwidth]{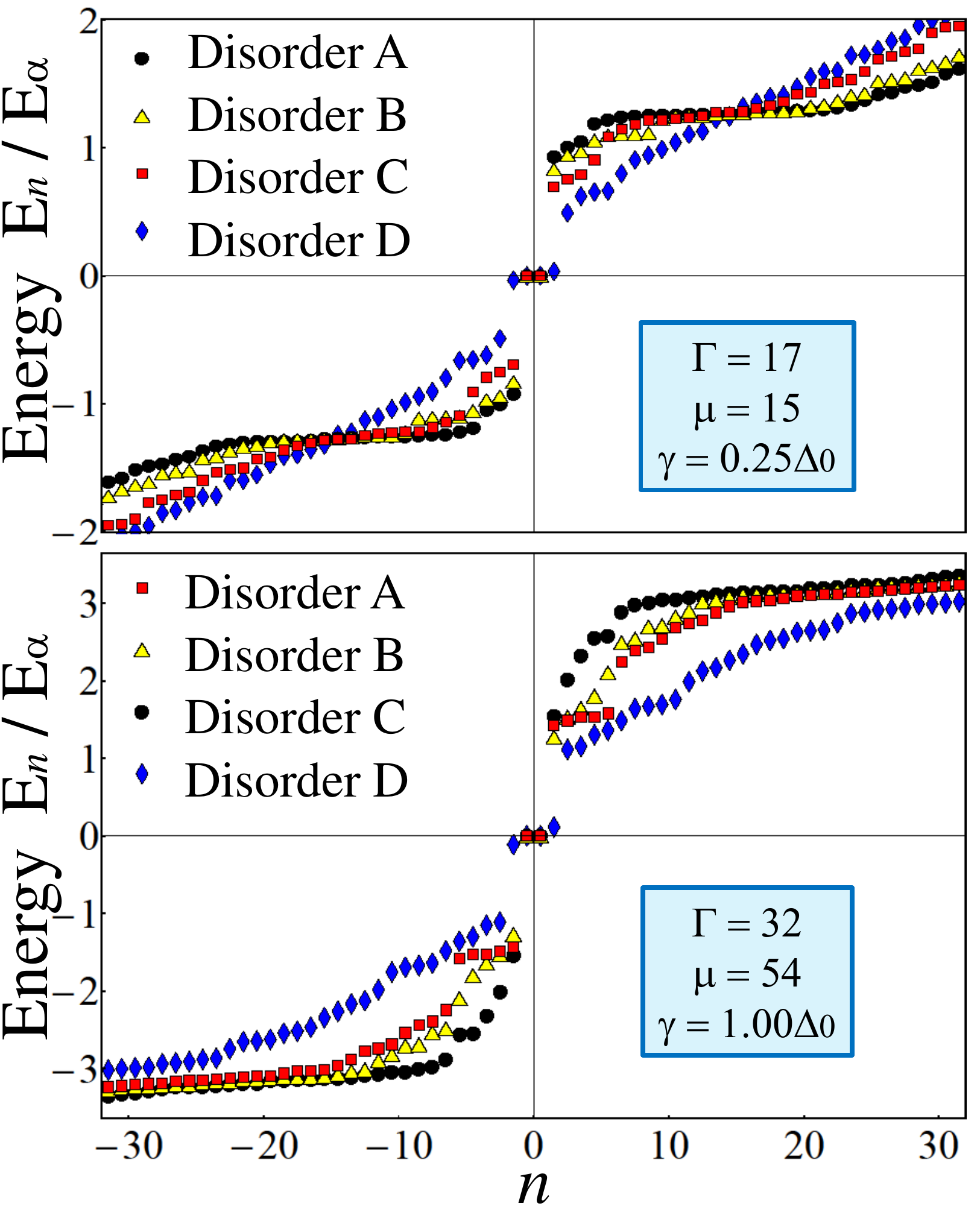}
\vspace{-7mm}
\end{center}
\caption{(Color online)  Low-energy spectra  of  a superconducting nanowire with charged disorder for two different sets of control parameters and four disorder realizations. The symbols correspond to the specific disorder realizations shown in Fig. \ref{Fig20}. Note that: i) The mini--gap $\Delta^*$  finite for impurity concentrations $n_{\rm imp}\leq 4 \mu m ^{-1}$ and collapses for $n_{\rm imp}= 8 \mu m ^{-1}$, and  ii) There is an overall tendency of the low--energy features to move at lower energies when increasing the impurity concentration, but the exact value of the mini--gap depends on the specific disorder realization, e.g., in the lower panel $\Delta^*(n_{\rm imp}= 4) > \Delta^*(n_{\rm imp}= 2)$.  } \label{Fig21}
\end{figure}

The low--energy spectra of a nanowire with random charged impurities distributed as in Fig. \ref{Fig20} are shown in Fig. \ref{Fig21} for two sets of control parameters. The general trend is that the  mini--gap decreases with increasing impurity concentration. However, for a given concentration $n_{\rm imp}$ the exact value of the mini-gap depends on the specific disorder realization. As mentioned above, averaging over disorder includes rare configurations characterized by small mini-gaps, hence  the averaged density of states is characterized  at low energies by a small weight that does not correspond to any physical state in a typical disorder realization. The signature of a  topological SC phase with $N=1$ (i.e., one pair of Majorana fermions) that distinguishes it from the trivial SC phase with $N=0$ is the presence of zero--energy quasiparticles separated by a finite gap from all other excitations. One key conclusion of our calculations is that the mini--gap that protects the topological SC phase remains finite for a significant range of realistic impurity concentrations.

\begin{figure}[tbp]
\begin{center}
\includegraphics[width=0.48\textwidth]{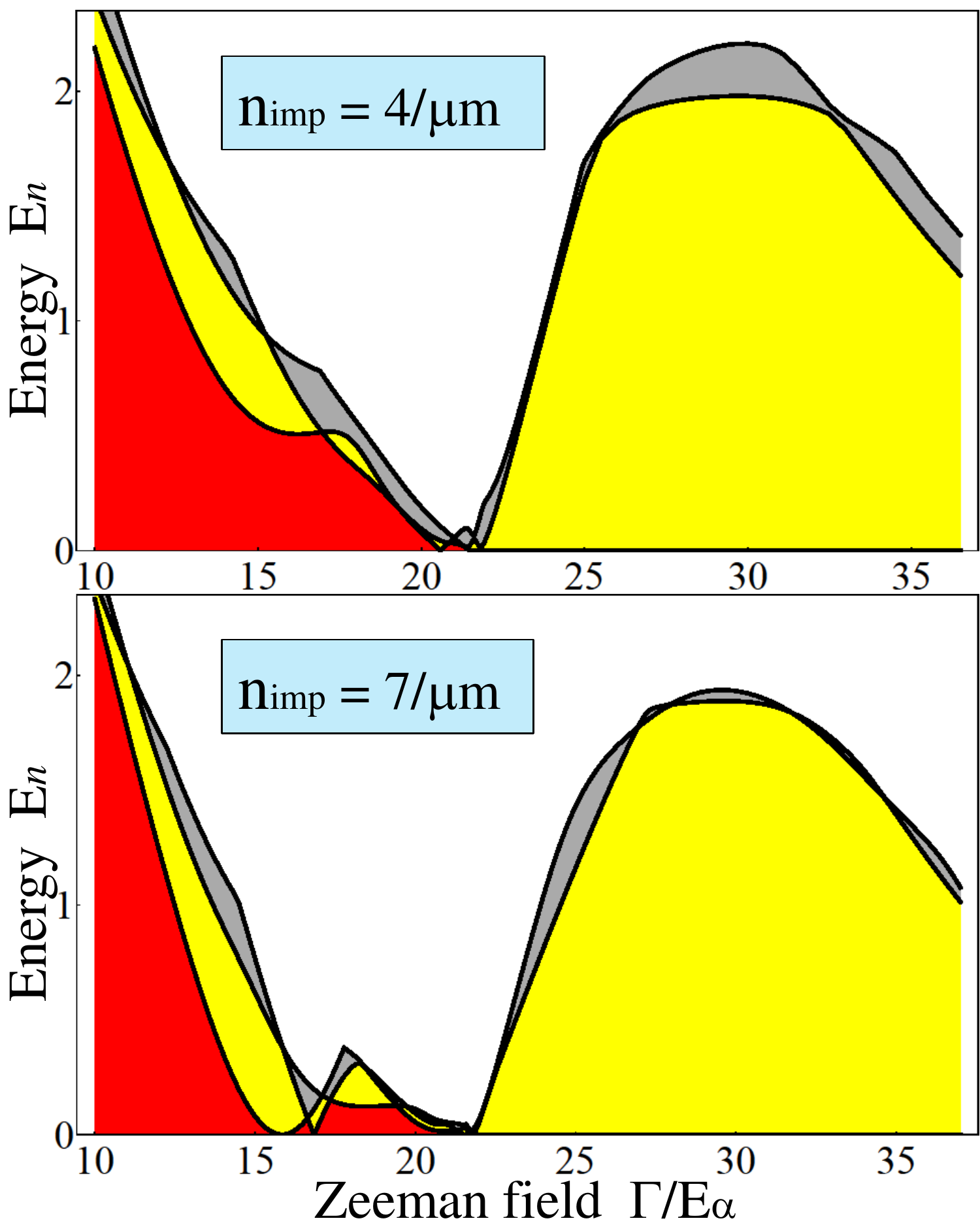}
\vspace{-7mm}
\end{center}
\caption{(Color online) Dependence of the lowest three energies $E_{\it n}$ (${\it n}= 1, 2, 3$) on the Zeeman field for a disordered nanowire with $n_{\rm imp}= 4 \mu m^{-1}$ (top) and $n_{\rm imp}= 7 \mu m^{-1}$ (bottom). The lowest energy state is characterized by a gap $\Delta_{\it 1}= E_{\it 1}$ (red/dark gray region) that vanishes for $\Gamma>21.6E_\alpha$, i.e., when the system enters the topological SC phase.  The gap between the first and the second levels, $\Delta_{\it 2}= E_{\it 2}-E_{\it 1}$ (yellow),  becomes the mini--gap in the topological phase. The transition zone is characterized by a high density of  low--energy modes and expands with increasing the impurity concentration. } \label{Fig22}
\end{figure}

\paragraph{Dependence of the low--energy spectrum on the Zeeman field.}

As shown in Fig. \ref{Fig21}, for certain high impurity concentrations (e.g., configuration D in fig. \ref{Fig20}) it is possible that, in addition to the Majorana zero mode,  another low--energy state has energy close to zero and is separated by a gap from the rest of the spectrum. How can one distinguish  experimentally this state from a topological superconductor characterized by a finite mini--gap, on one hand, and from a trivial superconductor with $N=2$, on the other? As the low--energy spectrum has a strong dependence on $\Gamma$, the key is to vary the Zeeman field. Tuning $\Gamma$ may push  the system into the topological SC phase and the energy of the additional state will increase. On the contrary, it is possible that varying the Zeeman field will lead to the appearance of  more low--energy excitations. This is the characteristic signature of the transition zone between phases with different topologies. To address this problem more systematically, let us consider a nanowire with the same parameters as in the bottom panel of Fig. \ref{Fig10}: $\mu=54.5 E_\alpha$, $\Gamma=\Delta_0$, and $\theta=0.8$.
The dependence of the quasiparticle gap on $\Gamma$ for the  infinite clean system in shown in Fig. \ref{Fig10}, and the  dependence of the  mini--gap  on $\Gamma$ for a finite wire is shown in Fig. \ref{Fig12}. The transition between the trivial SC phase with $N=0$ and the topological phase with $N=1$ is clearly marked by the vanishing of the gap at $\Gamma\approx 21.5 E_\alpha$.

Let us now add disorder and follow the evolution of the lowest three energy levels with the Zeeman field. The results for two different values of the impurity concentration are shown in Fig. \ref{Fig22}.  First, we note that the main features are similar with those observed in a clean system: at low values of the Zeeman field the system is in a trivial SC phase characterized by finite gap to all excitations, while for $\Gamma$ above a certain critical value $E_c\approx 22E_\alpha$ the system has a Majorana zero mode separated by a mini--gap from the  rest of the spectrum. The major difference from the clean case consists of the transition zone, which extends over a finite range of values of the Zeeman field and is characterized by multiple low-energy excitations.  This transition zone extends with increasing impurity concentration.  Assuming that we probe the low-energy properties of the system with a certain finite resolution, e.g., $\Delta E=0.2 E_\alpha$, the topological phase can be unambiguously distinguished from the trivial SC provided the mini--gap is larger that the energy resolution.
The trivial $N=0$ phase will be  characterized by a finite gap and no zero--energy excitation, while the topological phase will  have the characteristic zero--mode separated from the finite energy excitations by the mini--gap. In between the two phases there will be a transition zone characterized, within our finite energy resolution, by a continuum spectrum.  Starting with low values of the Zeeman field, i.e., deep inside the $N=0$ phase,  by increasing $\Gamma$  one first reaches the transition zone, the topological $N=1$ phase. We emphasize that for $\Delta^* < \Delta E$ the topological phase becomes indistinguishable from the transition zone. In addition, at large impurity concentrations the mini--gap will collapse completely.

\paragraph{Long range disorder potential.}

What is the effect of  a long-range disorder potential on the stability of the topological SC phase? Are there qualitative differences from the short--range   case discussed above? We are not interested here in the possible source of such long range disorder, but rather in identifying the magnitude of the amplitude of the random potential that would destroy the topological phase.

\begin{figure}[tbp]
\begin{center}
\includegraphics[width=0.48\textwidth]{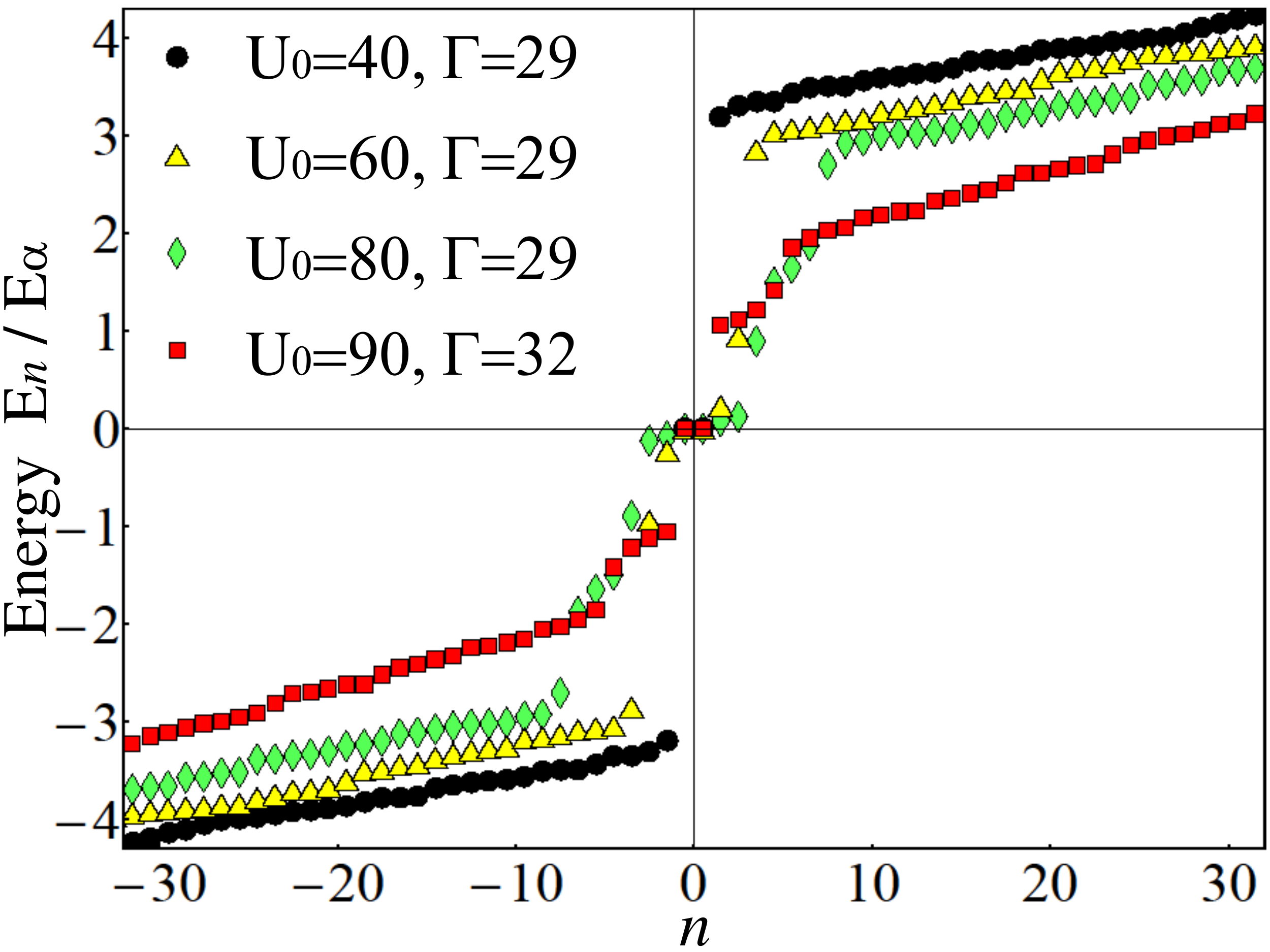}
\vspace{-7mm}
\end{center}
\caption{(Color online) Low-energy spectra of a nanowire with long--range disorder. The system in characterized by $\mu=54.5E_\alpha$ and $\gamma=\Delta_0$  and the impurity potential is given by Eq. (\ref{U0v}) with with a profile $v_{\rm imp}$ as in Fig. \ref{Fig15} (lower panel) and different amplitudes $U_0$. The parameters $U_0$ and $\Gamma$ are expressed in units of $E_\alpha$. For $\Gamma=29$ the  mini--gap collapses for $U_0\geq60$, while for a Zeeman field $\Gamma_32$ it survives up to an amplitude  $U_0\approx 100$.  } \label{Fig23}
\end{figure}

Let us consider a random potential with a characteristic length scale $\lambda > L_y$. Neglecting the dependence of the potential on $y$ and $z$, we have
\begin{equation}
V_{\rm imp}({\bm r}) = U_0 v_{\rm imp}(x/L_x), \label{U0v}
 \end{equation}
 where $U_0$ is the amplitude of the potential and $-1\leq v_{\rm imp}\leq 1$ is a random profile. For concreteness we consider the profile shown in the lower panel of Fig. \ref{Fig15} and a nanowire with $L_x=5\mu m$ and $L_y=0.12 \mu m$. These parameters correspond to $\lambda\approx 2L_y$. The corresponding low--energy spectra for different values of the disorder amplitude are shown in Fig. \ref{Fig23}. A striking feature is represented by the significant difference between the critical disorder amplitudes at which the mini--gap collapses for the two slightly different values of the Zeeman field. Note that both values are near the ``center'' of the topological $N=1$ phase, as one can see, for example, by inspecting the bottom panel of Fig. \ref{Fig12}.  What sets the scale for the critical amplitude? A hint can be obtained from the phase diagram shown in Fig. \ref{Fig9}. In essence, in the presence of a smoothly varying random potential topological superconductivity is stable as long as the local chemical potential at any point within the nanowire has values within the topological phase. The permissible range of variation for the local chemical potential depends strongly on the Zeeman field. From the phase diagram in Fig. \ref{Fig9} and the chemical potential dependence in Fig. \ref{Fig11} it is clear that $\Gamma\approx 32E_\alpha$ represents a value of the Zeeman field that allows for large amplitude chemical potential fluctuations.

We conclude that long range disorder can be treated as local chemical potential fluctuations. We emphasize that this is not the case for short range disorder. The topological phase is stable against  chemical potential  fluctuations up to a maximum amplitude that  depends on the Zeeman field, on the average chemical potential, and on the SM--SC coupling. Single band occupancy necessarily limits the critical amplitude due to the close proximity of the phase boundary. Multi--band systems avoid this problem and provide a more efficient screening for the short range potentials created by charged impurities. In addition, a strongly non--uniform coupling between the nanowire and the s--wave SC (with $\theta\sim 1$) with a coupling strength in the intermediate regime ($\gamma\sim \Delta_0$) provides  the optimal  shape of the phase diagram (see Figures \ref{Fig9} and \ref{Fig12}).

\subsection{Disorder at the semiconductor--superconductor interface}\label{sec:tunnelingdisorder}

\begin{figure}[tbp]
\begin{center}
\includegraphics[width=0.48\textwidth]{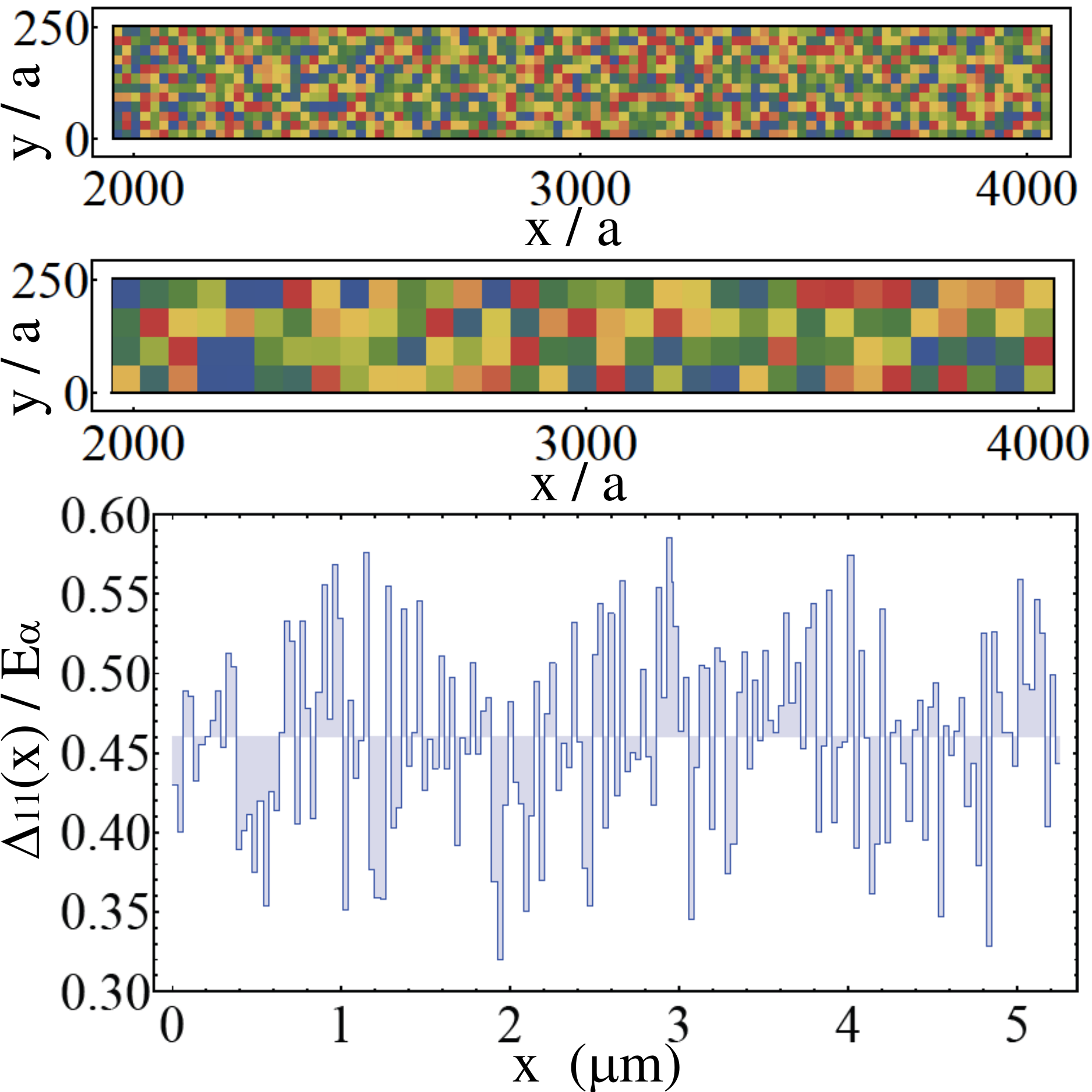}
\vspace{-7mm}
\end{center}
\caption{(Color online)  Random coupling at the SM--SC interface. The coupling $\tilde{t}$ contains a random component $\Delta\tilde{t}$ that is constant within patches of length {\it l}, but takes a random value within each patch. The patch sizes are ${\it l}=20a$ (top panel) and  ${\it l}=60a$ (middle panel). The strength of $\Delta\tilde{t}$ within the patches is color coded. Note that only part of the interface is shown, as the typical length of a nanowire used in the calculations is 0f the order $10^4a$. Bottom: Dependence of the induced gap $\Delta_{11}$ on the position along the wire for $\Gamma=32E_\alpha$, $\mu=54.5E_\alpha$, $\overline{\gamma}=\Delta_0$, $\theta=0.8$, and  a random coupling with ${\it l}=60a$ and an amplitude $(\Delta \tilde{t})_{max} = 0.25 \tilde{t}$. The huge local variations of the coupling are strongly reduced  by the integration over $y$ (see main text) and generate fluctuations of the order of $10\%$ in $\Delta_{11}(x)$. } \label{Fig24}
\end{figure}

The inhomogeneous random coupling at the semiconductor--superconductor interface is another significant source of disorder. In principle, the tunneling matrix elements $\widetilde{t}(i_x, i_y)$ between the nanowire and the s--wave SC are characterized by random real space variations due to inhomogeneities in the tunneling barrier. For example, realizing a nonuniform coupling $\widetilde{t}(i_y)$, which is critical for generating off--diagonal pairing and for stabilizing the topological phase near the sweet spots, may require the growth at the interface of an insulating layer with variable thickness across the wire. Any growth imperfection will translate into variations of $\widetilde{t}$. While ultimately the details of these variations will have to be determined by a careful experimental study of the interface, it is reasonable to assume that a typical interface is characterized by  atomic size variations with a characteristic length scale of a few lattice spacing, as well as longer range inhomogeneities with characteristic length scales comparable to the width of the wire, $L_y$, or larger.  The short range inhomogeneities could be generated by impurities or by point defects present at the interface, while longer range inhomogeneities could be due to extended defects. In the absence of a detailed microscopic description of the SM--SC interface, it is difficult to estimate the amplitude of these fluctuations. Here we consider a phenomenological model of the interface and we assume that the tunneling matrix is given by
\begin{equation}
\widetilde{t}(i_x, i_y) =\widetilde{t}(i_y) + \Delta\widetilde{t}(i_x, i_y),
\end{equation}
where $\widetilde{t}(i_y)$ is the smooth component of the nonuniform coupling, e.g., the interface transparency with the profile shown in Fig. \ref{Fig3}, and $\Delta\widetilde{t}(i_x, i_y)$ represents the random component. To model the short range disorder, we coarse grain the interface in square patches of side length ${\it l}$ and assume that $ \Delta\widetilde{t}$ is uniform within a patch, but varies randomly from patch to patch with an amplitude $(\Delta\widetilde{t})_{\rm max}$, i.e., with $-(\Delta\widetilde{t})_{\rm max}\leq \Delta\widetilde{t}(i_x, i_y) \leq (\Delta\widetilde{t})_{\rm max}$. In the numerical calculations we considered patches of sizes ${\it l}= 20a$  and ${\it l}= 60a\approx Ly/4$ and an amplitude $(\Delta\widetilde{t})_{\rm max}= 0.25 \widetilde{t}(0)$, where  $ \widetilde{t}(0)$ is the maximum value of the smooth nonuniform  tunneling component shown in Fig. \ref{Fig3}. We note that these are extremely large fluctuations of $\widetilde{t}(i_x, i_y)$, larger that the minimum value of the smooth tunneling component, $\widetilde{t}(Ly)= 0.2\widetilde{t}(0)$, which result in significant variations of the local effective coupling $\gamma(i_x, i_y) \propto |\widetilde{t}(i_x, i_y)|^2$. Examples of short--range random couplings within the patch model are shown in Fig. \ref{Fig24}.

Before we present the results of the numerical calculations, we would like to emphasize the specific way that interface disorder enters the effective Hamiltonian. While the effect of  charged impurities can be included through a random potential, variations in the SM-SC coupling generate randomness in the effective SC order parameter, $\Delta_{n_y n_y^\prime}$, as well as  fluctuations of the renormalization matrix $Z^{1/2}$. From equations (\ref{Z12}) and (\ref{Deltann}) we notice that the short range fluctuations of $\widetilde{t}$ and, implicitly,  the short range fluctuations of $\gamma$, are significantly reduced when taking the matrix elements with the eigenstates $|n_y\rangle$.  As we are mainly  interested in systems with only a few occupied sub--bands,  integration over $y$ effectively averages out fluctuations with characteristic  length scales ${\it l}\ll L_y$.
\begin{figure}[tbp]
\begin{center}
\includegraphics[width=0.48\textwidth]{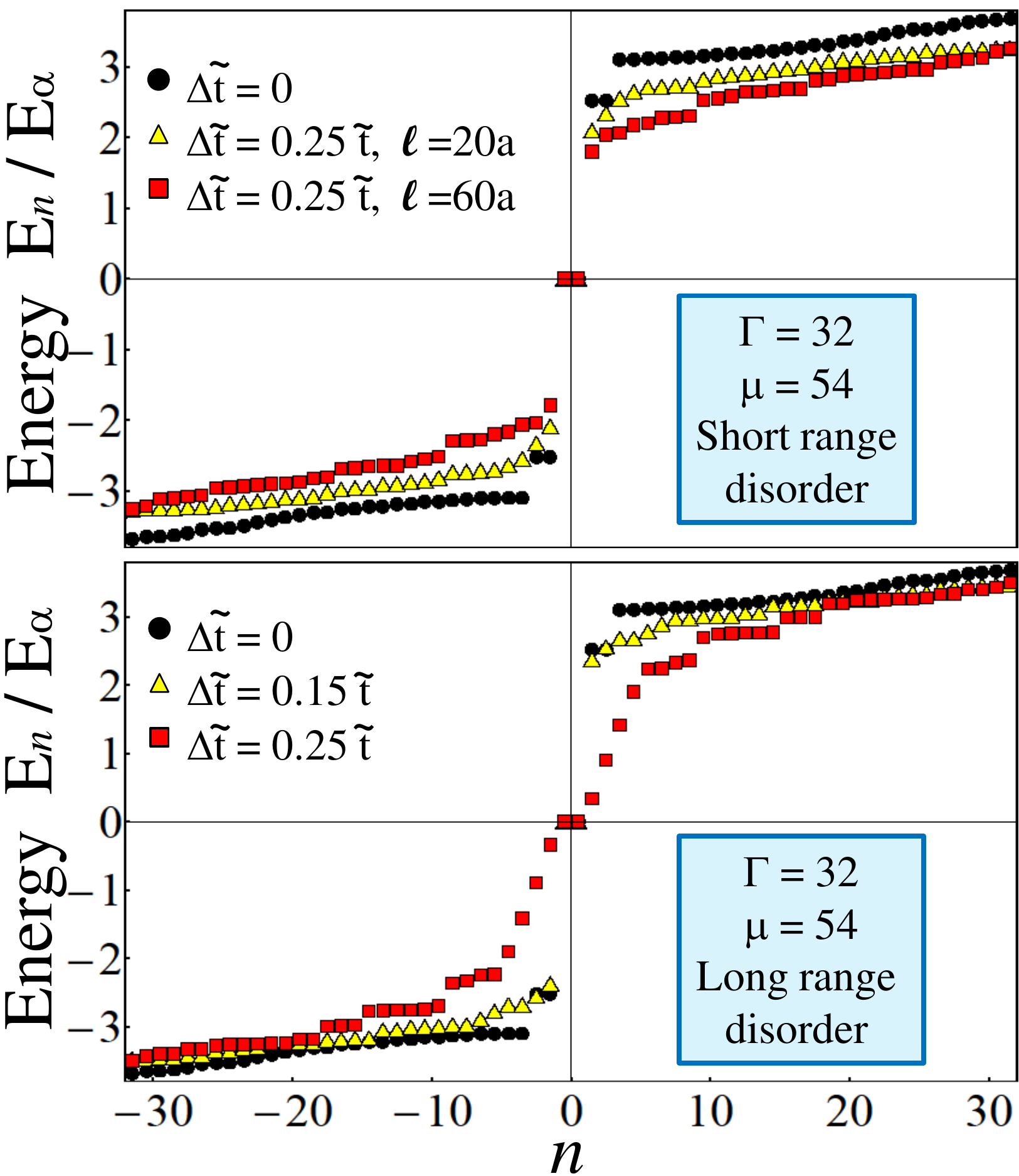}
\vspace{-7mm}
\end{center}
\caption{(Color online)  Spectra of  a nanowire with random SM--SC coupling. Upper panel: Nanowire with short range fluctuations of the SM--SC coupling. The random coupling is considered within the patch model described in the text and corresponds to the distributions shown in Fig. \ref{Fig24}. Lower panel: Nanowire with long--range SM--SC coupling fluctuations. The variations of $\widetilde{t}$ along the wire have a profile as shown in the upper panel of Fig. \ref{Fig15} and an amplitude $\Delta\widetilde{t}$. Note that for $\Delta\widetilde{t}=0.25\widetilde{t}(y=0)$ the gap collapses. } \label{Fig25}
\end{figure}
As an illustration of this property, we consider the case of random coupling with ${\it l}= 60 a$ and $(\Delta \widetilde{t})_{\rm max}=0.25\widetilde{t}(0)$ for a nanowire with $\mu = 54.5E_\alpha$, $\Gamma=32E_\alpha$,  and $\overline{\gamma}=\Delta_0$. In spite of the relatively large length scale of the fluctuation, ${\it l} = Ly/4$, the variations of the induced gap $\Delta_{11}(x)$ along the wire are only of the order of $10\%$ of the average value. The dependence  of the induced gap $\Delta_{11}$ on the position along the wire is shown in the lower panel of Fig. \ref{Fig18}.  The amplitude of the $\Delta_{11}$ fluctuations is further reduced if we consider shorter range coupling fluctuations. A similar behavior characterizes the renormalization matrix $Z^{1/2}$.  In addition,  as the low--energy properties of the system are determined by single particle states with small wave vectors $k_x$, we expect a further reduction of the effect of short range fluctuations as a result of the integration over $x$. In particular, if the clean, infinite wire is characterized by a maximum Fermi wave vector $k_F$, we expect the low-energy physics to be insensitive to random variations of the SM--SC coupling with characteristic length--scales ${\it l}<1/k_F$.

The effects of a random SM--SC coupling on the low--energy spectrum of the nanowire are illustrated in Fig. \ref{Fig25}. Remarkably, short range fluctuations with amplitudes up to $25\%$ of the average coupling at $y=0$ (i.e., the maximum value of the coupling in a nonuniform profile - see Fig. \ref{Fig3}) do not destroy the topological SC phase. The relatively weak effect of these strong fluctuations is due to the implicit averaging involved in the calculation of the matrix elements between single particle states with low wave vectors. Increasing the characteristic length ${\it l}$ makes this type of disorder more effective, as evident from the upper panel of Fig. \ref{Fig25}. Hence the natural question: what is the effect of long--range SM--SC coupling fluctuations? We consider a smooth variation of $\widetilde{t}$ along the wire with a profile as shown in the upper panel of Fig. \ref{Fig15}. An amplitude of these fluctuations equal to $25\%$ of $\widetilde{t}(y=0)$ results in the collapse of the  gap (see Fig. \ref{Fig25}, lower panel). However, the topological phase is robust against long range coupling fluctuations with amplitudes smaller than $20\%$. We note that long range variations of the coupling strength could result from the engineering process of the nonuniform interface. Limiting the amplitude of these fluctuations below a certain limit of about $10-15\%$ of the maximum coupling strength should be a priority of the experimental effort for realizing a topological SC state using semiconductor nanowires.

\begin{figure}[tbp]
\begin{center}
\includegraphics[width=0.48\textwidth]{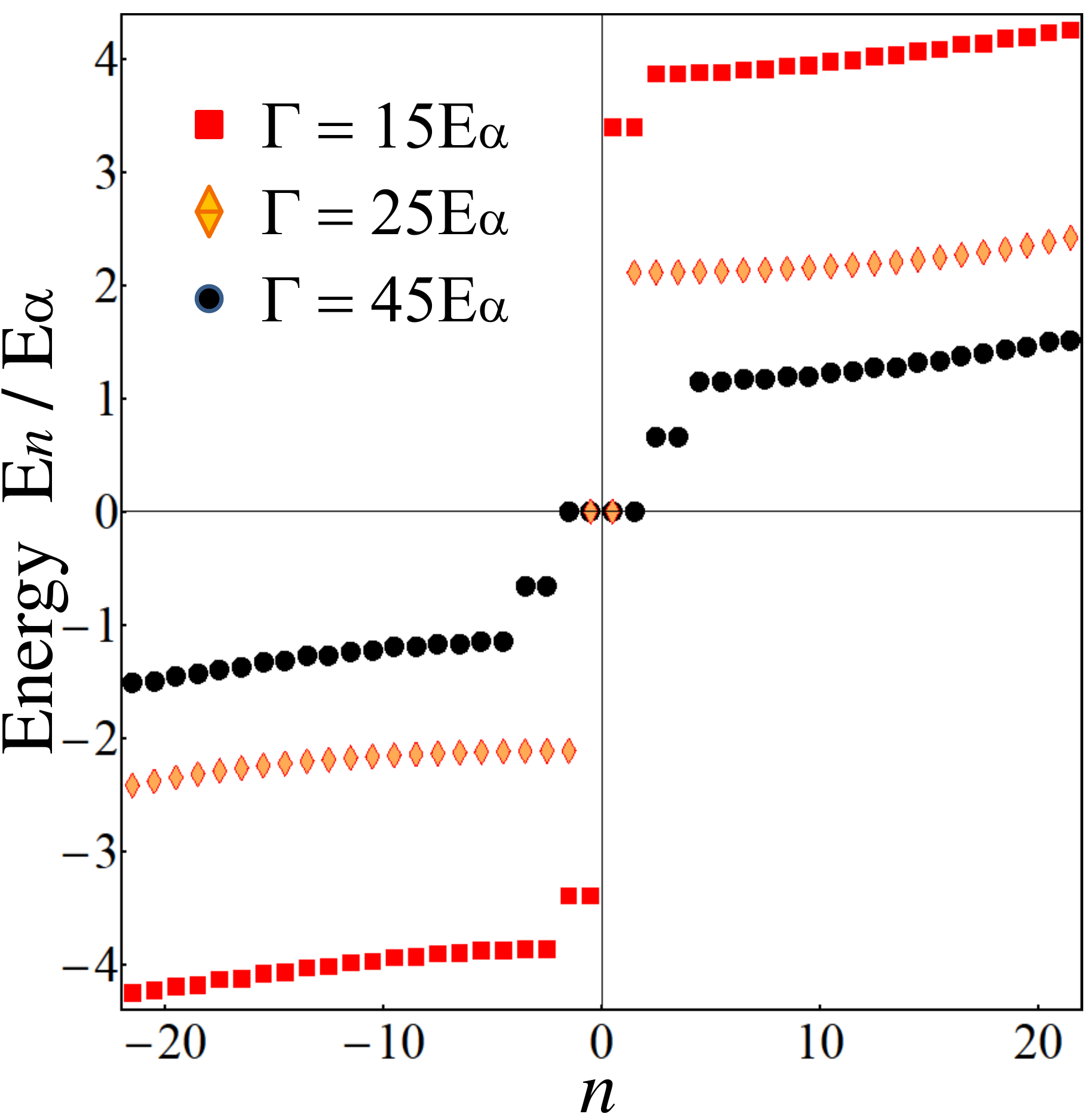}
\vspace{-7mm}
\end{center}
\caption{(Color online)  Typical spectra for a trivial SC with $N=0$ (red squares), a topological SC with $N=1$ (orange diamonds), and a trivial SC with $N=2$ (black circles). The system is characterized by $\mu = 54.5E_\alpha$,  $\theta=0.8$, and $\overline{\gamma}=\Delta_0$. The finite energy in--gap states  (for $Gamma=15E\alpha$ and $\Gamma=45E_\alpha$) together with the Majorana zero--modes are localized near the ends of the wire (see also Fig. \ref{Fig12}), while the rest of the states extend throughout the entire system. The corresponding LDOS is shown in Fig. \ref{Fig27}. } \label{Fig26}
\end{figure}

\section{Experimental signatures of Majorana bound states}\label{Sec:experimental}

Probing unambiguously the presence of Majorana bound states in the superconducting nanowire represents a critical task. In this section we show that local spectral measurements provide a simple and effective tool for accomplishing this task. We focus on the local density of states (LDOS), which could be measured using, for example,  scanning tunneling spectroscopy (STS),  and on the differential conductance associated with tunneling into the ends of the wire. We establish that these measurements should suffice in establishing the existence of the zero-energy Majorana edge modes in semiconductor nanowires.

\subsection{Local density of states in superconducting nanowires}

In the previous section we have shown that disorder generates low--energy states and reduces the mini--gap. Nonetheless, a small mini--gap does not implicitly mean that the Majorana bound state cannot be resolved in a spectroscopic measurement. The key observation is that the undesirable  low--energy states are generally localized near impurities and defects (see, foe example, Fig. \ref{Fig19}). A local measurement could easily distinguish between a zero--energy state localized at the end of the wire and a low--energy state localized somewhere inside the wire. To clarify the question regarding the real space distribution of the low--energy spectral weight, we calculate the local density of states (LDOS) for several relevant regimes and compare the LDOS of a clean ideal system with that of a disordered realistic nanowire.

\begin{figure}[tbp]
\begin{center}
\includegraphics[width=0.48\textwidth]{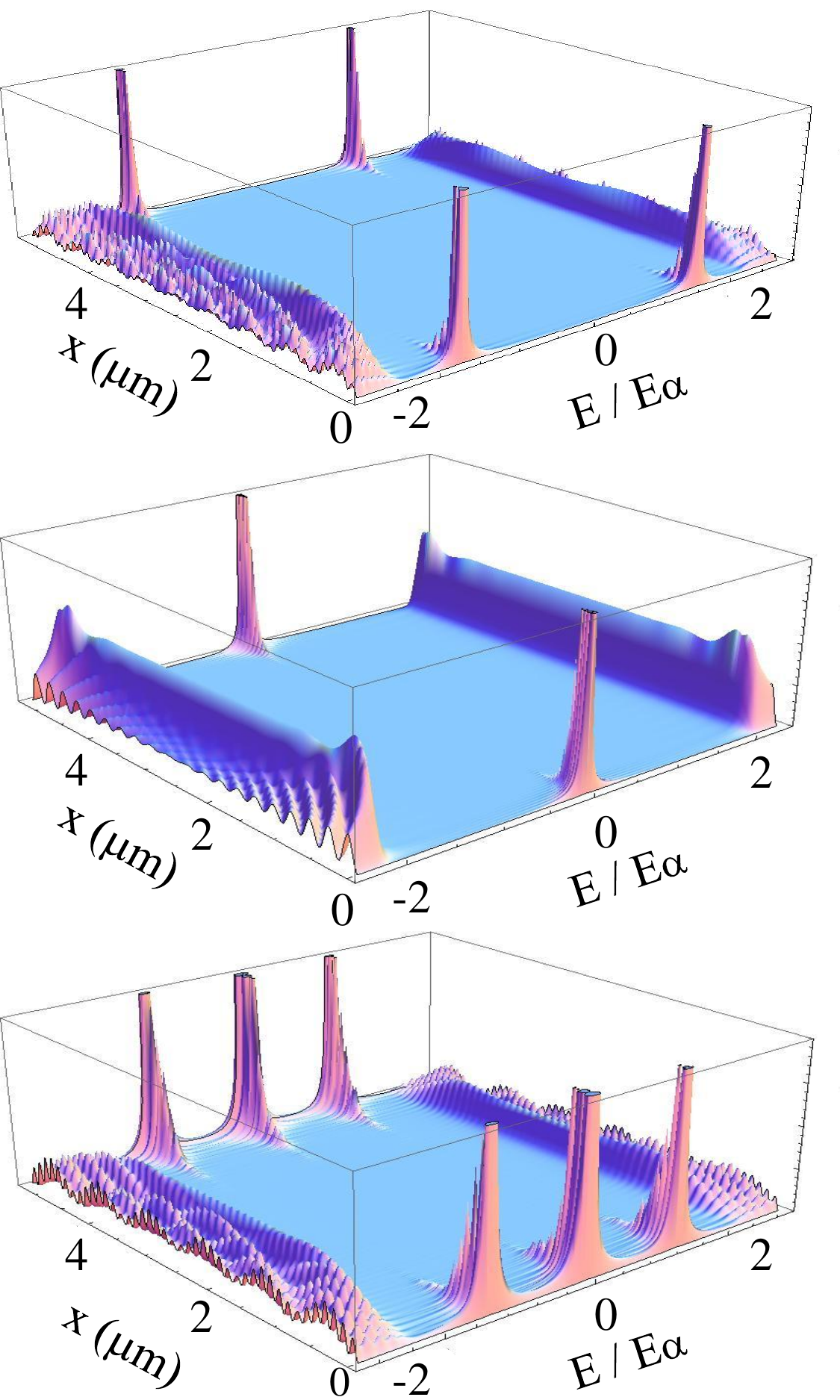}
\vspace{-7mm}
\end{center}
\caption{(Color online)  LDOS for a clean nanowire in three different phases:  trivial SC phase with $N=0$ ($\Gamma=15E_\alpha$, top),  topological SC phase with $N=1$ ($\Gamma=25E_\alpha$, middle), and  trivial SC phase with $N=2$ ($\Gamma=45E_\alpha$, bottom). The corresponding spectra are shown in Fig. \ref{Fig26}.  Notice the finite energy in--gap states localized near the ends of the wire (top and bottom) and the zero--energy Majorana modes (middle and bottom). The weight of the zero--energy modes in the $N=2$ phase (bottom) is twice the weight of the Majorana modes in the topological SC phase with $N=1$ (middle). However,  the clearest distinction can be made between the $N=0$ and the $N=1$ phases. The LDOS is integrated over the transverse coordinates $y$ and $z$.} \label{Fig27}
\end{figure}

We start with a clean  nanowire with three occupied sub--bands ($\mu = 54.5E_\alpha$ ) coupled non--uniformly to an s-wave superconductor. The coupling parameter is characterized by a non--homogeneity factor $\theta=0.8$ and an intermediate coupling strength $\overline{\gamma}=\Delta_0$, where $\Delta_0$ is the SC order parameter of the superconductor. In the absence of a Zeeman field ($\Gamma=0$), the nanowire is a trivial superconductor. Increasing the Zeeman field $\Gamma$ above a critical value induces a transition from the trivial SC phase with $N=0$ (no Majorana modes) to a topological SC phase with $N=1$ (one pair of Majorana modes). Further increasing $\Gamma$  drives the system though a series of alternating phases with trivial ($N$ even) and non-trivial ($N$ odd) topologies (see the phase diagram in Fig. \ref{Fig9}). Of major practical interest are the first two phases ($N=0$ and $N=1$), as stronger Zeeman fields involve smaller gaps (see Fig. \ref{Fig10}) or may destroy superconductivity altogether. Typical spectra from the first three phases ($N=0, 1, 2$) are shown in Fig. \ref{Fig26} and the corresponding LDOS is shown in Fig. \ref{Fig27}.

The main conclusion suggested by the results shown in Fig. \ref{Fig27} is that clear--cut  evidence for the existence of the Majorana zero modes can be obtained by driving the system from a trivial SC phase with $N=0$ to a topological SC state with $N=1$ by tuning the Zeeman field. In the trivial SC phase there is a well--defined gap for all excitations, including states localized near the ends of the wire. By contrast, the topological SC phase is characterized by sharp zero--energy peaks localized near the ends of the wire and separated from all other excitations (including possible localized in--gap states) by a well--defined mini--gap.

\begin{figure}[tbp]
\begin{center}
\includegraphics[width=0.48\textwidth]{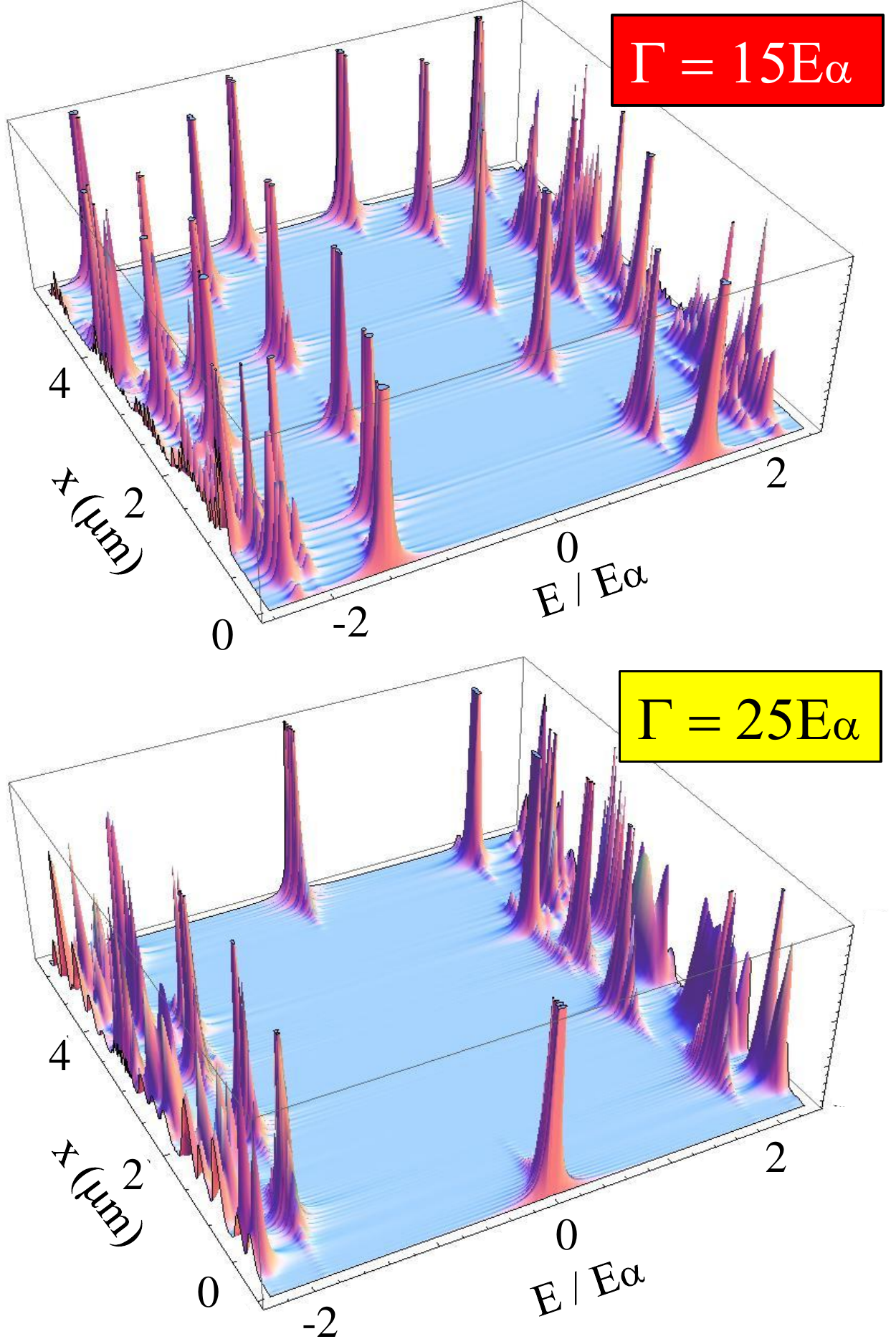}
\vspace{-7mm}
\end{center}
\caption{(Color online)  LDOS for a nanowire with charged impurities. The top picture corresponds to a trivial SC state with $N=1$, while the bottom picture  is for a system with $N=1$. The linear  impurity density is $n_imp=7/\mu m$. Notice that all the low-energy states are strongly localized, but the clear--cut distinction between the two phases holds. } \label{Fig28}
\end{figure}

Is it possible to clearly distinguish the two phases with different topologies in the presence of disorder? The answer is provided by the results shown in Fig. \ref{Fig28} for a nanowire with charged impurities. In contrast with the clean case, all the low-energy states are strongly localized. Nonetheless, the signature features of the two phases (the gaps and the zero--energy peaks) are preserved. At this point we emphasize two critical properties:
\begin{figure}[tbp]
\begin{center}
\includegraphics[width=0.48\textwidth]{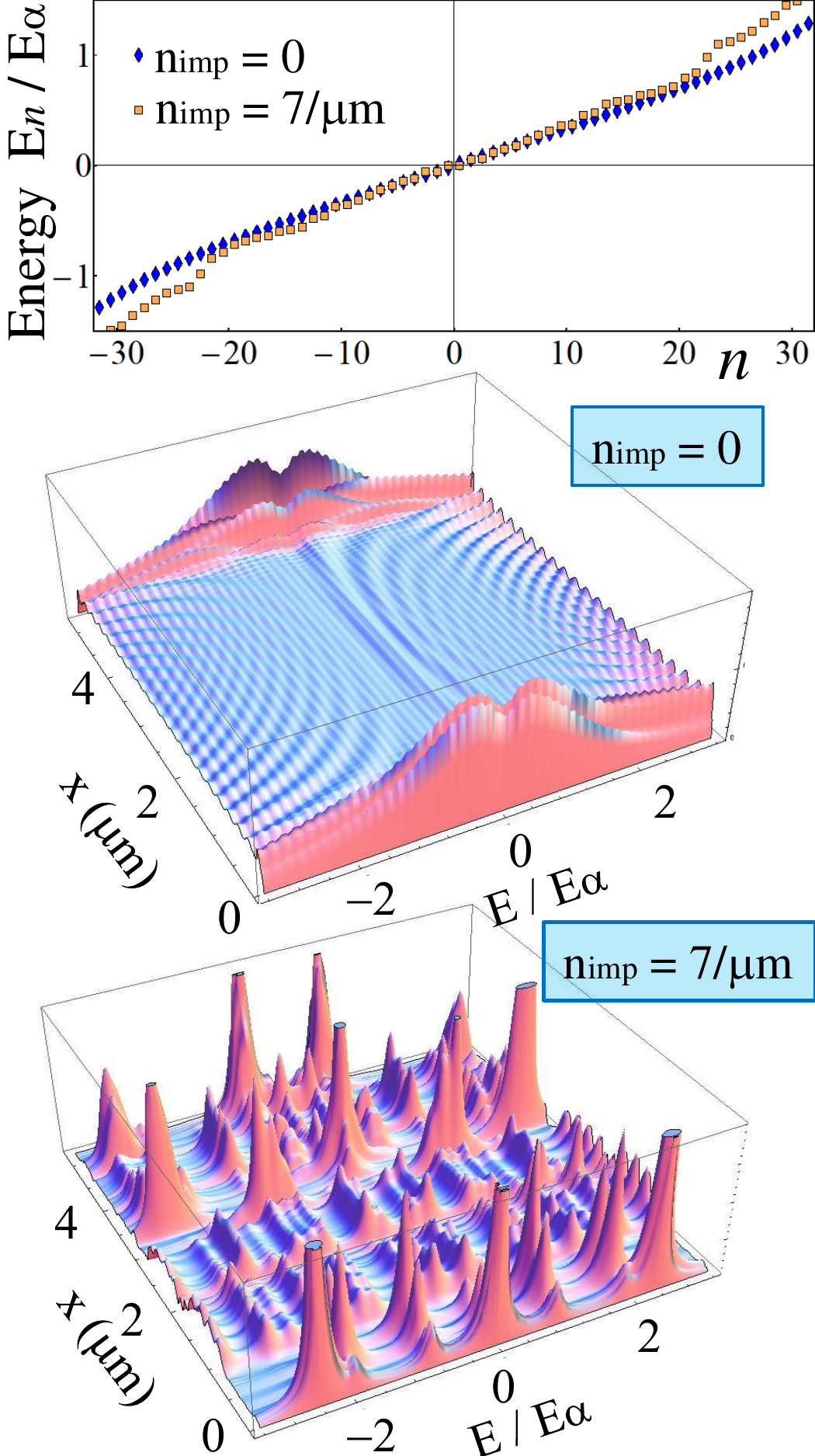}
\vspace{-7mm}
\end{center}
\caption{(Color online)  Energy spectrum and LDOS in the vicinity a topological phase transition. The top panel shows the spectra of a system with $\Gamma=21.5E_\alpha$ without disorder (blue diamonds) and in the presence of charged impurities ($n_{\rm imp}=0.7/\mu m$, orange squares).  Note the absence of a gap. The corresponding LDOS are shown in the middle (clean system) and bottom (disordered wire) panels. The spectral weight is distributed over the entire energy range, including at positions near the ends of the wire. } \label{Fig29}
\end{figure}
 i) The features illustrated in Fig. \ref{Fig28} are generic, i.e., they do not depend on the type or the source of disorder. Similar LDOS can be generated using any other significant type of disorder discussed in the previous section, or combinations of different types of disorder. ii) Observing a zero--energy peak at  a certain value of the Zeeman field  does not by itself prove  the realization of a topological SC phase. The trivial SC state with $N=2$ may also have a zero--energy peak separated from all other excitations by a mini-gap. To clearly identify the $N=1$ phase one must measure the LDOS as a function of the Zeeman field starting from $\Gamma=0$, i.e., from the trivial SC phase with $N=0$. Continuously increasing $\Gamma$ will generate a transition from a phase characterized by a well--defined gap to a phase with strong zero--energy peaks localized near the ends of the wire. But what is the signature of the transition?

Figure \ref{Fig29} shows the spectrum and the LDOS of a system  with a Zeeman field $\Gamma=21.5E_\alpha$. In a clean wire this corresponds to the transition between the $N=0$ and $N=1$ phases, which is marked by the vanishing of the quasi--particle gap, as shown in figures \ref{Fig10} and \ref{Fig12}. The LDOS is characterized by a distribution of the spectral weight over the entire low--energy range of interest. This property holds at any position along the wire. Adding disorder induces localization, but does not change this key property. In fact, based on the analysis of the results shown in Fig. \ref{Fig22}, we know that in disordered systems this type of critical behavior will characterize a  finite range of Zeeman fields. Observing the transition between the topologically trivial and nontrivial phases, which is characterized by the closing of the gap and by a spectral weight distributed over a wide energy range, is the final ingredient necessary for unambiguously  identifying the Majorana bound states using LDOS measurements.

\subsection{Tunneling differential conductance}

An ideal type of measurement that exploits the properties identified in the previous subsection consists of tunneling into the ends of the wire and measuring the differential conductance~\cite{Yamashiro_PRB'97, Bolech_PRL'08, Tewari_PRL'08, Nilsson_PRL'08, Law_PRL'09}. To a first approximation, $dI/dV$ is proportional to the local density of states at the end of the wire, so the general discussion presented above should apply. Here, we focus on certain specific aspects of a tunneling experiment, e.g., the specific form of the tunneling matrix elements and the role of finite temperature, that may limit the applicability of our conclusions. We find that values of the parameters consistent with an unambiguous identification of the Majorana bound state in the semiconductor nanowire are well within a realistic  parameter regime.

The tunneling current between a metallic tip and the nanowire can be evaluated within the Keldysh non-equilibrium formalism~\cite{Berthod'11}. In terms of real space Green's functions we have
\begin{widetext}
\begin{align}
I& = \frac{e}{h} \int d\omega {\rm Re} {\rm Tr} \left\{ \left[ 1 - G_0^R(\omega) \Gamma^R(\omega-eV)\right]^{-1}\left[(1-2f_{\omega-eV}) G_0^R(\omega)\Gamma^R(\omega-eV) \right.\right. \label{Itunnel}\\
&\left.\left.+2(f_{\omega-eV}-f_{\omega}) G_0^R\Gamma^A(\omega-eV)-(1-2f_\omega)G_0^A(\omega)\Gamma^A(\omega-eV)\right] \left[1 - G_0^A(\omega) \Gamma^A(\omega-eV)\right]^{-1}\right\}, \nonumber
\end{align}
\end{widetext}

\begin{figure}[tbp]
\begin{center}
\includegraphics[width=0.48\textwidth]{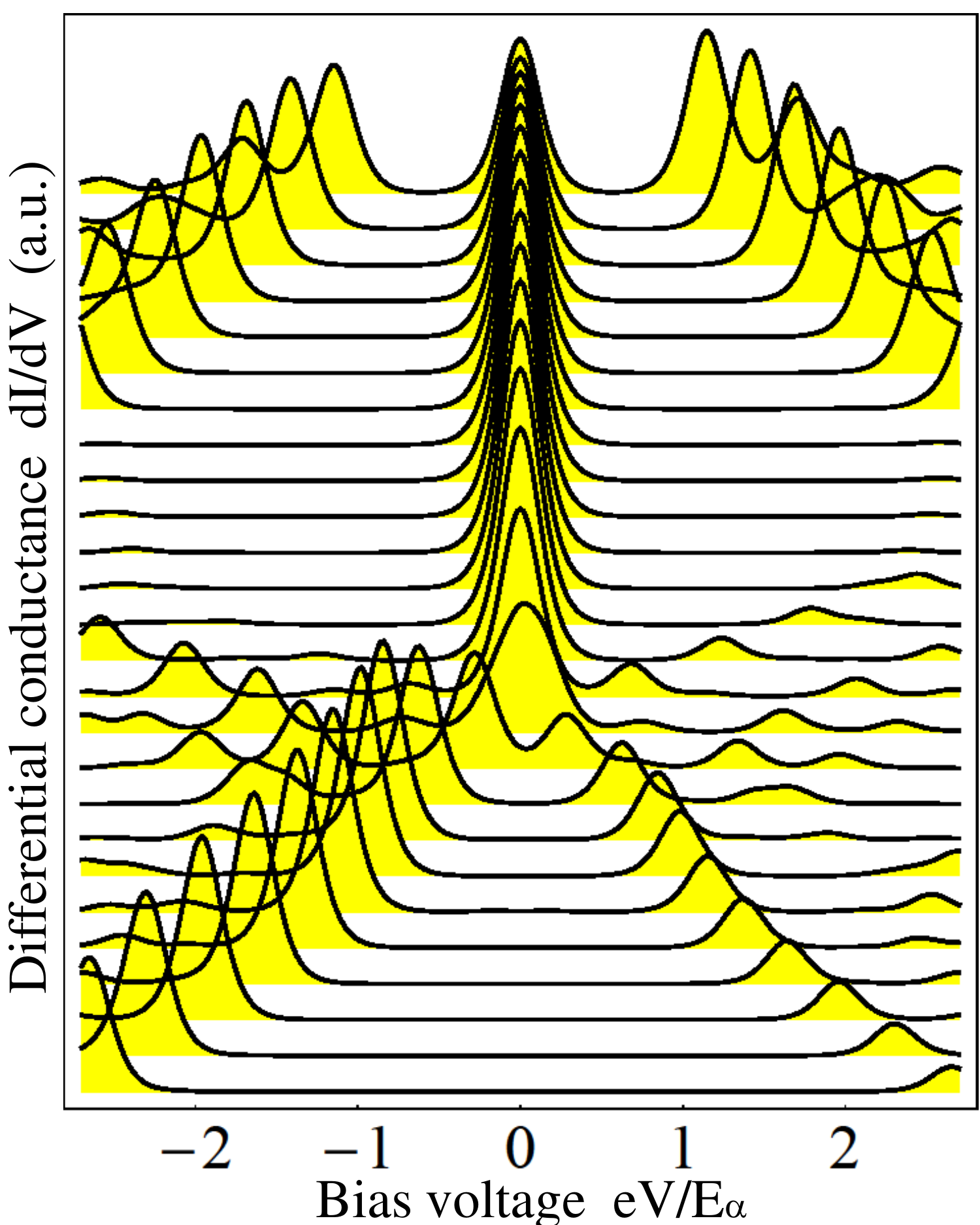}
\vspace{-7mm}
\end{center}
\caption{(Color online)  Differential conductance for tunneling into the end of a superconducting nanowire. The curves correspond to different values of the Zeeman field ranging from $\Gamma=11E_\alpha$ (bottom) to $\Gamma=36E_\alpha$ (top) in steps of $E_\alpha$. The curves were shifted vertically for clarity. The trivial SC phase ($\Gamma<21E_\alpha$)is characterized by a gap that vanishes in the critical region ($\Gamma\approx21E_\alpha$). The signature of the topological phase is the zero-energy peak resulting from tunneling into the Majorana mode. The differential conductance was calculated at a temperature $T\approx 50 mK$ for a disordered wire with a linear density of charged impurities $n_{imp}=7/\mu m$.} \label{Fig30}
\end{figure}

\noindent where $V$ is the bias voltage applied between the tip and the nanowire and $f_\omega=1/(e^{\beta \omega}+1)$ is the Fermi-Dirac distribution function corresponding to a temperature $k_bT=\beta^{-1}$. The retarded (advanced) Green's function for the nanowire has the expression
\begin{equation}
G_0^{R(A)}({\bm r}, {\bm r}^\prime, \omega) = \sum_{n} \left\{ \frac{u_{n}^*({\bm r}) u_{n}({\bm r}^\prime)}{\omega -E_n \pm i\eta} + \frac{v_{n}^*({\bm r}) v_{n}({\bm r}^\prime)}{\omega +E_n \pm i\eta} \right\},
\end{equation}
where $u_{n}$ and  $v_{n}$ are the particle and hole components of the wave function corresponding to the energy $E_n$. The wave functions and the energies are obtained by diagonalizing the effective BdG Hamiltonian for the nanowire, including the contibution from disorder, as described in the previous sections. The matrices  $\Gamma^{R(A)}$ contain information about the tip and the tip--nanowire coupling. Specifically, we have
\begin{equation}
\Gamma^{R(A)}({\bm r}, {\bm r}^\prime, \omega) = \gamma_{{\bm r}, {\bm r}^\prime}\int dx \frac{\nu(x)}{\omega-x\pm i \eta},
\end{equation}
where $\nu(x)$ represents the density of states of the metallic tip and $\gamma_{{\bm r}, {\bm r}}$ depends on the tunneling matrix elements between the tip and the wire. We note that in Eq. (\ref{Itunnel}) the trace is taken over the position vectors. We consider a tunneling model in which the amplitude of the tunneling matrix elements vary exponentially with the distance from the metallic tip. Specifically, we have  $\gamma_{{\bm r}, {\bm r}} \gamma_0 \theta_{\bm r}\theta_{{\bm r}^\prime}$, where $\gamma_0$ gives the overall  strength of the tip--nanowire coupling and the position--dependent factor is
\begin{equation}
 \theta_{\bm r} = e^{-\frac{1}{\xi}\left[\sqrt{(x-x_{tip})^2+(y-y_{tip})^2+(z-z_{tip})^2} - x_{tip} \right]},
\end{equation}
with $( x_{tip},  y_{tip},  z_{tip})$ being the position vector for the tip and $\xi$ a characteristic length scale associated with the exponential decay of the tip--wire coupling. In the numerical calculations we take $\xi=0.4 a$ and  $( x_{tip},  y_{tip},  z_{tip})=(-3a, L_y/2, L_z/2)$,  i.e., the tip is is at a distance equal with three lattice spacings away from the end of the wire.
With these choices, the differential conductance becomes
\begin{equation}
\frac{dI}{dV} \propto -\sum_{n}\left[f^{\prime}(E_n-eV)|\langle u_n|\theta\rangle|^2 + f^\prime(-E_n-eV)|\langle v_n|\theta\rangle|^2\right],
\end{equation}
where the matrix elements $\langle u_n|\theta\rangle$ and $\langle v_n|\theta\rangle$ involve summations over the lattice sites of the nanowire system and provide the amplitudes for tunneling into specific states. Finite temperature effects are  incorporated through the derivatives of the Fermi--Dirac function, $f^\prime$.

The dependence of the tunneling differential conductance on the bias voltage for a superconducting nanowire with disorder is shown in Fig. \ref{Fig30}. Different curves correspond to different values of the Zeeman field between $\Gamma=11E\alpha$ (bottom) and  $\Gamma=36E\alpha$ (top) and are shifted vertically for clarity. The temperature used in the calculation is $50mK$, a value that can be easily reached experimentally. Lower temperature values will generate sharper features, but the overall picture remains qualitatively the same. Note that the closing of the gap in the critical region between   the trivial SC phase and the topological SC phase can be clearly observed. In this region $dI/dV$ has features over the entire low-energy range, as discussed in the previous subsection. The Majorana bound state at $\Gamma>22E\alpha$ is clearly marked by a sharp peak at $V=0$, separated by a gap from other finite energy features. We conclude that measuring the tunneling differential conductance can provide a clear and unambiguous probe for Majorana bound states in semiconductor nanowires.

\section{Conclusions}\label{sec:discussion}

In conclusion, we have developed a comprehensive theory for the realization and the observation of the emergent non-Abelian Majorana mode in semiconductor (e.g.,  InAs, InSb) nanowires proximity coupled to an ordinary s--wave superconductor (e.g., Al, Nb) in the presence of a Zeeman splitting induced by an external magnetic field.  The importance of our work lies in the thorough investigation of the experimental parameter space, which is required in order to predict the optimal parameter regime to search for the Majorana mode in nanowires.  Since the number of possible physical parameters in the problem is large (e.g.,  electron density in the nanowire, geometric size of the wire, chemical potential, the strength of spin-orbit coupling, the superconducting gap, the hopping matrix elements between the semiconductor and the superconductor, the strengths of various disorder in the semiconductor, or in the superconductor, or at the interface between them), theoretical guidance, as provided in this work, is highly desirable for the success of the experimental search for the Majorana fermion in solid--state systems.  Aside from the obvious conclusions (e.g.,  strong spin-orbit coupling and large Lande $g$-factor in the semiconductor, large superconducting gap in the superconductor, and  low disorder everywhere stabilize the topological phase), we have discovered several unexpected results.  In particular, we find somewhat surprisingly that the strict one-dimensional limit with purely one--subband occupancy for the nanowire, as originally envisioned by Kitaev~\cite{Kitaev'01} and later used by many researchers~\cite{Lutchyn'10, Oreg'10, Aliceaetal'10}, is not only unnecessary, but is in fact detrimental to creating Majorana modes. In the presence of disorder, the optimal system should have a few ( 3-5) occupied subbands in the nanowire for the creation of the Majorana modes with maximal stability.  This is, of course, great news from the experimental perspective, because fabricating strictly 1D semiconductor nanowires with pure one subband occupancy is a challenging task.  Another important result of our analysis is the relative immunity of the  Majorana modes to the presence of disorder in the system.  The most dangerous disorder mechanisms are due to charged impurity centers in the semiconductor and to inhomogeneous hopping across the semiconductor-superconductor interface. Our work suggests that the optimal nanowires for observing the Majorana mode should not only have as little charge impurity disorder  as possible in the semiconductor, but they should also have a thin insulating layer separating the semiconductor and superconductor (so that the tunneling across the interface is not too strong), as well as some non--uniformity in the tunneling amplitude between the semiconductor and the superconductor across the width (but not the length) of the nanowire. Our detailed numerical calculations establish that the zero--energy Majorana modes should clearly show up in experiments,  even in the presence of considerable disorder in the nanowires.  In addition, we establish that the disorder in the superconductor has little effect on the Majorana mode in the nanowire.  Another salient aspect of our work is the detailed calculation of the expected tunneling spectroscopy spectra for observing the Majorana mode in the nanowire using realistic physical parameters.  Our results establish that the predicted topological quantum phase transition between the trivial phase with no Majorana mode to the topological phase with a well-defined zero energy Majorana bound state should be clearly observable as a striking zero-bias anomaly in the tunneling current when the Zeeman splitting is tuned through the quantum critical point separating the two phases.  More importantly, we calculate realistic tunneling spectra in the presence of uncontrolled spurious bound states in the system which are invariably present in real samples due to the localized random impurities in the semiconductor environment, clearly showing how to discern the topological features associated with the Majorana bound states from the background of contributions due to  trivial bound states caused by impurities in the system.

Our work  emphasizes the tunneling measurements, which would directly establish the existence of a robust zero energy mode in the system, providing the necessary condition for the existence of the Majorana fermion.  What we have done here is to develop a detailed theory for the existence of a topological phase in the semiconductor-superconductor heterostructure, by taking into account essentially all of the relevant physical effects.  Once the presence of a robust zero energy mode is established, hence the necessary condition for the existence of the Majorana is realized, one must move on to establish the sufficient condition, which would obviously be a harder task.  Several ideas for establishing definitively the existence of the Majorana mode (and its non-Abelian braiding statistics nature) have already been suggested in the literature, including experiments involving the fractional Josephson effect~\cite{Lutchyn'10, Kitaev'01, FuKane'09, Yakovenko'03, Aliceaetal'10}, the quantized differential conductance~\cite{Law_PRL'09, Wimmer'11}, and Majorana interferometry~\cite{FuKanePRL'11, Akhmerov_prl'09, Sau_prb'11}.

Our work establishes the realistic likelihood of the existence within laboratory conditions of the non-Abelian Majorana zero--energy mode in spin--orbit interacting semiconductor nanowires  proximity--coupled to ordinary superconductors.  We also establish that tunneling spectroscopy is  one of the easiest techniques to directly observe the elusive Majorana in realistic solid--state systems.  Greater challenges, such as carrying out topological quantum computation, lie ahead once the laboratory existence of the Majorana mode is established experimentally.

\begin{acknowledgments}

We would like to thank Leonid Glazman, Matthew Fisher and Chetan Nayak for discussions. This work is supported by the DARPA QuEST, JQI-NSF-PFC, Microsoft Q (SDS) and WVU startup funds (TS).

\end{acknowledgments}

\end{document}